\documentclass[oneside]{article}
\RequirePackage{amssymb,amsmath}
\RequirePackage{eufrak}
\RequirePackage{verbatim}
\RequirePackage[cp1251]{inputenc}
\RequirePackage[T2A]{fontenc}
\RequirePackage[english,russian]{babel}

\RequirePackage{graphicx}
\RequirePackage{ifpdf}
\ifpdf 
 \fi
\begin{document}

\oddsidemargin=0pt
\evensidemargin=1.4truecm
\voffset=0cm
\textwidth 147mm
\textheight 215mm
\renewcommand{\baselinestretch}{1.1}

\renewcommand \thesubsection {\arabic{section}.\arabic{subsection}.}
\renewcommand \thesubsubsection {\arabic{section}.\arabic{subsection}.\arabic{subsubsection}.}

\renewcommand \thesection {\thechapter\arabic{section}.}

\sloppy
\renewcommand{\thefootnote}{\fnsymbol{footnote}} 

\def\makeEngTit{\selectlanguage{english}
\medskip
\@Engtit
\nopagebreak
\@authorEng

\vskip 1mm
\hrule \vskip 1mm
\@englishabstract
\medskip
\@englishkeywords
}

\title{ О преобразовании Лоренца-Мёллера-Нэлсона в жёсткую \\неинерциальную систему отсчёта }
\author{В. В. Войтик\thanks {Башкирский Государственный
Педагогический Университет,
Октябрьской Революции 3-а, Уфа, 450077,
Россия,  voytik1@yandex.ru}}

\maketitle
\begin{abstract}
С помощью специального преобразования Лоренца-Мёллера-Нэлсона (ЛМН) получен закон преобразования скорости из лабораторной системы $S$ в произвольную поступательно двигающуюся жёсткую систему отсчёта $s$. Указан физический смысл параметра входящего в специальное преобразование ЛМН. Для случая малых собственных расстояний и плавного движения без рывков предложено обратное специальное преобразование ЛМН. Прямыми следствиями этого преобразования являются: а)рассинхронизация в системе $s$ координатных часов этой системы отсчёта, предварительно синхронизированных в лабораторной системе $S$ и б)сокращение Лоренца в $S$ линейки системы $s$. Также рассмотрены неоднородность движения точек системы координат $s$ и эффект неоднородности хода физического времени в системе $s$ с точки зрения лабораторной системы $S$. Получены уравнения, касающиеся матрицы поворота. Найден прямой и обратный закон преобразования аффинной скорости из $S$ в систему отсчёта $k$, которая сопутствует $s$, но не вращается. Показано, что собственное вращение осей $s$, рассматриваемое относительно $S$ не является жёстким. Также показано, что при неинерциальном движении локальной жёстко вращающейся системы отсчёта кинематическая деформация её системы координат относительно $S$ отсутствует в двух плоскостях. Рассмотрено применение преобразования аффинной скорости к жёсткому вращению тела. Вычислены локальные угловая скорость и скорость растяжения главных осей. Доказывается форминвариантность и единственность преобразования ЛМН. Выяснена тесная связь собственных прецессии Томаса и вращения Вигнера. Рассмотрена 4-мерная формулировк преобразований ЛМН. Найдены дифференциальные уравнения для обратной задачи релятивистской кинематики и их решение для случая постоянного собственного ускорения. Показано отличие равноускоренного движения от гиперболического. Двумя эквивалентными способами вычислены параметры равноускоренной системы отсчёта. Показана взаимная компенсация собственной прецессии Томаса и собственного вращения Вигнера для случая равноускоренного движения. Также указаны основные формулы обратного преобразования ЛМН выраженные через параметр, входящий в уравнения для обратной задачи кинематики.
\end{abstract}
 {Ключевые слова: \itshape обобщённое Лоренц-преобразование, преобразование Лоренца-Мёллера-Нэлсона, аффинное  движение, локальная угловая скорость, тензор скорости деформации, парадокс часов, парадокс Эренфеста, собственная прецессия Томаса, собственное вращение Вигнера, матрица поворота, собственное ускорение, собственная угловая скорость.}

\section*{Введение}

Неинерциальная система отсчёта 
\footnote[2]{Обычно под системой отсчёта в общей теории относительности (ОТО) понимается система тел (вместе со связанными с каждым из них произвольно идущими часами) заполняющих пространство наподобие некоторой среды \cite[с. 297]{1}). В этой статье система отсчёта понимается в самом раннем смысле специальной теории относительности (СТО)– как совокупность тел, покоящихся относительно наблюдателя, причём относительно друг друга (в направлении, перпендикулярном радиальному направлению от наблюдателя) эти тела не обязательно покоятся. Система координат предполагается прямоугольной. Точка, в которой покоится наблюдатель считается началом и полностью определяет все свойства жёсткой системы отсчёта. Таким образом говорят о единой системе отсчёта наблюдателя. При изменении положения наблюдателя в рамках системы координат первоначальной системы отсчёта из начала до удалённой точки, его система отсчёта изменяется за счёт изменения её собственного ускорения и собственной угловой скорости.} являющаяся жёсткой 
\footnote[3]{Обычно жёсткостью в ОТО называется свойство стационарности пространственной метрики неинерциальной системы отсчёта, т.е. её постоянство во времени в любой точке системы. Жёсткой при таком определении в СТО (по Борну) может быть только а) произвольно ускоренная система отсчёта, либо б) равномерно ускоренная и равномерно вращающаяся (т.е. стационарная) система отсчёта. В том случае, если состояние движения такой системы отсчёта изменяется, то пространственная метрика также меняется и система отсчёта теряет свойство жёсткости. В этой статье жёсткость системы отсчёта понимается как свойство сохранения расстояний по отношению к выделенной точке системы отсчёта: её началу (радиально жёсткая система отсчёта). Если это условие выполняется, то относительно других точек система отсчёта вообще говоря будет нежёсткой. Это приводит к расширению класса исследуемых систем отсчёта на (радиально) жёсткие произвольно ускоренные и произвольно вращающиеся (нестационарные) системы отсчёта. При этом следует учесть, что радиально жёсткая система отсчёта, несмотря на то, что в СТО нет идеально жёстких тел, всё-таки имеет право на существование, поскольку для любого закона движения её начала отсчёта относительно инерциальной системы всегда можно указать такой закон движения её других точек, что собственное расстояние от этой точки до начала отсчёта будет сохраняться.}
в СТО обычно изучается с помощью множества  методов. Один из них, очень популярный способ изучения жёсткой ускоренной системы отсчёта, использующийся вплоть до настоящего времени, заключается в том, что такая система представляется в виде бесконечной последовательности мгновенно сопутствующих инерциальных систем отсчёта. Предположение о такой эквивалентности называется гипотезой локальности (например \cite {2}), которая в её применении к линейкам неинерциальной системы отсчёта известна также как гипотеза линеек (например \cite[с. 72]{3}). Обычно гипотезой линеек и часов называется возможность того, что собственные линейки и часы некоторой ускоренной системы отсчёта в данный момент времени могут быть заменены соответствующими реквизитами мгновенно сопутствующей инерциальной системы отсчёта. Действительно, длина малой локальной области вблизи окрестности начала отсчёта т. О ускоренной системы совпадает с аналогичной длиной инерциальной системы отсчёта сопутствующей т. О независимо от состояния движения наблюдателя. Таким образом, в этом случае, гипотеза линеек и эквивалентность ускоренной и мгновенно сопутствующей инерциальной системы отсчёта выполняются. Достоинством такого рассмотрения (которое для краткости назовём методом МСИСО) является то, что анализ неинерциальной системы отсчёта сводится к изучению известных свойств инерциальных систем. Недостатки же такого рассмотрения весьма существенны. Во - первых, гипотеза линеек и гипотеза часов, измеряющих координатное время применима лишь для достаточно малой области \cite[формула (34)]{4}. Во - вторых, в отношении часов измеряющих физическое время (или измерения других производных величин) гипотеза локальности просто неверна. Например, промежуток собственного времени в ускоренной системе отсчёта (в отличие от инерциальной) неоднороден, пропорционален высоте точки измерения \cite[с. 303, формула (84.1)]{1}. Другим примером неприменимости постулата локальности даже для локальной области является различие в скоростях точек системы координат ускоренной и мгновенно ей сопутствующей инерциальной системы отсчёта  \cite[с. 65, формула (6)]{3}, \cite[формула (19)]{4}. Наконец ещё одним неудобством метода МСИСО является то, что для описания внутреннего движения неинерциальной системы отсчёта приходится рассматривать, по меньшей мере, две мгновенно сопутствующие инерциальные системы отсчёта для близких моментов времени.
 
Другой обычный метод изучения перехода между системами отсчёта - это использование локальных лоренцевых преобразований с коэффициентами, возможно зависящими от координат и являющимися релятивистским аналогом коэффициентов Ламе \cite {5}.  Преобразование между исходной и локальной инерциальными системами отсчёта является неголономным. Математическим и физическим достоинством этого метода является распространение аппарата локальных лоренцевых преобразований вообще на любые системы отсчёта необязательно являющиеся жёсткими. Кроме того этот метод применим как в СТО, так и в ОТО. 

Эти методы - не единственные методы изучения жёстких неинерциальных систем отсчёта в СТО. Ещё один способ  заключается в использовании особого глобального голономного преобразования оставляющего тензор кривизны 4-пространства неизменным и равным нулю. Данная статья применяет именно этот метод. Широко известным примером такого преобразования является преобразование Лоренца \cite {6} связывающее две инерциальные системы отсчёта. Другим примером является преобразование Мёллера в неравномерно движущуюся по прямой линии жёсткую неинерциальную систему отсчёта \cite {7}. Для произвольных жёстких систем отсчёта данное преобразование очевидно должно отличаться от этих преобразований. Оно представляет собой большой принципиальный интерес и было открыто Нэлсоном \cite {8}, \cite {9}. Его физический смысл заключается в переходе из лабораторной инерциальной системы отсчёта $S:(T,\mathbf{R})$   в жёсткую произвольно ускоренную и произвольно вращающуюся неинерциальную систему отсчёта $s':(t,\mathbf{r})$ . Первоначально оно получило название обобщённого Лоренц - преобразования \cite {8}. Но это преобразование, записанное в окончательной форме Нэлсоном, обобщает исследования его предшественников: Лоренца и Мёллера (см. далее начало п. 1) и по справедливости должно носить их имя. Таким образом, далее для краткости оно будет именоваться общим преобразованием ЛМН. 

Общее преобразование ЛМН проделывается в два этапа. 

На первом этапе параметром преобразования связывающего лабораторную инерциальную систему отсчёта $S$ и неинерциальную систему $s$ является всего одна величина $\mathbf{v}(t)$ (зависящая от времени $t$ системы $s$), поэтому орбитальное движение $s$ можно называть поступательным. Однако оказывается, что система отсчёта $s$ кроме собственного ускорения имеет ещё и некоторое собственное вращение \cite {8}, которое является собственной прецессией Томаса.  При этом  частота этой прецессии системы $s$ зависит от характера её орбитального движения. Преобразование соответствующее этому переходу, далее будет называться специальным (или частным) преобразованием ЛМН, чтобы отличить его от общего преобразования, в математическую форму которого входит ещё и матрица вращения. Забегая вперёд укажем, что параметр $\mathbf{v}(t)$,  является скоростью начала отсчёта $s$, но не остальных точек её системы координат. В связи с этим обстоятельством возникают два вопроса: 1) собственные точки какой системы отсчёта имеют скорость являющуюся параметром специального преобразования ЛМН и 2) какую же скорость в зависимости от парамера $\mathbf{v}(t)$  имеют собственные точки системы отсчёта $s$. Ответы на эти вопросы будут даны в п. 1, но предварительно для этого потребуется 1) найти прямой и обратный закон сложения скоростей и 2)доказать, что $\mathbf{v}(t)$ не является скоростью точек системы $s$.

Широко известно, что в случае, когда $s$ является инерциальной системой отсчёта существуют три основных кинематических эффекта. Это сокращение длины Фитцджеральда - Лоренца, замедление времени и относительность одновременности Эйнштейна. Неинерциальность системы отсчёта, вообще говоря, влияет на эти эффекты, несколько изменяя их величину \cite {4}, \cite {12}. Те же самые эффекты существуют и в неинерциальном случае, но для времени необходимо отдельно рассматривать координатную и физическую рассинхронизацию часов. Кроме того к этому списку ещё добавляется эффект неоднородности движения точек системы координат ускоренной системы отсчёта. Требуется также учесть, что каждый эффект необходимо рассматривать как относительно системы отсчёта $s$, так и относительно $S$, поскольку в релятивистском случае результаты получаются разными. В \cite[формулы (4),(13)]{11} для произвольной системы $s$ были расмотрены 2 эффекта: а) эффект рассинхронизации в лабораторной инерциальной системе отсчёта координатных часов покоящихся и синхронизированных в системе $s$ и б) эффект лоренцева сокращения длины линейки системы $S$ относительно системы $s$. В \cite {4}, \cite {12} для случая прямолинейного движения $s$ были рассмотрены 4 эффекта: а) неоднородность движения точек системы координат $s$, б) сокращение Лоренца в $S$ линейки $s$, в) рассинхронизация в системе $s$ координатных часов этой системы отсчёта предварительно синхронизированных в лабораторной системе, г) рассинхронизация в $S$ часов измеряющих физическое время в ускоренной системе отсчёта. Неоднородность движения точек системы координат $s$ рассматривалась также в \cite[с. 65, формула (6)] {3}. Цель п. 2 этой статьи заключается в определении тех же, что и в \cite {4}, \cite {12} эффектов в зависимости от собственного ускорения, но для случая произвольного движения системы s. Для достижения этой цели оказывается необходимым найти преобразование обратное к специальному преобразованию ЛМН. Уже вследствие этой причины данное преобразование вызывает большой интерес. Математическая форма частного преобразования ЛМН, в отличие от обычного преобразования Лоренца, имеет нелинейный вид, поэтому нахождение обратного к нему преобразования не является простым. Тем не менее, ограничиваясь разложением с точностью до вторых степеней собственных координат системы $s$ включительно найти обратное преобразование оказывается возможным.

Второй этап общего преобразования ЛМН заключается в переходе во вращающуюся с дополнительной угловой скоростью вокруг начала отсчёта $s$ систему отсчёта $s'$ \cite {9}. Это преобразование производится для того, чтобы обеспечить произвольность возможной собственной угловой скорости системы $s'$, т.е. её независимость от параметра характеризующего её движение по орбите. Движение осей координат системы отсчёта $s'$ относительно $S$ описывается с помощью преобразования ЛМН и происходит как далее в п. 4 будет показано сложным образом, с обобщённой угловой скоростью являющейся не антисимметричным пространственным тензором 2 ранга, а тензором общего вида.

Данное обстоятельство легко показать. Ясно, что собственное жёсткое вращение тела даже в инерциальной системе отсчёта (не говоря уже о неинерциальной системе $s'$) относительно $S$ невозможно. Действительно, если множество собственных точек не движущейся, но вращающейся системы координат относительно сопутствующей инерциальной системы отсчёта есть окружность, то относительно другой инерциальной системы отсчёта это множество точек вследствие сокращения Лоренца есть эллипс, в котором, как известно, расстояние от точки эллипса до его центра зависит от конкретного направления. На это обстоятельство в применении к прецессии системы $s'$ относительно $S$ почему-то до сих пор не обращается внимания (за исключением недавнего исследования С. С. Степанова \cite {30}). Если же ещё учесть, что система $s'$ может быть не обязательно жёсткой, то вывод о сложности движения точек системы координат $s'$  относительно $S$ почти очевиден. Таким образом цель п. 3 этой статьи  заключается в указании уравнений касающихся матрицы собственного вращения относительно лабораторной системы отсчёта $S$. Математически данная цель сводится к задаче нахождения первой производной по лабораторному времени от матрицы преобразования системы координат $S$  в $s'$. 

В статье \cite[формула (14)]{30} был выяснен вопрос о том, как изменяется относительно системы $S$ покоящийся в ускоренной системе отсчёта стержень. Иначе говоря, в этой работе была вычислена аффинная скорость стержня в системе $S$. Можно поставить более общий вопрос о том, как изменяется стержень в $S$, если в системе отсчёта связанной с его задней точкой он сам обладает аффинной скоростью. Выяснение этого вопроса, т. е. задачи заключающейся в вычислении связи произвольного аффинного нежёсткого движения внутри  системы $s$ и этого же движения, но относительно лабораторной системы отсчёта $S$ будет целью п. п. 4 и 5. Оказывается, однако, что соответствующие формулы несколько упрощаются, если рассматривать движение точки не относительно $s$, а относительно  системы отсчёта $k$, которая мгновенно сопутствует системе $s$ и оси которой мгновенно совпадают с осями $s$, но не имеют собственного вращения. 

Формулы преобразования аффинной скорости являются следствием преобразования ЛМН и полученное противоречие с реальностью, будучи установленным, прямо свидетельствовало бы о его некорректности. Следовательно, формулы  полученные в разделе 5 необходимо проверить на каком-нибудь достаточно простом примере. В качестве теста удобно использовать задачу о жёстком (в системе $S$) вращении тела.  Выбор именно этой задачи обусловлен тем, что угловая скорость и скорость растяжения главных осей для некоторой локальной области равномерно вращающегося диска давно известны. Локальная угловая скорость равномерно вращающейся жёсткой системы отсчёта была впервые вычислена в работе \cite{31} и далее неоднократно переоткрывалась (например \cite[с. 59]{36}). Коэффициент же растяжения известен из объяснения так называемого парадокса Эренфеста. Согласно этому парадоксу \cite{35} при увеличении угловой скорости вращения диска, его периметр в системе $S$ из-за сокращения Лоренца должен уменьшиться, что должно якобы привести к уменьшению его размеров. Разъяснение этого парадокса сводится к тому, что диск испытывает собственное локальное тангенциальное растяжение полностью компенсирующее лоренцево сокращение, так, что в результате размеры диска в системе $S$ остаются неизменными \cite[с. 68, с. 296]{1}). Вычисление данных величин осуществляется в п. 6. 

Зачем же вообще требуется рассмотрение такого, по сравнению с обычным преобразованием Лоренца, технически сложного преобразования как преобразование ЛМН? Во-первых, общее преобразование ЛМН интересно тем, что является основой релятивистской кинематики в некоторой привилегированной системе 4-координат (координат Нэлсона) описывающей жёсткую ускоренную и вращающуюся систему отсчёта. Такими координатами являются  декартовые трёхмерные координаты и некое координатное время, в которых 4-интервал имеет особо простой вид (см. далее начало 1 раздела, формула \eqref{3}). При этом метрика, следующая из этого преобразования, удовлетворяет принципу общей форминвариантности \cite {15}, т.е. как раз является метрикой жёсткой ускоренной и вращающейся системы отсчёта (см. \cite[c. 404, формула (13.71)]{16} с поправкой на отсутствие в СТО кривизны пространства-времени). Любое другое преобразование координат и времени не являющееся с точностью до замены системы 3-координат преобразованием ЛМН и понимаемое как преобразование связывающее движущуюся и покоящуюся системы отсчёта приведёт к интервалу для системы отсчёта $s$ не имеющему строго определённого вида и, следовательно, данная система отсчёта будет нежёсткой.

Другим аргументом "за"  \,является совпадение независимо вычисленного собственного ускорения с собственным ускорением следующим из специального преобразования ЛМН, что показал ещё Нэлсон \cite[формула (8)]{8}. 

Кроме того собственные ускорение и угловая скорость системы $s$ умноженные на малый промежуток времени $\delta t$ системы $s$ совпадают с независимо вычисленным методом МСИСО генератором инфинитезимального преобразования Лоренца связывающего две мгновенно сопутствующие $s$ инерциальные системы отсчёта в моменты $t$ и $t+\delta t$ как указано в \cite[задача 3.26]{19}. 

Этим данные причины, побуждающие считать преобразование ЛМН фундаментальным и надёжно установленным, не исчерпываются. Так, хорошо известно (например \cite[c. 44]{21}), что чистое преобразование Лоренца при бусте не образует группу. Группой является совокупность бустов и собственных поворотов. Для обычного преобразования Лоренца последовательное выполнение бустов (гиперболических поворотов) в различных направлениях не является бустом и для того, чтобы композиция бустов стала бустом, нужно её ещё домножить на некую собственную «подкручивающую» матрицу. Сам же дополнительный поворот (который следует называть собственным поворотом Вигнера) относителен, т.е. зависит от величины и направления буста, и давно вычислен \cite {27}, \cite{28}, \cite[формула (20)]{29}. В связи с этим требуется проверить будет ли форминвариантным при бусте общее преобразование ЛМН  и сравнить вычисленный угол собственного поворота при бусте с уже известным. Данные вопросы будут рассмотрены в разделе 7, а также будет показана единственность преобразования ЛМН. 

Наконец, ещё одним доводом в пользу справедливости преобразования  ЛМН является согласие с известным тетрадным методом \cite[c. 216, п. 6.5]{16}  и прекрасное совпадение общего преобразования ЛМН в четырёхмерном виде с уравнением \cite[c. 219, формула (6.16)]{16} (см. также недавние исследования \cite[раздел 2 до формулы (10)]{17}, \cite[раздел 2]{18}). В разделе 8 будет рассмотрено  представление преобразования ЛМН в 4-мерной форме. Другой целью этого раздела является указание тетрадного 4-мерного смысла собственных характеристик системы отсчёта. 

Кроме изучения составных движений (т. е. движений в двух взаимно перемещающихся системах отсчёта) у релятивистской кинематики как науки имеются 2 основные задачи. Если известны функции $\mathbf{v}(t)$ определяющая поступательное движение жёсткой неинерциальной системы отсчёта $s'$ и матрица собственного вращения как функции собственного времени  $t$, то тем самым известны и её собственные характеристики, т.е. собственное ускорение и угловая скорость. Эту задачу определения собственных характеристик произвольной системы отсчёта по известным параметрам назовём прямой или главной задачей кинематики. В разделе 1 будет показано как решается прямая задача кинематики для специального преобразования ЛМН (см. формулы \eqref{15}, \eqref{16}), а в п. 9 -  для общего преобразования ЛМН (см. формулы \eqref{546}, \eqref{547}). Тогда обратная задача кинематики состоит в нахождении параметров наиболее общего преобразования ЛМН в группу всех неинерциальных систем отсчёта имеющих заданное собственное ускорение и собственную угловую скорость. Роль данной задачи в физике заключается в том, что все такие найденные системы отсчёта с одинаковыми характеристиками будут обладать одинаковыми свойствами, т. е. для них будет выполняться обобщённый принцип относительности А. А. Логунова \cite[с. 126]{20}.

Обычное решение обратной задачи кинематики в классической механике заключается в первоначальном решении уравнения для матрицы собственного вращения и подстановке полученного решения в уравнение определяющее собственное ускорение. В релятивистском случае такой подход к обратной задаче неправилен, поскольку он не принимает во внимание наличие собственной прецессии Томаса у неинерциальной системы отсчёта $s$ двигающейся поступательно. Между тем, собственной прецессией Томаса при релятивистских скоростях пренебречь нельзя. Предлагаемый в разделе 9 порядок решения обратной задачи кинематики основывается на голономном общем преобразовании ЛМН автоматически учитывающим собственную прецессию Томаса. 

Уточним, что и в классическом, и в релятивистском случаях при решении задач кинематики имеется некоторое математическое неудобство, заключающееся в том, что характеристики неинерциальной системы отсчёта зависят от матрицы поворота. Это обстоятельство, хотя и является естественным, несколько затрудняет математическую запись уравнений для собственного ускорения и угловой скорости. В п. 9 будет показано, что используя определённую подстановку можно добиться того, что характеристики системы $s'$ будут записаны в удобном векторном виде. Также будет ясен и порядок решения обратной задачи.

 В следующем разделе 10 рассматривается наиболее общее (криволинейное) движение относительно лабораторной системы отсчёта $S$ равноускоренной системы $s'$ и определяются его параметры как частного случая преобразования ЛМН. Это движение представляет собой интерес, как с теоретической точки зрения, так и с практической. Теория равноускоренного движения применяется, например, при изучении электромагнитного излучения заряда. На практике движение равноускоренной системы отсчёта, кроме чисто физических приложений, может встретиться, например, в космонавтике, так как космические станции, совершая перелёты, обычно представляют собой системы с постоянным собственным ускорением. Другой причиной для анализа именно равноускоренного движения являются математические трудности при решении общих дифференциальных уравнений обратной задачи кинематики для произвольной жёсткой системы отсчёта. Забегая вперёд укажем, что решение найденных уравнений обратной задачи кинематики, даже для произвольной стационарной системы отсчёта, хотя и представляет собой большой интерес, но если и существует в известных функциях, то весьма громоздко. Поэтому в разделе 10 для проверки непротиворечивости полученных в разделе 9 дифференциальных уравнений удобно рассматривать  случай равноускоренного движения, т.е. когда собственное ускорение системы отсчёта постоянно, а собственное вращение отсутствует ($\mathbf{\Omega}=0$, $\mathbf{W}=const$). Данное движение является особо простым, поскольку в этом случае уравнения обратной задачи кинематики легко решаются. 

Другой кинематический способ рассмотрения в разделе 10 равноускоренного движения, кроме способа решением обратной задачи кинематики, заключается в определении параметров этого движения при бусте, аналогичном рассмотренному в разделе 7. Далее, в разделе 10 данные способы сравниваются между собой. 

Наконец, в разделе 11 представляется несколько видоизменённая форма общего преобразования ЛМН. Новый вид этого преобразования как будет видно из результатов раздела 9, имеет некоторое преимущество, связанное с тем, что собственные характеристики произвольной жёсткой системы отсчёта оказываются не зависящими от матрицы поворота и представимы в удобной векторной форме. Кроме того рассматриваются некоторые выведенные ранее формулы, в которых производится та же подстановка, какая была сделана в разделе 9. В частности указывается обратное общее преобразование ЛМН.

\subsection*{Результаты}

\subsection* {1. Физический смысл параметра входящего в специальное преобразование ЛМН и преобразование скорости}

Специальное преобразование ЛМН из лабораторной инерциальной системы отсчёта $S:(T,\mathbf{R})$  в жёсткую неинерциальную систему $s:(t,\mathbf{r})$, начало которой двигается, произвольно с параметром преобразования $\mathbf{v}(t)$ выглядит в виде \cite {8} 

\begin{equation} \label{10} 
     T=\frac{\mathbf{vr}}{\sqrt{1-v^{2} } } +\int _{0}^{t}\frac{dt}{\sqrt{1-v^{2} } }, 
\end{equation} 

\begin{equation} \label{11}    
 \mathbf{R=r}+\frac{1-\sqrt{1-v^{2} } }{v^{2} \sqrt{1-v^{2} } } \mathbf{(vr)v}+\int _{0}^{t}\frac{\mathbf{v}dt}{\sqrt{1-v^{2} } } .         
\end{equation}
Здесь $T$, $\mathbf{R}$ соответственно время и координаты лабораторной инерциальной системы отсчёта $S$; $t$ $\mathbf{r}$ соответственно время и координаты неинерциальной системы $s$.

Подчеркнём, что функция $\mathbf{v}$  не может зависеть от собственного времени $\tau$  начала системы отсчёта $s$  или, что то же самое от времени $T$ , поскольку в этом случае временные интегралы в преобразовании \eqref{10}, \eqref{11} не берутся однозначным образом (вследствие того, что подынтегральное выражение будет зависеть от двух переменных: $T=T(t,\mathbf{r})$ ).

Очевидно в случае $\mathbf{v}=const$  это преобразование переходит в обычное преобразование Лоренца. В случае же, когда $\mathbf{v}$  расположена вдоль оси $X$ и по величине равна 
\begin{equation} \label{3.1}  
v=th\theta \,  
\end{equation} 
преобразование \eqref{10}, \eqref{11} перейдёт в преобразование Мёллера \cite[c. 23, формула (64)]{7}, \cite[c. 206, формула (8. 160)]{21}
\begin{equation} \label{1}    
T=xsh{\theta}+\int _{0}^{t}ch{\theta}d\theta,\\\\\  X=xch{\theta}+\int _{0}^{t}sh{\theta}d\theta, \\\\\ 
Y=y, \\\\\ Z=z
\end{equation} 
где
\begin{equation} \label{2}    
\theta=\int_{0}^{t}Wdt, 
\end{equation} 
$W$ - собственное ускорение системы отсчёта.

    В преобразовании \eqref{10}, \eqref{11} время и компоненты радиус - вектора входят несимметрично. Это заранее может вызвать недоверие к данному преобразованию. Однако не следует придавать большого значения формулировке теории относительности симметричной по всем 4-координатам. Это обстоятельство связано с принципиальным различием между пространством и временем. Дифференцируя получим
    \begin{equation} \label{12}  
    dT=\left\{\frac{1}{\sqrt{1-v^{2} } } +\frac{\mathbf{\dot{v}r}}{\sqrt{1-v^{2} } } +\frac{(\mathbf{\dot{v}v)(vr)}}{\sqrt{1-v^{2} } ^{3} } \right\}dt+\frac{\mathbf{v}d\mathbf{r}}{\sqrt{1-v^{2} } } 
    \end{equation}
    
\[d\mathbf{R}=\left\{\frac{2\sqrt{1-v^{2} } ^{3} +3v^{2} -2}{v^{4} \sqrt{1-v^{2} } ^{3} } \mathbf{(vr)(\dot{v}v)v}+\frac{1-\sqrt{1-v^{2} } }{v^{2} \sqrt{1-v^{2} } } \left[\mathbf{(vr)\dot{v}+(\dot{v}r)v}\right]+\frac{\mathbf{v}}{\sqrt{1-v^{2} } } \right\}dt+d\mathbf{r}+\] 
\begin{equation} \label{13}  
 +\frac{1-\sqrt{1-v^{2} } }{v^{2} \sqrt{1-v^{2} } } \mathbf{(vdr)v}
 \end{equation}
 где
 \[\mathbf{\dot{v}}=\frac{d\mathbf{v}}{dt}\]
 Если подставить дифференциалы \eqref{12}, \eqref{13} в выражение для интервала инерциальной системы в прямоугольных координатах
 \begin{equation} \label{14} 
ds^{2} =dT^{2} -d\mathbf{R}^{2} ,  
\end{equation}
то получится интервал известной формы \cite{8},  \cite{15} для жёсткой ускоренной с собственным ускорением $\mathbf{W}$  и вращающейся системы отсчёта с собственной угловой скоростью $\mathbf{\Omega}$, 

\begin{equation} \label{3} 
ds^{2} =\left[\, (1+\mathbf{Wr})^{2} -(\mathbf{\Omega \, \, \times r})^{2} \right]^{} \, dt^{2} -2(\mathbf{\Omega \times r})d\mathbf{r}\, dt-d\mathbf{r}^{2} ,  
\end{equation} 
где
\begin{equation} \label{15} 
 \mathbf{W}=\frac{\mathbf{\dot{v}}}{\sqrt{1-v^{2} } } +\frac{1-\sqrt{1-v^{2} } }{v^{2} (1-v^{2} )} \mathbf{(\dot{v}v)v}, 
\end{equation}
а $\mathbf{\Omega}$ есть частота прецессии Томаса \footnote[4]{Обычно прецессией Томаса называется вращение системы отсчёта или вращение спина относительно лабораторной системы $S$. Здесь же данное вращение будет пониматься исключительно в системе $s$, т.е. это вращение является внутренним. Далее это обстоятельство не будет специально оговариваться.} и равна
\begin{equation} \label{16} 
\mathbf{\Omega} = \mathbf{\Omega}_{T}=\frac{1-\sqrt{1-v^{2} } }{v^{2} \sqrt{1-v^{2} } }\mathbf{v\times \dot{v}} 
\end{equation}
Эта угловая скорость зависит от характера орбитального движения системы отсчёта. 

Из \eqref{15} следует, что
\begin{equation} \label{17} 
\dot{\mathbf{v}}=\sqrt{1-v^{2} } \mathbf{W}-\frac{\sqrt{1-v^{2} } (1-\sqrt{1-v^{2} } )}{v^{2} } \mathbf{(vW)v} 
\end{equation} 

Подставляя это соотношение в \eqref{16} получим
\begin{equation} \label{18} 
\mathbf{\Omega} _{T} =\frac{1-\sqrt{1-v^{2} } }{v^{2} } \mathbf{v\times W} 
\end{equation} 

Учитывая \eqref{17}, \eqref{18}, формулы \eqref{12}, \eqref{13} можно переписать в более компактном виде
\begin{equation} \label{19} 
dT=\frac{1+(\mathbf{W+v\times \Omega} _{T} )\mathbf{r}}{\sqrt{1-v^{2} } } dt+\frac{\mathbf{v}d\mathbf{r}}{\sqrt{1-v^{2} } }  
\end{equation} 
\begin{equation} \label{20}
d\mathbf{R}=\left\{\frac{1+\mathbf{Wr}}{\sqrt{1-v^{2} } } \mathbf{v}\right. +\mathbf{\Omega} _{T} \times \mathbf{r}+\left. \frac{1-\sqrt{1-v^{2} } }{v^{2} \sqrt{1-v^{2} } } \left[\mathbf{v}\, (\mathbf{\Omega} _{T} \times \mathbf{r})\right]\;\mathbf{v}\right\}dt+d\mathbf{r}+\frac{1-\sqrt{1-v^{2} } }{v^{2} \sqrt{1-v^{2} } } (\mathbf{v}d\mathbf{r)v} 
\end{equation} 

Следовательно, скорости $\mathbf{U}$ и $\mathbf{u}$ 
\[\mathbf{U}=\frac{d\mathbf{R}}{dT} ,        \mathbf{u}_{s}=\frac{d\mathbf{r}}{dt} \] 
в системах отсчёта $S$ и $s$ связаны уравнением 
\begin{equation} \label{21} 
\mathbf{U}=\frac{(1+\mathbf{Wr)v}+\sqrt{1-v^{2} } (\mathbf{u}_{s}+\mathbf{\Omega} _{T} \times\mathbf{r})+\frac{1-\sqrt{1-v^{2} } }{v^{2} } \left[\mathbf{v}\; (\mathbf{u}_{s}+\mathbf{\Omega} _{T} \times \mathbf{r})\right]\; \mathbf{v}}{1+\mathbf{Wr+v\; (u}_{s}+\mathbf{\Omega} _{T} \times \mathbf{r})} .                  
\end{equation} 

Обращая это равенство, получим
\begin{equation} \label{22} 
\mathbf{u}_{s}=\frac{(1+\mathbf{Wr})\; \left[\sqrt{1-v^{2} } \mathbf{U-v}+\frac{1-\sqrt{1-v^{2} } }{v^{2} } \mathbf{v(vU})\right]}{1-\mathbf{vU}} -\mathbf{\Omega} _{T} \times \mathbf{r} 
\end{equation} 

Заметим, что в формулах \eqref{21} и \eqref{22} величина $\mathbf{u}_{s}+\mathbf{\Omega} _{T} \times\mathbf{r}$ является скоростью $\mathbf{u}_{k} $ точки в сопутствующей системе $s$ невращающейся системе отсчёта $k$, оси которой в каждый момент времени совпадают с осями $s$.
 Тогда эти формулы упрощаются и имеют вид
\begin{equation} \label{23} 
\mathbf{U}=\frac{(1+\mathbf{Wr)v}+\sqrt{1-v^{2} }\mathbf{u}_{k} +\frac{1-\sqrt{1-v^{2} } }{v^{2} } (\mathbf{vu}_{k} )\; \mathbf{v}}{1+\mathbf{Wr+vu}_{k} }  
\end{equation} 
\begin{equation} \label{24} 
\mathbf{u}_{k} =\frac{(1+\mathbf{Wr})\; \left[\sqrt{1-v^{2} } \mathbf{U-v}+\frac{1-\sqrt{1-v^{2} } }{v^{2} } \mathbf{(vU)v}\right]}{1-\mathbf{vU}}  
\end{equation} 
Отсюда очевидно, что для того, чтобы точки системы координат $k$ в этой системе отсчёта покоились $\mathbf{u}_{k} =0$, необходимо, чтобы относительно инерциальной системы отсчёта $S$ они двигались со скоростью $\mathbf{U=v}$. Таким образом $\mathbf{v}$ в специальном преобразовании ЛМН есть скорость точек системы координат $k$ относительно системы $S$. Далее в статье о параметре $\mathbf{v}$ специального преобразования ЛМН будет условно говориться как о просто скорости системы $s$, подразумевая под ней не скорость точек её системы коородинат, а скорость невращающейся системы $k$, оси которой сопутствуют  и сонаправлены осям системы $s$.

Возможно и другое понимание функции $\mathbf{v}$. Из \eqref{22} видно, что для того, чтобы точки 3-пространства системы $s$ покоились относительно её системы координат ($\mathbf{u}_{s}=0$) необходимо, чтобы они двигались относительно $S$ не со скоростью $\mathbf{U=v}$, а со скоростью $\mathbf{U=U}_{0} $
\begin{equation} \label{25} 
\mathbf{U}_{0} =\frac{(1+\mathbf{Wr)v}+\sqrt{1-v^{2} } \mathbf{\Omega} _{T} \times \mathbf{r}+\frac{1-\sqrt{1-v^{2} } }{v^{2} } \left[\mathbf{v\; (\Omega} _{T} \times \mathbf{r})\right]\; \mathbf{v}}{1+(\mathbf{W+v\times \Omega} _{T} )\; \mathbf{r}}  
\end{equation} 
Раскладывая выражение \eqref{25} по степеням $\mathbf{r}$ в первом приближении, получим
\begin{equation} \label{26} 
\mathbf{U}_{0} =\mathbf{v}+\sqrt{1-v^{2} }\mathbf{ \Omega} _{T} \times \mathbf{r}-\frac{\sqrt{1-v^{2} } (1-\sqrt{1-v^{2} } )}{v^{2} } \left[\mathbf{v(\Omega} _{T} \times \mathbf{r})\right]\mathbf{v} 
\end{equation} 
В частном случае прямолинейного движения системы отсчёта $s$ $(\mathbf{ \Omega} _{T} =0)$ из \eqref{25}, \eqref{26} видно, что в этом случае точки её системы координат движутся в ту же сторону со скоростью $\mathbf{U}_{0} =\mathbf{v}$. Это конечно не означает, что скорость всех точек системы координат $s$ будет одинаковой и равна скорости $\mathbf{V}$, с которой движется начало отсчёта $s$.  
Уравнения \eqref{25}, \eqref{26} наводят на мысль, что с точки зрения наблюдателя в $S$ система координат неинерциальной системы $s$ двигается неоднородно и, значит, не будет жёсткой. Оказывается, что скорость $\mathbf{v}$ сама зависит от $\mathbf{r}$, поэтому окончательно, неоднородность движения точек $s$ будет доказана в следующем разделе. Таким образом, другое понимание функции $\mathbf{v}$ в инерциальной системе отсчёта заключается в том, что через эту функцию определяется поле скоростей $\mathbf{U}_{0} $ точек системы отсчёта $s$ относительно $S$ в данный момент $T$. Очевидно данная скорость конкретной точки $\mathbf{r}$ может зависеть только от скорости $\mathbf{V}$ начала отсчёта и его собственного ускорения $\mathbf{W}$.

\subsection*  {2. Обратное специальное преобразование ЛМН. Основные кинематические эффекты}

 Нетривиальность преобразования \eqref{10},\eqref{11}  в случае движения системы отсчёта $s$ с ускорением приводит к любопытным эффектам, которые ярче всего проявляются, если предварительно найти преобразование к нему обратное. Преобразование обратное \eqref{10},\eqref{11} в общем случае в аналитической форме можно получить только приближённо. Обозначим интегралы времени как

\begin{equation} \label{27} 
\int _{0}^{t}\frac{dt}{\sqrt{1-v^{2} } }  =\theta _{t} ,  
\end{equation} 
\begin{equation} \label{28} 
\int _{0}^{t}\frac{\mathbf{v}dt}{\sqrt{1-v^{2} } }=\boldsymbol{\lambda} _{t} .  
\end{equation} 
При этом зависимость функции от какой-либо величины будем показывать нижним индексом. Тогда выражая из равенства \eqref{27} $t$ и подставляя его в $\mathbf{v}(t)$ и $\boldsymbol{\lambda} _{t} $  можно представить их как функции  $\theta $
\begin{equation} \label{29} 
\mathbf{v=V}_{\theta } ,  \boldsymbol{\lambda}_{t} =\mathbf{\Lambda}_{\theta } .  
\end{equation} 
Преобразование \eqref{10},\eqref{11} будет выглядеть в виде

\begin{equation} \label{30} 
T=\theta +\frac{\mathbf{V}_{\theta }\mathbf{r}}{\sqrt{1-V_{\theta } ^{2} } } ,  
\end{equation} 
\begin{equation} \label{31} 
\mathbf{R}={\mathbf{\Lambda}_\theta } +\mathbf{r}+\frac{1-\sqrt{1-V_{\theta }^{2} } }{V_{\theta }^{2} \sqrt{1-V_{\theta }^{2} } } (\mathbf{rV}_{\theta } )\mathbf{V}_{\theta } ,  
\end{equation} 
где ${\mathbf{\Lambda}_\theta }$ есть 
\begin{equation} \label{32} 
\mathbf{\Lambda}_{\theta } =\int _{0}^{\theta }\mathbf{V}_{\theta } d\theta  .  
\end{equation} 
 Функция $\mathbf{V}_{\theta } $   является скоростью движения точки $\bf{r}$ системы $s$ в момент $\theta $   относительно инерциальной системы отсчёта $S$ и является таким же параметром преобразования, как и $\mathbf{v}(t)$.  Уравнение обратное \eqref{27}  будет
\begin{equation} \label{33} 
t=\int _{0}^{\theta }\sqrt{1-V_{\theta } ^{2} } d\theta  ,  
\end{equation} 
где $\theta $   является корнем уравнения \eqref{30}. Для решения этого уравнения будем использовать разложение по степеням $\mathbf{r}$ . В первом приближении  $\theta $  будет равно
\begin{equation} \label{34} 
\theta =T-\frac{\mathbf{Vr}}{\sqrt{1-V^{2} } } ,  
\end{equation} 
где $\mathbf{V=V}_{T} $ и является скоростью начала отсчёта системы $s$ зависящей от лабораторного времени $T$. Разлагая ${\mathbf{V}_{\theta }  \mathord{\left/{\vphantom{V_{\theta }  \sqrt{1-V_{\theta }^{2} } }}\right.\kern-\nulldelimiterspace} \sqrt{1-V_{\theta }^{2} } } $ в ряд Тейлора получим
\begin{equation} \label{35} 
\frac{\mathbf{v}}{\sqrt{1-v^{2} } } =\frac{\mathbf{V}}{\sqrt{1-V^{2} } } -\frac{\mathbf{(Vr)\dot{V}}}{1-V^{2} } -\frac{(\mathbf{\dot{V}V)(Vr)V}}{(1-V^{2} )^{2} } ,  
\end{equation} 
где
\begin{equation} \label{36} 
\dot{\mathbf{V}}=\frac{d\mathbf{V}}{dT} .                                                      
\end{equation} 
Следовательно в первом приближении по степеням $\mathbf{r}$ параметр $\mathbf{v}$ равен   
\begin{equation} \label{37} 
\mathbf{v=V}-\frac{\mathbf{(Vr)\dot{V}}}{\sqrt{1-V^2}} , \,\,\,\,   \mathbf{\dot{V}}=\frac{d\mathbf{V}}{dT} .           
\end{equation} 
Данная формула определяет скорость точки системы $s$ с координатой $\mathbf{r}$. 
Из уравнений \eqref{25}, \eqref{26}, \eqref{37} видно, что для того чтобы в собственной системе отсчёта положение какой либо точки её системы координат не изменялось (т.е. система отсчёта была жёсткой по Борну), необходимо, чтобы эта точка двигалась в лабораторной инерциальной системе отсчёта согласованно, в некоторой корреляции с движением начала отсчёта. 

Подставляя разложение \eqref{35} в \eqref{30} получим, опуская нижний индекс, показывающий зависимость данной величины от $T$ , что во втором приближении 
\begin{equation} \label{38} 
\theta =T-\frac{\mathbf{Vr}}{\sqrt{1-V^{2} } } +\frac{\mathbf{(Vr)(\dot{V}r)}}{1-V^{2} } +\frac{\mathbf{(\dot{V}V)(Vr)}^{2} }{(1-V^{2} )^{2} } .  
\end{equation} 
Здесь и далее в формулах для обратного специального преобразования мы ограничимся в разложениях членами содержащими вторые степени координат, т.е. только членами содержащими ускорение в первой степени, но не его производные. Условия, когда можно пренебречь кубами $\mathbf{r}$ находятся из вычисления членов содержащих третьи степени координат, но проще руководствоваться соображениями размерности. В результате имеем
\begin{equation} \label{38.83} 
r<<\frac{W}{\dot{W}} ,\,\,\, r<<\frac{1}{W}
\end{equation}

С той же точностью произвольная скалярная функция $y_{\theta } $ будет равна

\begin{equation} \label{39} 
y_{\theta } =y-\left(\frac{\mathbf{Vr}}{\sqrt{1-V^{2} } } -\frac{(\mathbf{Vr)(\dot{V}r)}}{1-V^{2} } -\frac{(\mathbf{Vr})^{2} (\mathbf{\dot{V}V})}{(1-V^{2} )^{2} } \right)\dot{y}+\frac{(\mathbf{Vr})^{2} }{2(1-V^{2} )} \ddot{y}.  
\end{equation} 
и произвольный вектор $\mathbf{x}_{\theta } $ будет равен
\begin{equation} \label{40} 
\mathbf{x}_{\theta}=\mathbf{x}-\dot{\mathbf{x}}\frac{\mathbf{Vr}}{\sqrt{1-V^2}} 
+\dot{\mathbf{x}}\frac{\mathbf{(Vr)(\dot{V}r)}}{1-V^2} +
\frac{1}{2} \ddot{\mathbf{x}}\frac{\mathbf{(Vr)}^{2} }{1-V^{2}}+\dot{\mathbf{x}}\frac{(\mathbf{\dot{V}V)(Vr)}^{2} }{(1-V^{2} )^{2} },    
\end{equation} 
где
$$\mathbf{\dot{\mathbf{x}}}=\frac{d\mathbf{x}}{dT} ,  \ddot{\mathbf{x}}=\frac{d^{2} \mathbf{x}}{dT^{2} }$$  
и все величины в правой стороне равенств берутся в момент времени $T$.

В качестве произвольного вектора $\mathbf{x}_{\theta } $ в \eqref{40}    можно, например, взять вектор  $\mathbf{\Lambda}_{\theta}$ из \eqref{32}. Следовательно, можно записать, опуская нижний индекс, показывающий зависимость от $T$ ($\mathbf{\Lambda}=\int _{0}^{T}\mathbf{V}dT $)
\begin{equation} \label{42} 
\boldsymbol{\lambda} =\mathbf{\Lambda}_{\theta } =\mathbf{\Lambda}-\frac{\mathbf{(Vr)V}}{\sqrt{1-V^{2} } } +\frac{\mathbf{(Vr)(\dot{V}r)V}}{1-V^{2} } +\frac{\mathbf{(Vr)}^{2} \dot{\mathbf{V}}}{2(1-V^{2} )} +\frac{\mathbf{(Vr)}^{2} (\mathbf{\dot{V}V)V}}{(1-V^{2} )^{2} } .  
\end{equation} 
Подставив в \eqref{39} вместо скалярной функции $y_{\theta } $ функцию $t_{\theta } $ из \eqref{33} получим 
\begin{equation} \label{43} 
t=\int _{0}^{T}\sqrt{1-V^{2} } dT-\left\{\mathbf{Vr}-\frac{\mathbf{(Vr)(\dot{V}r)}}{\sqrt{1-V^{2} } } -\frac{\mathbf{(Vr)}^{2} (\mathbf{\dot{V}V)}}{2\sqrt{1-V^{2} } ^{3} } \right\} .  
\end{equation} 

Мировое время $t$ в \eqref{43} не является произвольной функцией $T$, $\mathbf{r}$ . Вид этой функции определился условием, чтобы метрика имела определённую форму \eqref{3}. Первый член в правой части этого выражения представляет собой собственное время начала отсчёта $\tau$ ($\mathbf{r}=0$). Это делает очевидным замедление времени в неинерциальной системе отсчёта и решает так называемый парадокс часов в общем виде. Собственное время начала отсчёта никак не совпадает с мировым временем $t$   из-за существования эффекта относительности одновременности. Это обстоятельство постфактум оправдывает применение в преобразовании \eqref{10}, \eqref{11} параметра преобразования $\mathbf{v}$ как функции $t$, а не $\tau$ . Второй член в фигурных скобках в \eqref{43} определяет рассинхронизацию двух координатных часов в неинерциальной системе $s$, если они первоначально были синхронизированы в лабораторной системе $S$ ($dT=0$). Очевидно, величина рассинхронизации в рассматриваемом приближении не зависит от закона движения системы отсчёта $s$, а определяется её мгновенным значением скорости и ускорения. Само существование рассинхронизации связано с относительностью времени в системах отсчёта $S$ и $s$. 

Подставляя \eqref{42} в \eqref{31} и учитывая, что
$$\frac{1-\sqrt{1-v^{2} } }{v^{2} \sqrt{1-v^{2} } } v_{\alpha } v_{\beta } =\frac{1-\sqrt{1-V^{2} } }{V^{2} \sqrt{1-V^{2} } } V_{\alpha } V_{\beta } -\frac{1-\sqrt{1-V^{2} } }{V^{2} (1-V^{2} )} \mathbf{(Vr)}(\dot{V}_{\alpha } V_{\beta } +V_{\alpha } \dot{V}_{\beta } )-$$
 \begin{equation} \label{44} 
-\frac{(1-\sqrt{1-V^{2} } )(1+3\sqrt{1-V^{2} } )}{V^{4} (1-V^{2} )^{2} } \mathbf{(\dot{V}V)(Vr)}V_{\alpha } V_{\beta } .  
\end{equation} 
получим после некоторых преобразований, что
\[\mathbf{R-\Lambda=L=r}-\frac{1-\sqrt{1-V^{2} } }{V^{2} }\mathbf{(Vr)V}+\frac{1-\sqrt{1-V^{2} } }{V^{2} \sqrt{1-V^{2} } }\mathbf{(Vr)(\dot{V}r)V}-\]
\begin{equation} \label{45} 
-\frac{(1-\sqrt{1-V^{2} } )^{2} }{2V^{2} (1-V^{2} )} \mathbf{(Vr)}^{2} \mathbf{\dot{V}}+\frac{(1-\sqrt{1-V^{2} } )^{2} }{V^{4} (1-V^{2} )} \mathbf{(Vr)}^{2} \mathbf{(\dot{V}V)V}.  
\end{equation} 
Здесь $\mathbf{L}$ является длиной идеально жёсткого стержня в момент $T$ в лабораторной инерциальной системе отсчёта $S$, чья собственная длина равна $\mathbf{r}$ и начальная точка которого движется относительно $S$ со скоростью $\mathbf{V}$ и ускорением $\dot{\mathbf{V}}$. Согласно этому уравнению прямой стержень длины $\mathbf{r}$ выглядит в $S$ искривлённым. Получившееся векторное уравнение \eqref{45} можно решить относительно $\mathbf{r}$, умножая обе части \eqref{45} скалярно на $\mathbf{V}$ и на $\mathbf{\dot{V}}$. Получим
\begin{equation} \label{46} 
\mathbf{LV}=\sqrt{1-V^{2} }\mathbf{Vr}+\frac{1-\sqrt{1-V^{2} } }{\sqrt{1-V^{2} } } \mathbf{(Vr)(\dot{V}r)}+\frac{(1-\sqrt{1-V^{2} } )^{2} }{2V^{2} (1-V^{2} )} \mathbf{(Vr)}^{2} \mathbf{(\dot{V}V)},  
\end{equation} 
\[\mathbf{L\dot{V}=\dot{V}r}-\frac{1-\sqrt{1-V^{2} } }{V^{2} } \mathbf{(Vr)(\dot{V}V)}+\frac{1-\sqrt{1-V^{2} } }{V^{2} \sqrt{1-V^{2} } } \mathbf{(Vr)(\dot{V}r)(\dot{V}V)}-\]
\begin{equation} \label{47} 
-\frac{(1-\sqrt{1-V^{2} } )^{2} }{2V^{2} (1-V^{2} )} \mathbf{(Vr)}^{2} \mathbf{\dot{V}}^{2}  
+\frac{(1-\sqrt{1-V^{2} } )^{2} }{V^{4} (1-V^{2} )} \mathbf{(Vr)}^{2} \mathbf{(\dot{V}V)}^{2} .  
\end{equation} 
Во втором и третьем членах в правой части \eqref{46}, а также в третьем, четвёртом и пятом членах уравнений \eqref{45}, \eqref{47} с требуемой точностью множитель $\mathbf{Vr}$ можно полагать согласно \eqref{46} равным
\begin{equation} \label{48} 
\mathbf{Vr}=\frac{\mathbf{LV}}{\sqrt{1-V^{2} } } .  
\end{equation} 
С той же точностью из \eqref{47} следует
\begin{equation} \label{49} 
\mathbf{\dot{V}r=L\dot{V}}+\frac{1-\sqrt{1-V^{2} } }{V^{2} \sqrt{1-V^2}} \mathbf{(LV)\dot{V}V}.  
\end{equation} 
Имея ввиду эти равенства, получим из \eqref{45}, что
\begin{equation} \label{50} 
\mathbf{Vr}=\frac{\mathbf{LV}}{\sqrt{1-V^{2} } } -\frac{1-\sqrt{1-V^{2} } }{\sqrt{1-V^{2} } ^{\,\,3} } \mathbf{(LV)(L\dot{V}})-\frac{(1-\sqrt{1-V^{2} } )^{2} (1+2\sqrt{1-V^{2} } )}{2V^{2} \sqrt{1-V^{2} } ^{\,\,5} } \mathbf{(LV)}^{2} (\mathbf{\dot{V}V}) 
\end{equation} 
Подставим сейчас \eqref{50} в \eqref{45}, тогда с помощью \eqref{48}, \eqref{49} получим с точностью до вторых степеней по $\mathbf{L=R-\Lambda=R}-\int _{0}^{T}\mathbf{V}dT $ включительно, что
\[\mathbf{r=L}+\frac{1-\sqrt{1-V^{2} } }{V^{2} \sqrt{1-V^{2} } } \mathbf{(LV)V}-\frac{(1-\sqrt{1-V^{2} } )^{2} (1+3\sqrt{1-V^{2} } )}{2V^{4} \sqrt{1-V^{2} } ^{\,\,5} } (\mathbf{\dot{V}V)(LV)}^{2} \mathbf{V}+\]
\begin{equation} \label{51} 
+\frac{(1-\sqrt{1-V^{2} } )^{2} }{2V^{2} (1-V^{2} )^{2} } \mathbf{(LV)}^{2}\mathbf{\dot{V}}-\frac{1-\sqrt{1-V^{2} } }{V^{2} \sqrt{1-V^{2} } ^{\,\,3} } \mathbf{(LV)(L\dot{V})V}=invariant.  
\end{equation} 
Подставив в \eqref{43} значение $\mathbf{r}$ из \eqref{51} окончательно получим
\begin{equation} \label{52} 
t=\int _{0}^{T}\sqrt{1-V^{2} } dT -\left\{\frac{\mathbf{LV}}{\sqrt{1-V^{2} } } -\frac{\mathbf{(LV)(L\dot{V})}}{\sqrt{1-V^{2} } ^{\,\,3} } -\frac{\mathbf{(LV)}^{2} \mathbf{(\dot{V}V)}}{\sqrt{1-V^{2} } ^{\,\,5} } \right\},  
\end{equation} 
Наконец, подставляя в \eqref{37} формулу \eqref{48} получим с той же точностью
\begin{equation} \label{53} 
\mathbf{v=V}-\frac{\mathbf{(VL)\dot{V}}}{1-V^{2} } .  
\end{equation} 

Уравнения \eqref{43}, \eqref{45} или \eqref{51}, \eqref{52} являются искомым обратным преобразованием с точностью до второго порядка по $\mathbf{r}$  (или $\mathbf{L}$) включительно. Очевидно, прямое и обратное преобразование ЛМН отличаются друг от друга. Это особенность связана с радикальным различием  инерциальной и неинерциальной систем отсчёта. Для постоянной скорости $\mathbf{V}$  обратное преобразование переходит в обычное преобразование Лоренца. 

\subsubsection*{Обратное преобразование ЛМН выраженное через собственное ускорение системы $s$}
Представляет интерес также выразить обратное преобразование через собственное ускорение системы отсчёта $s$. Для этого надо предварительно выразить уравнения \eqref{15}, \eqref{16} через  $\mathbf{V}$ и $\mathbf{\dot{V}}$ . Считая  $\mathbf{r}=0$ получим  формулы 
\begin{equation} \label{54} 
\mathbf{W}=\frac{\mathbf{\dot{V}}}{1-V^{2} } +\frac{1-\sqrt{1-V^{2} } }{V^{2} \sqrt{1-V^{2} } ^{\,\,3} } (\mathbf{\dot{V}V)V} 
\end{equation} 
\begin{equation} \label{55} 
\mathbf{\Omega} _{T} =\frac{1-\sqrt{1-V^{2} } }{V^{2} (1-V^{2} )} \mathbf{V\times \dot{V}} 
\end{equation} 
Из \eqref{54} следует, что                            
\begin{equation} \label{56} 
\mathbf{\dot{V}}=(1-V^{2} )\mathbf{W}-\frac{(1-V^{2} )(1-\sqrt{1-V^{2} } )}{V^{2} }\mathbf{(VW)V} 
\end{equation} 
\begin{equation} \label{57}
\mathbf{V\dot{V}}=\sqrt{1-V^{2} } ^{\,\,3} \mathbf{VW}. 
\end{equation} 
Следовательно, подставляя эти равенства в \eqref{43},\eqref{45}, \eqref{51}, \eqref{52} получим
\begin{equation} \label{58} 
t=\int_{0}^{T}\sqrt{1-V^{2}}dT-\left\{\mathbf{Vr}-\sqrt{1-V^{2} }\mathbf{(Vr)(Wr)} 
-\frac{(1-\sqrt{1-V^2})^2}{2V^2}(\mathbf{Vr})^{2} (\mathbf{VW}) \right\} 
\end{equation} 
\[\mathbf{L=r}-\frac{1-\sqrt{1-V^{2} } }{V^{2} } \mathbf{(Vr)V}+\frac{(1-\sqrt{1-V^{2}})\sqrt{1-V^2}} {V^{2} }  \mathbf{(Vr)(Wr)V}-\] 
\begin{equation} \label{59} 
-\frac{(1-\sqrt{1-V^{2} } )^{2} }{2V^{2} }\mathbf{(Vr)}^{2}\mathbf{W}+\frac{(1-\sqrt{1-V^{2} } )^{3} }{2V^{4} }\mathbf{(Vr)}^{2}\mathbf{(VW)V}.  
\end{equation} 

\begin{equation} \label{60} 
t=\int _{0}^{T}\sqrt{1-V^{2} } dT -\left\{\frac{\mathbf{LV}}{\sqrt{1-V^{2} } } -\frac{\mathbf{(LV)(LW)}}{\sqrt{1-V^{2} } } -\frac{(1-\sqrt{1-V^{2} } )}{V^{2} (1-V^{2} )} \mathbf{(LV)}^{2} \mathbf{(VW)}\right\} 
\end{equation} 
\[\mathbf{r=L}+\frac{1-\sqrt{1-V^{2} } }{V^{2} \sqrt{1-V^{2} } }\mathbf{(LV)V}-\frac{1-\sqrt{1-V^{2} } }{V^{2} \sqrt{1-V^{2} } } \mathbf{(LV)(LW)V}+\]
\begin{equation} \label{61} 
+\frac{(1-\sqrt{1-V^{2} } )^{2} }{2V^{2} (1-V^{2} )} \mathbf{(LV)}^{2} \mathbf{W}- \frac{(1-\sqrt{1-V^{2} } )^{2} }{V^{4} (1-V^{2} )} \mathbf{(VW)(LV)}^{2}\mathbf{V}.  
\end{equation} 

Из \eqref{58}, \eqref{60} видно, что рассинхронизация двух часов: в начале отсчёта и в точке с координатой $\mathbf{r}$  не зависит от $\mathbf{\dot{W}}$ . Аналогично, из \eqref{51}, \eqref{59} видно, что длина эталонной линейки системы координат нестационарной системы отсчёта в сделанном приближении не зависит от скорости изменения её собственного ускорения $\mathbf{\dot{W}}$  и при $\mathbf{V}=0$  совпадает с $\mathbf{r}$.
 
\subsubsection*{Скорость точек системы координат $s$}
Вычислим сейчас в явном виде скорость точек $\mathbf{U}_{0}$ системы $s$. Подставив \eqref{56} в \eqref{37} получим 
\begin{equation} \label{62} 
\mathbf{v=V}-\sqrt{1-V^{2} }\mathbf{(Vr)W}+\frac{\sqrt{1-V^{2} } (1-\sqrt{1-V^{2} } )}{V^{2} }\mathbf{(Vr)(VW)V} 
\end{equation} 
где $\mathbf{V}$- скорость начала отсчёта s относительно S. Подставив в \eqref{26} уравнение \eqref{62} получим 
\[\mathbf{U}_{0} =\mathbf{V}-\sqrt{1-V^{2} }\mathbf{(Vr)W}+\frac{\sqrt{1-V^{2} } (1-\sqrt{1-V^{2} } )}{V^{2} } \mathbf{(Vr)(VW)V}+\] 
\begin{equation} \label{63} 
+\sqrt{1-V^{2} } \mathbf{\Omega} _{T} \times\mathbf{r}-\frac{\sqrt{1-V^{2} } (1-\sqrt{1-V^{2} } )}{V^{2} } \left[\mathbf{V(\Omega} _{T} \times \mathbf{r})\right]\mathbf{V} 
\end{equation} 
Наконец подставляя сюда значения \eqref{54}, \eqref{55} получим
\begin{equation} \label{64}
\mathbf{U}_{0} =\mathbf{V}-\frac{1-\sqrt{1-V^{2} } }{V^{2} } \mathbf{(Vr)\dot{V}}-\frac{(1-\sqrt{1-V^{2} } )^{2} }{V^{4} \sqrt{1-V^{2} } } \mathbf{(Vr)(\dot{V}V)V}-\frac{1-\sqrt{1-V^{2} } }{V^{2} } \mathbf{(\dot{V}r)V} 
\end{equation}
 
 \subsubsection*{Рассинхронизация физических часов закреплённых в $s$}
Интерес представляет ещё рассинхронизация физических часов расположенных в разных точках системы координат $s$ рассматриваемая в лабораторной системе отсчёта. Этот эффект появляется из-за неоднородности движения разных точек системы отсчёта $s$. Например для рассинхронизации часов за время $T$ в начале отсчёта и в точке с координатой $\mathbf{r}$ величина рассинхронизации будет равна
\begin{equation} \label{65}
\Delta\tau=\int^{T}_{0}\left(\sqrt{1-U^{2}_{0}}-\sqrt{1-V^2}\right)dT,
\end{equation} 
что при учёте равенства \eqref{63}, составит величину 
\begin{equation} \label{66}
\Delta \tau=\mathbf{r} \int^{T}_{0}\sqrt{1-V^2}\left[\mathbf{(VW)V-V \times \Omega }_{T}\right]dT,
\end{equation} 
или подставляя сюда \eqref{54}, \eqref{55}
\begin{equation} \label{67}
\Delta\tau=\mathbf{r} \int^{T}_{0}\left(\frac{1-\sqrt{1-V^2}}{V^2(1-V^2)} \mathbf{(V \dot{V})V}+\frac{1-\sqrt{1-V^2}}{\sqrt{1-V^2}} \mathbf{\dot{V}}\right)dT,
\end{equation} 
 \subsubsection*{Сравнение полученных формул с известными результатами}

Прежде всего требуется выяснить справедливы ли формулы \eqref{54} и \eqref{55}. Формула \eqref{54} давно известна \cite[с.109, формула (194)]{33}. Однако, \eqref{55} противоречит ныне принятой точке зрения  \cite[с.123]{32}, согласно которой собственная прецессия Томаса зависит от лабораторных скорости  $\mathbf{V}$ и ускорения $\mathbf{\dot{V}}$ как
\[\mathbf{\Omega} _{T} =\frac{1-\sqrt{1-V^{2} } }{V^{2} \sqrt{1-V^{2} }} \mathbf{V\times \dot{V}}\] 
Впрочем формула \eqref{55} хорошо согласуется с некоторыми другими источниками, например \cite[формула (34)]{34}, поэтому данному расхождению не следует придавать значения.

Рассмотрим теперь легко проверяемые следствия указанного обратного преобразования в случае прямолинейного движения $s$ без вращения в направлении осей X и x. Прежде всего, рассматривая прямолинейное ускоренное движение системы $s$ вдоль оси 1 можно убедиться, что \eqref{62} совпадает с уже известной формулой для скорости точек системы координат $s$ \cite[c. 65, формула (6)]{3}, \cite[формула (19)]{4}.

Пусть теперь два события произошли в одно и то же время по часам лабораторной инерциальной системы отсчёта $S$, но в разных точках $x_{1} $ и $x_{2} $   ускоренной системы отсчёта $s$. Определим промежуток мирового времени между событиями 1 и 2 в ускоренной системе отсчёта $s$. Из \eqref{58} имеем, что моменты времени  $t_{1} $ и $t_{2}$ по часам системы $s$ будут равны 

\begin{equation} \label{68} 
t_{\alpha } =\int _{0}^{T}\sqrt{1-V^{2} } dT-\; \left\{Vx_{\alpha } -V(1-\frac{V^{2} }{2} )\; Wx_{\alpha }^{2}\right\} ,       (\alpha =1,\; 2 )
\end{equation} 
Следовательно
\begin{equation} \label{69} 
\Delta t=t_{2} -t_{1} =-\left(Vx-V(1-\frac{V^{2} }{2} )\, Wx^{2} \right)\mathop{{ \mathord{\left/{\vphantom{ }}\right.\kern-\nulldelimiterspace} } }\limits_{x=x_{1} }^{x=x_{2} }  
\end{equation} 
Аналогично, из \eqref{59} следует, что длина стержня собственной длины $x$  расположенного вдоль направления его прямолинейного движения равна
\begin{equation} \label{70} 
L=\sqrt{1-V^{2} } x+\frac{V^{2} \sqrt{1-V^{2} } }{2} Wx^{2}  
\end{equation} 
Легко также видеть, что в случае прямолинейного движения формула  \eqref{62} упрощается и переходит в 
\begin{equation} \label{67.5}
\Delta \tau =\mathbf{r}\ \int_{0}^{T}{\frac{{{V}^{2}}}{1-{{V}^{2}}}\mathbf{\dot{V}}}dT=x\left( arthV-V \right)
\end{equation}

Формулы \eqref{69}, \eqref{70} и \eqref{67.5} уже известны из прямого расчёта использующего уравнение прямолинейного равноускоренного движения - соответственно \cite[формула (18)]{12}, \cite[формула (30)]{4} и \cite[формула (29)]{12}.

 \subsubsection* {О пригодности полученных формул для реальной системы отсчёта с предельно большой жёсткостью}

  Преобразование ЛМН является точным для жёсткой по Борну нестационарной системы отсчёта. Однако реальные системы отсчёта реализованные реальным телом с заданным движением его начальной точки и свободными остальными точками её координат с произвольными возможными в СТО размерами являются вообще говоря нежёсткими, и преобразование ЛМН не будет являться для них верным. Тем не менее, для такой реальной системы отсчёта с предельно большой жёсткостью (под ней понимается система отсчёта, в которой скорость звука в материале линеек равна скорости света) условие \eqref{38.83} является одновременно и необходимым условием, при котором уравнения \eqref{60}, \eqref{61} имеют смысл. Действительно, в случае такой нестационарной системы отсчёта $s$, её координаты $\mathbf{r}$ и время $t$,  чтобы перейти  в соответствующие формулы \eqref{60}, \eqref{61} для стационарной системы отсчёта кроме скорости начала отсчёта $\mathbf{V}$ могут зависеть только от собственного ускорения (и, возможно, его производных) в предшествующий момент времени $T'$ равный (для небольшого участка вблизи начала отсчёта) 
\begin{equation} \label{70.5} 
 T'=T-L.
\end{equation} 
 Но для достаточно малых собственных координат и достаточно плавного изменения собственного ускорения (при условии \eqref{38.83}) этим запаздыванием можно пренебречь и, следовательно, формулы \eqref{60}, \eqref{61} (и другие им эквивалентные) в сделанном приближении пригодны.

\subsection* {3. Уравнения для матрицы собственного вращения}

Перейдём теперь из системы отсчёта $s$ обладающей собственными характеристиками $\mathbf{W}$  и $\mathbf{\Omega}_{T}$  в систему отсчёта $s'$ жёстко вращающуюся с дополнительной угловой скоростью $\boldsymbol{\omega}$  относительно $s$ вокруг общего центра. Это означает, что в системе $s$ необходимо произвести стандартную замену координат и времени 

 \begin{equation} \label{71} 
r^{\alpha } =a^{\beta \alpha} r'^{\beta } , 
\end{equation} 
\[t=t'\],  
что приводит к общему преобразованию \cite{9}
\begin{equation} \label{72} 
T=\frac{v^{\alpha } a^{\beta \alpha} r'^{\beta } }{\sqrt{1-v^{2} } } +\int _{0}^{t}\frac{dt'}{\sqrt{1-v^{2} } }  
\end{equation}
\begin{equation} \label{73} 
 R^{\alpha } =\left\{a^{ \beta \alpha} r'^{\beta } +\frac{1-\sqrt{1-v^{2} } }{v^{2} \sqrt{1-v^{2} } } v^{\alpha } v^{\gamma } a^{\beta \gamma} r'^{\beta } \right\}+\int _{0}^{t}\frac{v^{\alpha } dt'}{\sqrt{1-v^{2} } } 
  \end{equation}
 Матрицу $a^{\alpha \beta}$ называют матрицей поворота. Эта матрица является ортогональной, т.е. справедливо соотношение
\begin{equation} \label{74}
a^{\beta \alpha} a^{\gamma \alpha} =a^{\alpha \beta} a^{\alpha\gamma} =\delta ^{\beta\gamma}
\end{equation}
 Далее, если строки (или столбцы) матрицы $a^{\alpha \beta}$ рассматривать как координаты единичных векторов $\mathbf{i}^{\alpha}$, то верны следующие соотношения \cite[c. 42]{22}
 \begin{equation} \label{75}
\mathbf{i}^{\alpha } \times \mathbf{i}^{\beta } =e^{\alpha \beta \gamma } \mathbf{i}^{\gamma }  
 \end{equation}
В представлении вместо $\mathbf{i}^{\alpha } $ использующим матрицу поворота $a^{\alpha \beta } $, эти равенства выглядят так \cite[формула (14)]{15}
\begin{equation} \label{76}
 e^{\alpha \mu \nu } a^{\mu \beta } a^{\nu \gamma } =e^{\mu \beta \gamma } a^{\alpha \mu } ,\,\,\,\,
 e^{\alpha \mu \nu } a^{\beta \mu } a^{\gamma \nu } =e^{\mu \beta \gamma } a^{\mu \alpha } 
 \end{equation}
 Справедливость данных равенств "уничтожения" \, нетрудно проверить для конкретного их представления, например для углов Эйлера. 
 
 В системе $s'$ изменение произвольного вектора $\mathbf{r'}$ связанного с системой отсчёта $s$ подчиняется уравнению
\begin{equation} \label{77}
\frac{d\mathbf{r'}}{dt} =-\, \boldsymbol{\omega'} \times \mathbf{r'}.  
 \end{equation} 

 Здесь $\omega'^{\alpha}$ есть вектор угловой скорости вращения $s'$ относительно системы $s$ выраженный в системе координат $s'$. Подставляя сюда $r'^{\beta}=a^{\beta\alpha}r^{\alpha}$ и считая , что $r^{\alpha } =const$ получим, что эта матрица должна удовлетворять уравнению

\begin{equation} \label{79,2}
 \frac {da^{\beta \alpha}}{dt} =e^{\xi\eta\beta} \omega'^{\eta} a^{\xi\alpha}
 \end{equation} 

 Уравнения \eqref{74}, \eqref{76} полностью описывают матрицу $a^{\alpha \beta}$ как матрицу вращения, а уравнение \eqref{79,2} фактически является определением угловой скорости выраженным через $a^{\alpha \beta }$. 
 
 \begin{equation} \label{79,3}
e^{ \mu \nu \beta} \omega'^{\nu }=a^{\mu\alpha } \frac{da^{\beta \alpha } }{dt} ,\,\,\,\,\, \omega'^{\gamma }=\frac {1}{2}\,e^{\mu \gamma\beta }a^{ \mu\alpha } \frac{da^{ \beta \alpha } }{dt}
\end{equation} 

Если подставить \eqref{71} в интервал системы отсчёта $s$ \eqref{3}, то с помощью требований \eqref{74}, \eqref{76}, \eqref{79,2} будет видно, что в системе $s'$ форма интервала также примет вид \eqref{3}, как и должно быть \cite{15}.  Характеристики новой системы отсчёта станут \cite[формула (16), с учётом того, что в этой статье матрица поворота входила в транспонированном виде]{15}.
\begin{equation} \label{80}
 W'^{\alpha } =a^{ \alpha \beta} W^{\beta }  
 \end{equation}
\begin{equation} \label{81}
 \Omega '^{\alpha } = a^{ \alpha \beta}\Omega _{T}^{\beta } +\omega'^{\alpha } 
\end{equation}

Выясним теперь как же преобразуется собственное вращение \eqref{71} в лабораторную систему отсчёта. Длина $\mathbf{L}$ малой собственной линейки $\mathbf{r}$ системы $s$ в лабораторной системе будет 
\begin{equation} \label{82}
\mathbf{L=r}-\frac{1-\sqrt{1-V^{2} } }{V^{2} } \mathbf{(Vr)V}
\end{equation},
где $\mathbf{V}$ -скорость начала отсчёта системы $s'$ относительно $S$.
Подставим в это уравнение \eqref{71}, тогда получим
\begin{equation} \label{83}
L^{\alpha}=a^{\beta\alpha} r'^{\beta}-\frac{1-\sqrt{1-V^2}}{V^2}V^{\gamma}V^{\alpha}a^{\beta\gamma}r'^{\beta}
\end{equation}
 Данное уравнение можно рассматривать как некоторое преобразование с матрицей $ A^{\alpha \beta }$ из лабораторной 
 системы отсчёта в новую систему $s'$
\begin{equation} \label{83.25}
L^{\alpha}=A^{\beta\alpha}r'^{\beta} 
\end{equation}
Следовательно матрица поворота $a^{\alpha \beta }$ в лабораторной системе имеет вид $A^{\alpha \beta }$ , причём
\begin{equation} \label{84}
A^{\beta \alpha}= a^{\beta \alpha }-\frac{1-\sqrt{1-V^{2} } }{V^{2} } V^{\gamma}V^{\alpha}a^{\beta\gamma }
\end{equation}

Если матрица $A^{\alpha\beta}$ представляет собой матрицу поворота, то для неё должен выполняться закон изменения этой матрицы со временем аналогичный формуле \eqref{79,2}. Можно ли  использовать   уравнение 
\begin{equation} \label{85}
 \frac{dA^{ \beta \alpha } }{dT} =e^{\alpha \eta\xi} \varpi^{\eta} A^{\beta \xi},
 \end{equation} 
где $\boldsymbol{\varpi}$ имеет смысл угловой скорости в лабораторной системе $S$? Для выяснения этого вопроса продифференцируем уравнение \eqref{84} по времени $T$ лабораторной системы отсчёта, заменим всюду в правой части получившегося уравнения производные матрицы $a^{\beta \alpha}$ по времени $T$ на 
\begin{equation} \label{86}
\frac{da^{\beta \alpha  } }{dT} =\sqrt{1-V^2}\,\frac{da^{\beta \alpha } }{dt}\,,
\end{equation}
и далее заменим эти производные $da^{ \beta \alpha }/dt$ на правую часть уравнения \eqref{79,2}. В результате получим

\[\frac{dA^{ \beta \alpha} }{dT} = \sqrt{1-V^{2} } e^{\xi\eta\beta} \omega'^{\eta} a^{\xi\alpha}- \frac{\sqrt{1-V^2}(1-\sqrt{1-V^2})}{V^2}V^{\gamma}V^{\alpha}e^{\xi\eta\beta} \omega'^{\eta} a^{\xi\gamma} - \]
 \begin{equation} \label{87}
-\frac{\sqrt{1-V^2}(1-\sqrt{1-V^2})}{V^2}(V^{\alpha}\dot{V}^{\gamma}+ V^{\gamma}\dot{V}^{\alpha})a^{\beta\gamma }- \frac{(1-\sqrt{1-V^2})^2}{V^4 }(\mathbf{V \dot{V}})V^{\gamma}V^{\alpha}a^{\beta\gamma }  
\end{equation}
Далее, выразим $a^{\beta\alpha}$ из \eqref{84}. Для этого достаточно умножить это уравнение на $V^{\alpha}$ и ответ снова подставить в \eqref{84}. Получим
\begin{equation} \label{88}
a^{\beta \alpha}= A^{\beta\alpha}+\frac{1-\sqrt{1-V^{2} } }{V^{2} \sqrt{1-V^2}} V^{\gamma}V^{\alpha}A^{\beta\gamma }
\end{equation}
Сейчас в уравнении \eqref{87} осталось заменить $a^{\beta\alpha }$ согласно \eqref{88}. В итоге возникает следующее уравнение общего вида (сравни с \eqref{85} и \cite[формулы (4), (10)]{9} )

\[ \frac{dA^{ \beta \alpha} }{dT} =\sqrt{1-V^2} e^{\xi\eta\beta}\omega'^{\eta} A^{\xi\alpha}-\]
\begin{equation} \label{89}
-\left(\frac{1-\sqrt{1-V^2}}{V^2\sqrt{1-V^2}}\mathbf{(V \dot{V})}V^{\alpha}V^{\nu}+\frac{\sqrt{1-V^2}(1-\sqrt{1-V^2})}{V^2} V^{\alpha}\dot{V}^{\nu}+\frac{1-\sqrt{1-V^2}}{V^2} V^{\nu}\dot{V}^{\alpha}\right)A^{\beta\nu}
\end{equation}
Очевидно, что только в том случае, когда движение $s'$ является равномерным, уравнение \eqref{89} становится похожим на \eqref{79,2}. Проверив же для матрицы $A^{ \beta \alpha}$ условие ортогональности и равенство уничтожения  аналогичные \eqref{74} и \eqref{76}, можно убедиться, что эти уравнения не выполняются. Таким образом матрица линейного преобразования $A^{ \beta \alpha}$ не является матрицей поворота. 

Обычно под прецессией Томаса понимается вращение относительно лабораторной системы отсчёта \cite{23}. Однако, как было ясно выше, это вращение из-за сокращения Лоренца не является жёстким, поэтому термин "прецессия" \, понимаемый в этом смысле незаконен.

\subsection* {4. Преобразование собственной аффинной скорости в лабораторную систему отсчёта}

 Центроаффинное движение точки в системе отсчёта $k$ подчиняется уравнению
\begin{equation} \label{225} 
u_{k}^{\alpha } =\omega ^{\alpha \beta } r^{\beta } ,                                                          
\end{equation} 
где $u_{k}^{\alpha } $- скорость точки, $r^{\alpha } $- её координата. Величину $\omega ^{\alpha \beta } $ обычно называют собственной обобщённой угловой скоростью точки. Однако это название не является удачным, например, для аффинного растяжения $\omega ^{\alpha \beta } $ отлично от нуля, но это преобразование не является вращением, так, что название «угловая скорость» неприменимо. Правильнее называть тензор $\omega ^{\alpha \beta } $ аффинной или аффиннорной скоростью в системе $k$. Интересно движение этой точки относительно лабораторной инерциальной системы отсчёта. Довольно очевидно, что оно вблизи начала координат будет удовлетворять некоторому уравнению
\begin{equation} \label{226} 
U^{\alpha } =V^{\alpha } +\Omega ^{\alpha \beta } L^{\beta } ,                                                  
\end{equation} 
где $V^{\alpha } $- скорость начала координат системы s, $U^{\alpha } $- скорость точки в системе отсчёта $S$, $\Omega ^{\alpha \beta } $- некоторая матрица аффинной скорости, $L^{\alpha } $- расстояние от точки до начала координат k рассматриваемое относительно $S$. В первом порядке оно определяется из \eqref{82}. Какова же зависимость $\Omega^{\alpha\beta}$ от $\omega ^{\alpha \beta } $ согласно теории относительности? Выведём её из формулы сложения скоростей \eqref{23}. Очевидно, что во втором и третьем слагаемом числителя и в знаменателе с требуемой точностью $\mathbf{v}$ можно заменить на $\mathbf{V}$. Тогда подставив в \eqref{23} формулу \eqref{225} и в первое слагаемое числителя - \eqref{37}, получим, раскладывая дробь с точностью до первых степеней $\mathbf{r}$
\begin{equation} \label{228} 
U^{\alpha } =V^{\alpha } -\frac{\mathbf{(Vr)}\, \dot{V}^{\alpha } }{\sqrt{1-V^{2} } } +\sqrt{1-V^{2} } \omega ^{\alpha \beta } r^{\beta } -\frac{\sqrt{1-V^{2} } (1-\sqrt{1-V^{2} } )}{V^{2} } \omega ^{\gamma \beta } V^{\gamma } V^{\alpha } r^{\beta }  
\end{equation} 
Подставим в \eqref{228} вместо $\mathbf{r}$ выражение 
\begin{equation} \label{229} 
\mathbf{r=L}+\frac{1-\sqrt{1-V^{2} } }{V^{2} \sqrt{1-V^{2} } } \mathbf{(LV)V}.                                                         
\end{equation} 
, которое является обратным к \eqref{82}. Отсюда    
\[U^{\alpha } -V^{\alpha } =\sqrt{1-V^{2} } \omega ^{\alpha \beta } L^{\beta } +\frac{1-\sqrt{1-V^{2} } }{V^{2} } \mathbf{(VL)}\, \omega ^{\alpha \gamma } V^{\gamma } -\] 
\[-\frac{\sqrt{1-V^{2} } (1-\sqrt{1-V^{2} } )}{V^{2} } (\omega ^{\gamma \beta } V^{\gamma } L^{\beta } )V^{\alpha } -\frac{(1-\sqrt{1-V^{2} } )^{2} }{V^{4} } \mathbf{(VL)}(\, \omega ^{\gamma \mu } V^{\gamma } V^{\mu } )V^{\alpha } -\] 
\begin{equation} \label{230} 
-\frac{\mathbf{(VL)}\dot{V}^{\alpha } }{1-V^{2} }  
\end{equation} 
Выражение \eqref{230} нельзя представить в виде 
\begin{equation} \label{231} 
U^{\alpha } -V^{\alpha } =\dot{L}^{\alpha } =e^{\alpha \beta \gamma } \Omega ^{\beta } L^{\gamma } ,   
\end{equation} 
с некоторым $\Omega ^{\alpha } $, значит, аффинное движение точек в системе k относительно лабораторной системе S не является жёстким. Сравнивая это выражение с \eqref{226} окончательно получим   
\[\Omega ^{\alpha \beta } =\sqrt{1-V^{2} } \omega ^{\alpha \beta } -\frac{\sqrt{1-V^{2} } (1-\sqrt{1-V^{2} } )}{V^{2} } \omega ^{\gamma \beta } V^{\gamma } V^{\alpha } -\] 
\begin{equation} \label{232} 
-\frac{(1-\sqrt{1-V^{2} } )^{2} }{V^{4} } (\omega ^{\gamma \mu } V^{\gamma } V^{\mu } )V^{\alpha } V^{\beta } +\frac{1-\sqrt{1-V^{2} } }{V^{2} } \omega ^{\alpha \gamma } V^{\beta } V^{\gamma } -\frac{V^{\beta } \dot{V}^{\alpha } }{1-V^{2} }  
\end{equation} 

Уравнение \eqref{232} состоит из пяти слагаемых, четыре из которых существуют даже при равномерном движении $k$, а последнее является следствием влияния ускорения начала отсчёта. Очевидно, что $\Omega ^{\alpha \beta }$ не антисимметричен, а является тензором общего вида. Это обстоятельство резко противоречит существующей точке зрения \cite{23}, но, тем не менее, полностью удовлетворяет требованиям теории относительности. Из этой формулы в частности следует, что покоящийся в системе отсчёта k вектор $\mathbf{r}$ изменяется в лабораторной системе S с мгновенной аффинной скоростью
\begin{equation} \label{233} 
\Omega ^{\alpha \beta } =-\frac{V^{\beta } \dot{V}^{\alpha } }{1-V^{2} },\,\,\,\, \dot{V}^{\alpha}=\frac{dV^{\alpha}}{dT} 
\end{equation} 
Это впрочем, видно непосредственно из формулы \eqref{37}, в которую необходимо подставить \eqref{229} и результат сравнить с \eqref{226}. Данная формула хорошо согласуется с результатом статьи С. С. Степанова \cite[формула (14)]{30}.

 Формулу \eqref{232} можно несколько упростить вводя симметричный тензор $S^{\alpha \beta}$ и вектор $\Omega ^{\alpha}$ дуальный антисимметричной части $\Omega ^{\alpha \beta } $. По определению
\begin{equation} \label{234} 
\Omega ^{\alpha } =\frac{1}{4} e^{\alpha \mu \nu } \left(\Omega ^{\nu \mu } -\Omega ^{\mu \nu } \right).       
\end{equation} 
\begin{equation} \label{235} 
S^{\alpha \beta } =\frac{1}{2} \left(\Omega ^{\alpha \beta } +\Omega ^{\beta \alpha } \right),                                               
\end{equation} 
значит
\begin{equation} \label{236} 
\Omega ^{\alpha \beta } =S^{\alpha \beta } +e^{\alpha \gamma \beta } \Omega ^{\gamma }  
\end{equation} 
Физический смысл этих величин таков. Если выбрать такую систему отсчёта, в которой тензор $S^{\alpha \beta } $ принимает постоянную диагональную форму, то $\Omega ^{\alpha } $ представляет собой угловую скорость вращения его главных осей. При этом диагонализированный $S^{\alpha \beta } $ представляет собой их скорость растяжения. Подстановка \eqref{234}-\eqref{236} в \eqref{232} даёт
\[\Omega ^{\alpha } =\frac{\sqrt{1-V^{2} } e^{\alpha \mu \nu } (\omega ^{\nu \mu } -\omega ^{\mu \nu } )}{4} -\frac{\sqrt{1-V^{2} } (1-\sqrt{1-V^{2} } )}{4V^{2} } e^{\alpha \mu \nu } (\omega ^{\gamma \mu } V^{\nu } -\omega ^{\gamma \nu } V^{\mu } )V^{\gamma } +\] 
\begin{equation} \label{237} 
+\frac{1-\sqrt{1-V^{2} } }{4V^{2} } e^{\alpha \mu \nu } (\omega ^{\nu \gamma } V^{\mu } -\omega ^{\mu \gamma } V^{\nu } )V^{\gamma }  -\frac{e^{\alpha \mu \nu } }{4(1-V^{2} )} (V^{\mu } \dot{V}^{\nu } -V^{\nu } \dot{V}^{\mu } ) 
\end{equation} 
\[S^{\alpha \beta } =\frac{\sqrt{1-V^{2} } }{2} (\omega ^{\alpha \beta } +\omega ^{\beta \alpha } )-\frac{\sqrt{1-V^{2} } (1-\sqrt{1-V^{2} } )}{2V^{2} } (\omega ^{\gamma \beta } V^{\alpha } +\omega ^{\gamma \alpha } V^{\beta } )V^{\gamma } -\] 
\[-\frac{(1-\sqrt{1-V^{2} } )^{2} }{V^{4} } (\omega ^{\gamma \mu } V^{\gamma } V^{\mu } )V^{\alpha } V^{\beta } +\frac{1-\sqrt{1-V^{2} } }{2V^{2} } (\omega ^{\alpha \gamma } V^{\beta } +\omega ^{\beta \gamma } V^{\alpha } )V^{\gamma } -\] 
\begin{equation} \label{238} 
-\frac{V^{\beta } \dot{V}^{\alpha } +V^{\alpha } \dot{V}^{\beta } }{2\sqrt{1-V^{2} } }  
\end{equation} 

Подставляя в эти выражения разложение аналогичное формуле \eqref{236}
\begin{equation} \label{239} 
\omega ^{\alpha \beta } =s^{\alpha \beta } +e^{\alpha \gamma \beta } \omega ^{\gamma } ,                                           
\end{equation} 

где величины $s^{\alpha \beta } $, $\omega ^{\gamma } $ есть
\begin{equation} \label{240} 
\omega ^{\alpha } =\frac{1}{4} e^{\alpha \mu \nu } \left(\omega ^{\nu \mu } -\omega ^{\mu \nu } \right).                                            
\end{equation} 
\begin{equation} \label{241} 
s^{\alpha \beta } =\frac{1}{2} \left(\omega ^{\alpha \beta } +\omega ^{\beta \alpha } \right),                                                
\end{equation} 

и имеют тот же смысл, получим
\[\Omega ^{\alpha } =\sqrt{1-V^{2} } \omega ^{\alpha } -\frac{1}{2} e^{\alpha \mu \nu } s^{\gamma \mu } V^{\nu } V^{\gamma } +\frac{(1-\sqrt{1-V^{2} } )^{2} }{2V^{2} } e^{\alpha \mu \nu } e^{\gamma \lambda \mu } \omega ^{\lambda } V^{\nu } V^{\gamma } -\] 
\begin{equation} \label{242} 
-\frac{1}{2(1-V^{2} )} e^{\alpha \mu \nu } V^{\mu } \dot{V}^{\nu }  
\end{equation} 
\[S^{\alpha \beta } =\sqrt{1-V^{2} } s^{\alpha \beta } -\frac{1}{2} \left(e^{\gamma \lambda \beta } V^{\alpha } +e^{\gamma \lambda \alpha } V^{\beta } \right)\; \omega ^{\lambda } V^{\gamma } +\frac{(1-\sqrt{1-V^{2} } )^{2} }{2V^{2} } (s^{\alpha \gamma } V^{\beta } +s^{\beta \gamma } V^{\alpha } )V^{\gamma } -\] 
\begin{equation} \label{243} 
-\frac{(1-\sqrt{1-V^{2} } )^{2} }{V^{4} } (s^{\gamma \mu } V^{\gamma } V^{\mu } )V^{\alpha } V^{\beta } -\frac{V^{\beta } \dot{V}^{\alpha } +V^{\alpha } \dot{V}^{\beta } }{2(1-V^{2} )}  
\end{equation} 

Ранее рассматривалось движение точки относительно системы отсчёта k, система координат которой мгновенно совпадает с системой координат s, но не имеет собственного вращения. Если же возникнет необходимость рассмотреть аффинное движение относительно системы s, то получить формулы для аффинной скорости соответствующие \eqref{232}, \eqref{237}, \eqref{238}, \eqref{242}, \eqref{243}, будет нетрудно совершив в этих формулах замену
\begin{equation} \label{271}
\omega ^{\alpha \beta }=\omega ^{\alpha \beta }_{s} +e^{\alpha \gamma \beta } \Omega ^{\gamma }_{T} ,  
\end{equation}
В итоге формулы станут немного сложнее. Например эта подстановка в формулу \eqref{232} даёт
\[\Omega ^{\alpha \beta } =\sqrt{1-V^{2} } \omega ^{\alpha \beta }_{s} -\frac{\sqrt{1-V^{2} } (1-\sqrt{1-V^{2} } )}{V^{2} } \omega ^{\gamma \beta }_{s} V^{\gamma } V^{\alpha } - 
\frac{(1-\sqrt{1-V^{2} } )^{2} }{V^{4} } (\omega ^{\gamma \mu }_{s} V^{\gamma } V^{\mu } )V^{\alpha } V^{\beta }+\]
\[ +\frac{1-\sqrt{1-V^{2} } }{V^{2} } \omega ^{\alpha \gamma }_{s} V^{\beta } V^{\gamma }
 -\frac{1-\sqrt{1-V^{2} } }{V^{2}} V^{\alpha}\dot{V}^{\beta}-\frac{1-\sqrt{1-V^{2} } }{V^{2}\sqrt{1-V^2}} V^{\beta}\dot{V}^{\alpha}- \]
 \begin{equation} \label{272} 
 -\frac{1-\sqrt{1-V^{2} }  }{V^{2} (1-V^{2}) }(\mathbf{\dot{V}V})V^{\alpha}V^{\beta}
\end{equation} 
При этом новая величина $\omega ^{\alpha \beta }_{s}$ имеет смысл аффинной скорости относительно системы s.  Если в системе s $\omega ^{\alpha \beta }_{s}=0$, то после некоторых вычислений можно получить, что
\begin{equation} \label{275} 
\mathbf{\Omega} =-\frac{(1-\sqrt{1-V^2})^2}{2V^2\sqrt{1-V^2}} \mathbf{V \times\dot{V}}  
\end{equation} 
\begin{equation} \label{276} 
S^{\alpha \beta } =-\frac{1-\sqrt{1-V^{2} }  }{V^{2}(1-V^2) } (\mathbf{V \dot{V}} )V^{\alpha } V^{\beta } -\frac{V^{\beta } \dot{V}^{\alpha } +V^{\alpha } \dot{V}^{\beta } }{2\sqrt{1-V^2}}  
\end{equation} 

Выражение \eqref{230} можно представить и в виде другой, более наглядной суммы 
\begin{equation} \label{244} 
 \mathbf{U-V=\dot{L}=\Omega_{n} \times L}+k_{\mathbf{n}} \,\mathbf{L},  
\end{equation} 
где $\mathbf{\Omega _{n} L}=0$.
Здесь $\mathbf{\Omega _{n}} $ является угловой скоростью вращения вектора $\mathbf{n}$, а $k_{\mathbf{n}} $ - коэффициент растяжения вдоль направления $\mathbf{n}$. Приравнивая \eqref{244} к \eqref{230}, умножая обе части получившегося равенства векторно на $\mathbf{L}$ и разделив на $L^{2} $ можно найти, что 
\[\Omega _{\bf{n}}^{\sigma } =\sqrt{1-V^{2} } e^{\sigma \tau \alpha } \omega ^{\alpha \beta } n^{\tau } n^{\beta } +\frac{1-\sqrt{1-V^{2} } }{V^{2}} \mathbf{(nV)}\; e^{\sigma \tau \alpha } \omega ^{\alpha \gamma } n^{\tau } V^{\gamma } -\] 
\[-\frac{\sqrt{1-V^{2} } (1-\sqrt{1-V^{2} } )}{V^{2} } (\omega ^{\gamma \beta } V^{\gamma } n^{\beta } )\; e^{\sigma \tau \alpha } \, n^{\tau } \, V^{\alpha } -\] 
\begin{equation} \label{245} 
-\frac{(1-\sqrt{1-V^{2} } )^{2} }{V^{4} } \mathbf{(nV)}(\omega ^{\gamma \mu } V^{\gamma } V^{\mu } )\; e^{\sigma \tau \alpha } \, n^{\tau } V^{\alpha } -\frac{e^{\sigma \tau \alpha } n^{\tau } \dot{V}^{\alpha} \mathbf{(nV)}} {1-V^{2} } ,\,\,\,\,\,\,  \mathbf{n}=\frac{\mathbf{L}}{L}  
\end{equation} 
Таким образом, очевидно, что каждое направление $\mathbf{n}$ в данный момент вращается относительно $S$ со своей угловой скоростью. Из \eqref{245} кроме того следует, что если движение в системе $k$ отсутствует ($\omega ^{\alpha \beta} =0$), то любое направление $n^{\tau}$ перпендикулярное вектору скорости $\mathbf{V}$ или параллельное вектору ускорения $\mathbf{\dot{V}}$ не вращается в лабораторной системе отсчёта $S$. 

Если же в подстановке \eqref{244} в \eqref{230} умножить обе части получившегося равенства скалярно на $\mathbf{L}$ и разделить на $L^{2} $, то получится, что коэффициент растяжения $k_{\mathbf{n}} $ равен
\[k_{\mathbf{n}} =\sqrt{1-V^{2} } \omega ^{\alpha \beta } n^{\alpha } n^{\beta } +\frac{1-\sqrt{1-V^{2} } }{V^{2} } \mathbf{(nV)}\, (\omega ^{\alpha \gamma } n^{\alpha } V^{\gamma } )-\] 
\[-\frac{\sqrt{1-V^{2} } (1-\sqrt{1-V^{2} } )}{V^{2} } (\omega ^{\gamma \beta } V^{\gamma } n^{\beta } )\mathbf{(Vn)}-\frac{(1-\sqrt{1-V^{2} } )^{2} }{V^{4} } \mathbf{(Vn)}^{2} (\, \omega ^{\gamma \mu } V^{\gamma } V^{\mu } )-\] 
\begin{equation} \label{246} 
-\frac{\mathbf{(Vn)(\dot{V}n)}}{1-V^{2} }  
\end{equation} 
Подставив в это выражение разложение \eqref{239} получим
\[k_{\mathbf{n}} =\sqrt{1-V^{2} } s^{\alpha \beta } n^{\alpha } n^{\beta } +\mathbf{(Vn)}\, \left[\mathbf{n}(\boldsymbol{\omega} \times \mathbf{V})\right]-\] 
\[+\frac{(1-\sqrt{1-V^{2} } )^{2} }{V^{2} } (s^{\gamma \beta } V^{\gamma } n^{\beta } )\mathbf{(Vn)}-\frac{(1-\sqrt{1-V^{2} } )^{2} }{V^{4} } (\mathbf{Vn})^{2} (\, s^{\gamma \mu } V^{\gamma } V^{\mu } )-\] 
\begin{equation} \label{247} 
-\frac{\mathbf{(Vn)(\dot{V}n)}}{1-V^{2} }  
\end{equation} 

В том случае, если аффинное движение в системе $k$ является чистым вращением ($s^{\alpha \beta } =0$) данная формула примет вид
\begin{equation} \label{248} 
k_{\bf{n}} =\mathbf{(nV)}\left[\mathbf{n}\, \left(\boldsymbol{\omega} \times\mathbf{V}-\frac{\dot{\mathbf{V}}}{1-V^{2} } \right)\right], \,\,\,\, \mathbf{n}=\frac{\mathbf{L}}{L}  
\end{equation} 
Отсюда очевидно, что система координат жёсткой неинерциальной системы отсчёта при её движении относительно лабораторной системы $S$ не деформируется только в плоскости перпендикулярной вектору $\mathbf{V}$ и в плоскости перпендикулярной вектору $\boldsymbol{\omega} \times\mathbf{V}- \dot{\mathbf{V}}/(1-V^{2})$. Таким образом, точка системы координат $k$ движется относительно $S$ не только поступательно со скоростью $\mathbf{V}$ и вращаясь вокруг центра с некоторой угловой скоростью, но и имеет составляющую скорости соответствующую растяжению относительно тела отсчёта. 

\subsection* {5. Преобразование лабораторной аффинной скорости в сопутствующую невращающуюся систему отсчёта}

Зависимость $\omega ^{\alpha \beta }$ от  $\Omega^{\alpha \beta}$ можно было бы найти непосредственно из \eqref{232}, но можно поступить по-другому и вывести её из \eqref{24}. Подставим в \eqref{24} для малых $\mathbf{r}$ выражение \eqref{226} и соотношение \eqref{37} между функцией преобразования $\mathbf{v}(t)$ и скоростью $\mathbf{V}$ начала координат системы s. Ответ необходимо искать с точностью до членов пропорциональных первой степени расстояния. При этом заметим, что второй сомножитель числителя правой части \eqref{24} уже пропорционален компоненте $r^{\alpha } $, поэтому с точностью до первых степеней $r^{\alpha } $ первый сомножитель можно положить равным 1, а знаменатель равным $1-V^{2} $. Далее используя \eqref{37} нетрудно подсчитать, что 
\begin{equation} \label{249} 
\sqrt{1-v^{2} } =\sqrt{1-V^{2} } +\frac{(\mathbf{\dot{V}V)(Vr)}}{1-V^{2} }  
\end{equation} 
\[\frac{1-\sqrt{1-v^{2} } }{v^{2} } v^{\alpha } v^{\beta } =\frac{1-\sqrt{1-V^{2} } }{V^{2} } V^{\alpha } V^{\beta } -\frac{(1-\sqrt{1-V^{2} } )^{2} }{V^{4} (1-V^{2} )} (\mathbf{V\dot{V})(Vr)}V^{\alpha } V^{\beta } -\] 
\begin{equation} \label{250} 
-\frac{1-\sqrt{1-V^{2} } }{V^{2} \sqrt{1-V^{2} } } \mathbf{(Vr)}(\dot{V}^{\alpha } V^{\beta } +\dot{V}^{\beta } V^{\alpha } ) 
\end{equation} 

Учитывая эти соображения можно найти, что
\[u^{\alpha } =\frac{1-\sqrt{1-V^{2} } }{V^{2} \sqrt{1-V^{2} } ^{3} } (\mathbf{\dot{V}V)(Vr)}\, V^{\alpha } +\frac{\mathbf{(Vr)}}{1-V^{2} } \dot{V}^{\alpha } +\frac{\Omega ^{\alpha \beta } L^{\beta }}{\sqrt{1-V^{2} } }  +\] 
\begin{equation} \label{251} 
+\frac{1-\sqrt{1-V^{2} } }{V^{2} (1-V^{2} )} (\Omega ^{\beta \gamma } V^{\beta } L^{\gamma } )V^{\alpha }  
\end{equation} 
где $L^{\alpha } $ определяется из \eqref{82}. Следовательно, если подставить эту формулу в \eqref{251} и сравнить получившееся выражение с \eqref{225} получим, что
\[\omega ^{\alpha \beta } =\frac{\Omega ^{\alpha \beta } }{\sqrt{1-V^{2} } } +\frac{1-\sqrt{1-V^{2} } }{V^{2} (1-V^{2} )} \Omega ^{\gamma \beta } V^{\alpha } V^{\gamma } -\frac{(1-\sqrt{1-V^{2} } )^{2} }{V^{4} (1-V^{2} )} (\Omega ^{\gamma \mu } V^{\gamma } V^{\mu } )V^{\alpha } V^{\beta } -\] 
\begin{equation} \label{252} 
-\frac{1-\sqrt{1-V^{2} } }{V^{2} \sqrt{1-V^{2} } } \Omega ^{\alpha \gamma } V^{\gamma } V^{\beta } +\frac{1-\sqrt{1-V^{2} } }{V^{2} \sqrt{1-V^{2} } ^{3} } (\mathbf{\dot{V}V)}\, V^{\alpha } V^{\beta } +\frac{V^{\beta } \dot{V}^{\alpha } }{1-V^{2} } =invariant 
\end{equation} 
Данная величина должна быть инвариантом при выполнении буста.

Из \eqref{252} можно найти угловую скорость $\omega ^{\alpha } $ главных осей аффинного тензора \eqref{240} и тензор скорости центрального растяжения $s^{\alpha \beta } $ \eqref{241}. Вычисления дают
\[\omega ^{\alpha } =\frac{e^{\alpha \mu \nu } (\Omega ^{\nu \mu } -\Omega ^{\mu \nu } )}{4\sqrt{1-V^{2} } } -\frac{1-\sqrt{1-V^{2} } }{4V^{2} \sqrt{1-V^{2} } } e^{\alpha \mu \nu } V^{\gamma } (\Omega ^{\nu \gamma } V^{\mu } -\Omega ^{\mu \gamma } V^{\nu } )+\] 
\begin{equation} \label{253} 
+\frac{1-\sqrt{1-V^{2} } }{4V^{2} (1-V^{2} )} e^{\alpha \mu \nu } V^{\gamma } (\Omega ^{\gamma \mu } V^{\nu } -\Omega ^{\gamma \nu } V^{\mu } )+\frac{e^{\alpha \mu \nu } }{2(1-V^{2} )} V^{\mu } \dot{V}^{\nu }  
\end{equation} 
\[s^{\alpha \beta } =\frac{\Omega ^{\alpha \beta } +\Omega ^{\beta \alpha } }{2\sqrt{1-V^{2} } } - \frac{1-\sqrt{1-V^{2} } }{2V^{2} \sqrt{1-V^{2} } } V^{\gamma } (\Omega ^{\alpha \gamma } V^{\beta } 
+\Omega ^{\beta \gamma } V^{\alpha } )+\]
\[+\frac{1-\sqrt{1-V^{2} } }{2V^{2} (1-V^{2} )} V^{\gamma } (\Omega ^{\gamma \beta } V^{\alpha } +\Omega ^{\gamma \alpha } V^{\beta } )-\frac{(1-\sqrt{1-V^{2} } )^{2} }{V^{4} (1-V^{2} )} (\Omega ^{\gamma \mu } V^{\gamma } V^{\mu } )V^{\alpha } V^{\beta}+\] 
\begin{equation} \label{254}
+\frac{1-\sqrt{1-V^{2} } }{V^{2} \sqrt{1-V^{2} } ^{3} } (\mathbf{V\dot{V}})V^{\alpha } V^{\beta } +\frac{\dot{V}^{\alpha } V^{\beta } +V^{\alpha } \dot{V}^{\beta } }{2(1-V^{2} )} .       
\end{equation} 

Используя симметричный тензор $S^{\alpha \beta } $ и угловую скорость вращения $\Omega ^{\alpha } $ согласно \eqref{234}-\eqref{236} находим
\begin{equation} \label{255} 
\omega ^{\alpha } =\frac{\Omega ^{\alpha } }{\sqrt{1-V^{2} } } +\frac{(1-\sqrt{1-V^{2} } )^{2} }{2V^{2} (1-V^{2} )} e^{\alpha \nu \mu } e^{\mu \lambda \gamma } \Omega ^{\lambda } V^{\nu } V^{\gamma } +\frac{e^{\alpha \mu \nu } V^{\mu } \dot{V}^{\nu } }{2(1-V^{2} )} +\frac{e^{\alpha \mu \nu } V^{\gamma } V^{\nu } S^{\gamma \mu } }{2(1-V^{2} )}  
\end{equation} 
\[s^{\alpha \beta } =\frac{S^{\alpha \beta } }{\sqrt{1-V^{2} } } +\frac{(1-\sqrt{1-V^{2} } )^{2} }{2V^{2} (1-V^{2} )} V^{\gamma } (S^{\alpha \gamma } V^{\beta } +S^{\beta \gamma } V^{\alpha } )-\frac{(1-\sqrt{1-V^{2} } )^{2} }{V^{4} (1-V^{2} )} (S^{\gamma \mu } V^{\gamma } V^{\mu } )V^{\alpha } V^{\beta } +\] 
\begin{equation} \label{256} 
+\frac{(\dot{V}^{\alpha } -e^{\alpha \lambda \gamma } \Omega ^{\lambda } V^{\gamma } )V^{\beta } +V^{\alpha } (\dot{V}^{\beta } -e^{\beta \lambda \gamma } \Omega ^{\lambda } V^{\gamma } )}{2(1-V^{2} )} +\frac{1-\sqrt{1-V^{2} } }{V^{2} \sqrt{1-V^{2} } ^{3} } (\mathbf{V\dot{V}})V^{\alpha } V^{\beta }  
\end{equation} 

\subsection*  {6. Жёстко вращающееся тело}

Применим полученные формулы к достаточно малой области $A$ около координаты $\mathbf{R}$ жёстко вращающегося тела, все точки которого вращаются вокруг начала координат лабораторной инерциальной системы отсчёта со скоростью 
\begin{equation} \label{257} 
\mathbf{V=\Omega \times R} 
\end{equation} 

Жёсткость вращения относительно лабораторной системы отсчёта означает отсутствие скорости центрального растяжения области $A$ ($S^{\alpha \beta } =0$). Поэтому последний член в \eqref{255} исчезнет. В результате получим в векторном виде, что относительно системы отсчёта k
\begin{equation} \label{258} 
\boldsymbol{\omega} =\frac{\mathbf{\Omega}}{\sqrt{1-V^{2} } } +\frac{(1-\sqrt{1-V^{2} } )^{2} }{2V^{2} (1-V^{2} )} \mathbf{V\times (\Omega \times V)}+\frac{\mathbf{V\times \dot{V}}}{2(1-V^{2} )}  
\end{equation} 

Величина $\mathbf{\dot{V}}$ равна
\begin{equation} \label{259} 
\mathbf{\dot{V}=\Omega \times V+\dot{\Omega }\times R} 
\end{equation} 
Подставляя её в \eqref{258}, раскрывая скобки, приводя подобные члены и учитывая, что  $\mathbf{V\Omega} =0$ получим
\begin{equation} \label{260} 
\boldsymbol{\omega} =\frac{\mathbf{\Omega}}{1-V^{2} } +\frac{\mathbf{V\times (\dot{\Omega }\times R)}}{2(1-V^{2} )}  
\end{equation} 
 Эта формула даёт мгновенную локальную угловую скорость вращения главных осей области вращающегося тела вблизи начала отсчёта системы k.

Интересно специально рассмотреть случай раскручивающегося тела, когда вектор углового ускорения тела параллелен вектору его угловой скорости. Тогда $\mathbf{V\times (\dot{\Omega }\times R)}=0$ и формула \eqref{260} радикально упрощается
\begin{equation} \label{261} 
\boldsymbol{\omega} =\frac{\mathbf{\Omega}}{1-V^{2} }  
\end{equation} 
Данное значение уже известно из различных источников \cite{31}, \cite [с. 59]{36}, \cite [формула (31)]{15}.

Рассмотрим теперь деформацию системы координат. Скорость центрального растяжения определяется тензором $s^{\alpha \beta } $ из формулы \eqref{256}. В этой формуле первые три члена в правой части из-за жёсткости движения исчезают. Подставляя сюда \eqref{259} в результате получим
\begin{equation} \label{262} 
s^{\alpha \beta } =\frac{1-\sqrt{1-V^{2} } }{V^{2} \sqrt{1-V^{2} } ^{3} } (V^{\mu } e^{\mu \nu \sigma } \dot{\Omega }^{\nu } R^{\sigma } )\; V^{\alpha } V^{\beta } +\frac{e^{\alpha \delta \gamma } \dot{\Omega }^{\delta } R^{\gamma } V^{\beta } +V^{\alpha } e^{\beta \delta \gamma } \dot{\Omega }^{\delta } R^{\gamma } }{2(1-V^{2} )}  
\end{equation} 
Таким образом, если тело относительно осевого наблюдателя вращается жёстко и равномерно с постоянной угловой скоростью ($\mathbf{\dot{\Omega }}=0$), то тензор скорости деформации $s^{\alpha \beta } $ в т. А равен нулю. Это означает, что ближайшие к наблюдателю точки тела вращаются без центрального растяжения, т. е. тоже как твёрдое тело. В том же случае, если тело вращается с угловым ускорением, скорость деформации, вообще говоря, ненулевая. 

Рассматривая тензор скорости деформации, предварительно ориентируем оси системы координат так, чтобы ось 2 была в направлении вектора угловой скорости $\mathbf{\Omega} $, вектор $\mathbf{R}$ был направлением оси 3, при этом вектор скорости $\mathbf{V=\Omega \times R}$ будет направлением оси 1. Тогда тензор $s^{\alpha \beta } $ диагонализируется и единственной ненулевой компонентой будет его главное значение $s^{1} $ равное из \eqref{262}
\begin{equation} \label{263} 
s^{1} =s^{11} =\frac{V\dot{\Omega }R}{\sqrt{1-V^{2} } ^{3} } =\frac{R^{2} \Omega \dot{\Omega }}{\sqrt{1-R^{2} \Omega ^{2} } ^{3} }  
\end{equation} 

В системе отсчёта связанной с главными осями центроаффинное движение точки с начальной координатой до начала вращения $r^{1}$ подчиняется уравнению
\begin{equation} \label{264} 
u^{1} =s^{1} r^{1} .                                                      
\end{equation} 
Интегрируя это уравнение, получим, что конечное положение этой точки $r^{1}(t) $ определяется коэффициентом относительного растяжения оси 1 $\lambda ^{1} $
\begin{equation} \label{265} 
r^{1}(t) -r^{1} =\lambda ^{1} r^{1} ,                                                 
\end{equation} 
где 
\begin{equation} \label{266} 
\lambda ^{1} =\frac{r^{1}(t) -r^{1} }{r^{1} } =\int _{0}^{t}s^{1} dt  
\end{equation} 

Вычислим собственное относительное растяжение области $A$ $\lambda ^{1}$, за всё время вращения из состояния покоя до угловой скорости $\Omega (t)$ согласно формуле \eqref{266}.  Получим
\begin{equation} \label{267} 
\lambda ^{1} =\frac{\Delta r^{1} }{r^{1} } =\int _{0}^{t}\frac{R^{2} \Omega \dot{\Omega }}{\sqrt{1-R^{2} \Omega ^{2} } ^{3} } dt =\frac{1}{\sqrt{1-R^{2} \Omega ^{2} (t)} } -1,                           
\end{equation} 
\begin{equation} \label{268} 
\frac{\Delta r^{2} }{r^{2} } =\frac{\Delta r^{3} }{r^{3} } =0 
\end{equation} 
Отсюда 
\begin{equation} \label{269} 
r^{1}(t) =\frac{r^{1} }{\sqrt{1-R^{2} \Omega ^{2} (t)} }  
\end{equation} 
Этот расчёт прекрасно согласуется с известным фактом, что у вращающегося диска в процессе ускоренного вращения его материал растягивается в направлении оси 1 как раз в $1/\sqrt{1-R^2 \Omega^2(t)}$ раз.

\subsection*  {7. Некоторые свойства общего преобразования ЛМН. Вращение Вигнера}
\subsubsection* {Единственность специального преобразования ЛМН}
Специальное преобразование ЛМН является единственным возможным голономным преобразованием при условии его соответствия неголономному дифференциальному преобразованию Лоренца, если движение в данный момент времени стало равномерным.  Рассмотрим это обстоятельство чуть более подробно. 

Действительно, в том случае, когда система $s$ становится инерциальной системой отсчёта ( т. е. $\mathbf{v}=const$ ), малые приращения координат $dR^{\alpha}$  и времени $dT$  системы S должны перейти в  приращения определяемые из неголономного дифференциального преобразования Лоренца 
\begin{equation} \label{91} 
dX^{i} =\Lambda _{\; k}^{i} dx^{k}   
\end{equation} 
или
\begin{equation} \label{92}  
dT=\frac{dt}{\sqrt{1-v^{2} } } +\frac{\mathbf{v}d\mathbf{r}}{\sqrt{1-v^{2} } } 
\end{equation}
\begin{equation} \label{93}     
d\mathbf{R}=\frac{\mathbf{v}dt}{\sqrt{1-v^{2} } } +d\mathbf{r}+\frac{1-\sqrt{1-v^{2} } }{v^{2} \sqrt{1-v^{2} } } (\mathbf{v}d\mathbf{r)v}
\end{equation}

Если считать, что функция  $\mathbf{v}$ есть функция времени s, то наиболее просто специальное преобразование ЛМН \eqref{10}, \eqref{11} можно получить производя тривиальное интегрирование \eqref{92}, \eqref{93}. При интегрировании надо учесть, что коэффициенты $\Lambda _{\; k}^{i}$ зависят только от собственного мирового времени $t$. Постоянные интегрирования можно считать равными нулю, если в нулевой момент времени по часам $s$ и $S$ начала координат обеих систем отсчёта совпадают. 

Если теперь произвести обратное дифференцирование уравнений \eqref{10}, \eqref{11}, то, полагая затем $\mathbf{\dot{v}}=0$ , эти дифференциалы перейдут в \eqref{92}, \eqref{93}. Частное преобразование ЛМН является единственным преобразованием, обладающим таким свойством. Это обстоятельство легко доказать, прибавляя к правой части \eqref{10} некую скалярную функцию $f(t,\mathbf{r})$ , а к правой части \eqref{11} векторную функцию $\mathbf{f}(t,\mathbf{r})$ , дифференцируя и полагая затем  $\mathbf{\dot{v}}=0$. В результате получим, что 
\begin{equation} \label{94}  
f(t,\mathbf{v,r})=\mathbf{hv}+const,\,\,\,\, f^{\alpha}(t,\mathbf{v,r})=h^{\alpha\beta}v^{\beta}+const
\end{equation}
где  $\mathbf{h}$, $h^{\alpha\beta}$ - некоторые постоянные вектор и тензор соответственно. То, что они вообще равны нулю видно из сравнения получившегося преобразования с голономным преобразованием Лоренца. Следовательно данные функции сводятся к постоянным, что соответствует простым изменениям начала отсчёта времени и начала отсчёта системы координат. Таким образом, специальное, а значит и общее преобразование ЛМН является единственным. 

\subsubsection* {Форминвариантность общего преобразования ЛМН}
Ещё одним свойством общего преобразования ЛМН является его форминвариантность при произвольном повороте и при переходе в другую инерциальную систему отсчёта, как это и должно быть. Ввиду того, что координаты $R^{\alpha}$ представляют собой вектор, форминвариантность преобразования относительно поворота очевидна. Чтобы доказать неизменяемость его относительно буста, условимся, что система $s'$:$(t,\mathbf{r'})$ движется с параметрами преобразования $\mathbf{v}^*$, $a^{*\alpha\beta}$ относительно некоторой инерциальной системы отсчёта $S^*$:$(T^*,\mathbf{R^*})$, совершим переход из системы отсчёта $S^*$ в движущуюся с постоянной скоростью $\mathbf{u}$ относительно неё инерциальную систему отсчёта $S$:$(T,\mathbf{R})$ и покажем, что преобразование 
\begin{equation} \label{402} 
T^*=\frac{v^{*\alpha } a^{*\beta \alpha} r'^{\beta } }{\sqrt{1-v^{*2} } } +\int _{0}^{t}\frac{dt}{\sqrt{1-v^{*2} } }  
\end{equation}
\begin{equation} \label{403} 
 R^{*\alpha } =\left\{a^{ *\beta \alpha} r'^{\beta } +\frac{1-\sqrt{1-v^{*2} } }{v^{*2} \sqrt{1-v^{*2} } } v^{*\alpha } v^{*\gamma } a^{*\beta \gamma} r'^{\beta } \right\}+\int _{0}^{t}\frac{v^{*\alpha } dt}{\sqrt{1-v^{*2} } } 
  \end{equation}
в новой системе отсчёта $S$, которую можно принять за лабораторную, не изменится. Координаты и время в системах отсчёта $S$, $S^*$ связаны обычным преобразованием Лоренца         
\begin{equation} \label{95} 
T=\frac{T^*-\mathbf{uR^*}}{\sqrt{1-u^{2} } }  
\end{equation} 
\begin{equation} \label{96} 
\mathbf{R=R^*}+\frac{1-\sqrt{1-u^{2} } }{u^{2} \sqrt{1-u^{2} } }\mathbf{(uR^*)u}-\frac{\mathbf{u}T^*}{\sqrt{1-u^{2} } } .                                 
\end{equation} 

Рассмотрим сначала специальное преобразование ЛМН в систему $s$: $(t,\mathbf{r}) $($a^{*\alpha\beta}=\delta^{\alpha\beta}$). Подстановка уравнений \eqref{95}, \eqref{96} в \eqref{402}, \eqref{403} даёт ($\mathbf{r'=r}$)
\begin{equation} \label{97} 
T=\left\{\frac{\mathbf{v^*}}{\sqrt{1-u^{2} } \sqrt{1-v^{*2} } } -\frac{\mathbf{u}}{\sqrt{1-u^{2} } } -\frac{1-\sqrt{1-v^{*2} } }{v^{*2} \sqrt{1-v^{*2} } } \frac{\mathbf{(uv^*)v^*}}{\sqrt{1-u^{2} } } \right\}\mathbf{r}+\int _{0}^{t}\left(\frac{1-\mathbf{uv^*}}{\sqrt{1-u^{2} } \sqrt{1-v^{*2} } } \right) \; dt 
\end{equation} 
\[\mathbf{R}=\left\{\mathbf{r}+\frac{1-\sqrt{1-u^{2} } }{u^{2} \sqrt{1-u^{2} } } \right. \mathbf{(ur)u}+\frac{1-\sqrt{1-v^{*2} } }{v^{*2} \sqrt{1-v^{*2} } } \mathbf{(v^*r)v^*}+\frac{(1-\sqrt{1-v^{*2} } )(1-\sqrt{1-u^{2} } )}{v^{*2} u^{2} \sqrt{1-v^{*2} } \sqrt{1-u^{2} } } \mathbf{(uv^*)(v^*r)u}-\] 
\begin{equation} \label{98} 
-\left. \frac{\mathbf{(v^*r)u}}{\sqrt{1-u^{2} } \sqrt{1-v^{*2} } } \right\}+\int _{0}^{t}\left(\frac{\mathbf{v^*}}{\sqrt{1-v^{*2} } } -\frac{\mathbf{u}}{\sqrt{1-u^{2} } \sqrt{1-v^{*2} } } +\frac{1-\sqrt{1-u^{2} } }{u^{2} \sqrt{1-u^{2} } \sqrt{1-v^{*2} } } \mathbf{(uv^*)u}\right) \; dt 
\end{equation} 

Обозначим 
\begin{equation} \label{99} 
\frac{\mathbf{v^*}}{\sqrt{1-v^{*2} } } -\frac{\mathbf{u}}{\sqrt{1-u^{2} } \sqrt{1-v^{*2} } } +\frac{1-\sqrt{1-u^{2} } }{u^{2} \sqrt{1-v^{*2} } } \frac{\mathbf{(uv^*)u}}{\sqrt{1-u^{2} } } =\frac{\mathbf{v}}{\sqrt{1-v^{2} } } ,                         
\end{equation} 

где
\begin{equation} \label{100} 
\mathbf{v}=\frac{\sqrt{1-u^{2} }\mathbf{v^*-u}}{1-\mathbf{uv^*}} +\frac{(1-\sqrt{1-u^{2} } )\mathbf{(uv^*)u}}{u^{2} (1-\mathbf{uv^*})} ,                          
\end{equation} 

а
\begin{equation} \label{101} 
\frac{\mathbf{v^*}}{\sqrt{1-v^{*2} } \sqrt{1-u^{2} } } -\frac{\mathbf{u}}{\sqrt{1-u^{2} } } -\frac{1-\sqrt{1-v^{*2} } }{v^{*2} \sqrt{1-v^{*2} } } \frac{\mathbf{(uv^*)v^*}}{\sqrt{1-u^{2} } } =\frac{\mathbf{v'}}{\sqrt{1-v'^{2} } } ,               
\end{equation} 

где
\begin{equation} \label{102} 
\mathbf{v'}=\frac{\mathbf{v^*}-\sqrt{1-v^{*2} } \mathbf{u}}{1-\mathbf{uv^*}} -\frac{(1-\sqrt{1-v^{*2} })\mathbf{(uv^*)v^*}}{v^{*2}(1-\mathbf{uv^*})} ,
\end{equation} 

\[\mathbf{r}+\frac{1-\sqrt{1-u^{2} } }{u^{2} \sqrt{1-u^{2} } }  \mathbf{(ur)u}+\frac{1-\sqrt{1-v^{*2} } }{v^{*2} \sqrt{1-v^{*2} } } \mathbf{(v^*r)v^*}+\frac{(1-\sqrt{1-v^{*2} } )(1-\sqrt{1-u^{2} } )}{v^{*2} u^{2} \sqrt{1-v^{*2} } \sqrt{1-u^{2} } } \mathbf{(uv^*)(v^*r)u}-\] 
\begin{equation} \label{103} 
- \frac{\mathbf{(v^*r)u}}{\sqrt{1-u^{2} } \sqrt{1-v^{*2} } }=\mathbf{A}
\end{equation}
Физический смысл $\mathbf{v}$ заключается в том, что эта величина является скоростью другой системы отсчёта $s$ (начало которой двигается также как и $s'$, а оси которой во время её движения двигаются условно поступательно) относительно системы $S$ (см. замечание после формулы \eqref{24}). Величина же $\mathbf{v'}$, как далее будет видно, является этой скоростью в системе координат $s'$, имеющей другую ориентацию относительно $s$.  

\noindent Тогда преобразование \eqref{97}, \eqref{98} можно привести к виду
\begin{equation} \label{104} 
T=\frac{\mathbf{v'r}}{\sqrt{1-v'^{2} } } +\int _{0}^{t}\frac{dt}{\sqrt{1-v^{2} } }   
\end{equation} 
\begin{equation} \label{105} 
\mathbf{R}=\mathbf{A}+\int _{0}^{t}\frac{\mathbf{v}dt}{\sqrt{1-v^{2} } }   
\end{equation} 
где 
\begin{equation} \label{106}
v^{2} =v'^{2} =\frac{(\mathbf{u-v^*})^{2} -(\mathbf{u\times v^*})^{2} }{(1-\mathbf{uv^*})^{2} },\,\,\, \sqrt{1-v^{2} } =\frac{\sqrt{1-u^{2} } \sqrt{1-v^{*2} } }{1-\mathbf{uv^*}}  
\end{equation} 
Примем теперь $\mathbf{A}$ равным
\begin{equation} \label{107}
\mathbf{A}=\boldsymbol{\rho} +\frac{1-\sqrt{1-v^{2} } }{v^{2} \sqrt{1-v^{2} } } (\mathbf{v\boldsymbol{\rho} )v} 
\end{equation} 
где $\boldsymbol{\rho} $ есть некоторая величина подлежащая определению, причём из этого уравнения и \eqref{103} ясно, что
\begin{equation} \label{114} 
\rho ^{\alpha } =b^{ \beta \alpha } r^{\beta }  
\end{equation} 
с некоторым $b^{ \beta \alpha }$. 
Из \eqref{107} $\boldsymbol{\rho} $  будет равно 
\begin{equation} \label{108} 
\boldsymbol{\rho} =\mathbf{A}-\frac{1-\sqrt{1-v^{2} } }{v^{2} } \mathbf{(vA)v}, 
\end{equation} 
а из \eqref{108} видно, что
\begin{equation} \label{118} 
\mathbf{v}\boldsymbol{\rho} =\sqrt{1-v^{2} }\,\mathbf{vA}. 
\end{equation} 
С  другой стороны, перемножая $\mathbf{v}$ и $\mathbf{A}$, используя \eqref{100} и \eqref{103}, раскрывая скобки и приводя подобные члены получим
\begin{equation} \label{111} 
\mathbf{vA}=\frac{\mathbf{v^*r}}{\sqrt{1-u^{2} } \sqrt{1-v^{*2} } } -\frac{\mathbf{ur}}{\sqrt{1-u^{2} } } -\frac{1-\sqrt{1-v^{*2} } }{v^{*2} \sqrt{1-v^{*2} } \sqrt{1-u^{2} } }\mathbf{(uv^*)(v^*r)}. 
\end{equation} 
Из \eqref{111} видно, учитывая значение $\mathbf{v'}/\sqrt{1-v'^2}$ из \eqref{101}, что 
\begin{equation} \label{119} 
\mathbf{vA}=\frac{1}{\sqrt{1-v^{2} } } \,\,\mathbf{v'\,r} 
\end{equation} 
Подставляя \eqref{119} в \eqref{118}, видно, что 
\begin{equation} \label{120} 
\mathbf{v'r=v}\boldsymbol{\rho}  
\end{equation} 
и вектор $\mathbf{v'}$ связан со скоростью соотношением
\begin{equation} \label{121} 
v'^{\beta}=b^{\beta\alpha}v^{\alpha},  
\end{equation}
т.е. является вектором скорости в новой системе координат.

С учётом \eqref{107} и равенства \eqref{120} преобразование \eqref{104}, \eqref{105} окончательно выглядит в виде
\begin{equation} \label{122} 
T=\frac{\mathbf{v}\boldsymbol{\rho}}{\sqrt{1-v^{2} } } +\int _{0}^{t}\frac{dt}{\sqrt{1-v^{2} } }   
\end{equation} 
\begin{equation} \label{123} 
\mathbf{R}=\left(\boldsymbol{\rho} +\frac{1-\sqrt{1-v^{2} } }{v^{2} \sqrt{1-v^{2} } } (\mathbf{v}\boldsymbol{\rho} )\mathbf{v}\right)+\int _{0}^{t}\frac{\mathbf{v}dt}{\sqrt{1-v^{2} } }   
\end{equation} 
В матричном виде это преобразование имеет вид общего преобразования ЛМН
\begin{equation} \label{124} 
T=\frac{v^{\alpha } b^{ \beta \alpha} r^{\beta } }{\sqrt{1-v^{2} } } +\int _{0}^{t}\frac{dt}{\sqrt{1-v^{2} } }   
\end{equation} 
\begin{equation} \label{125} 
R^{\alpha } =\left\{b^{ \beta \alpha } r^{\beta } +\frac{1-\sqrt{1-v^{2} } }{v^{2} \sqrt{1-v^{2} } } v^{\alpha } v^{\gamma } b^{\beta \gamma} r^{\beta } \right\}+\int _{0}^{t}\frac{v^{\alpha } dt}{\sqrt{1-v^{2} } }  ,              \end{equation} 

Таким образом при переходе из лабораторной инерциальной системы отсчёта в другую инерциальную систему двигающуюся со скоростью $\mathbf{u}$ специальное преобразование ЛМН \eqref{10}, \eqref{11} переходит в общее преобразование \eqref{124}, \eqref{125}. Общее же преобразование ЛМН \eqref{72}, \eqref{73} с матрицей вращения $a^{*\alpha \beta } $ при переходе в другую систему отсчёта перейдёт также в общее преобразование (т.е. будет форминвариантно), но с новой скоростью $\mathbf{v}$ равной \eqref{100} и новой матрицей вращения $a^{\alpha \beta } $ равной
\begin{equation} \label{126} 
a^{ \beta \alpha} =b^{\gamma \alpha} a^{* \beta \gamma}.                                    
\end{equation} 
При совершении буста собственное ускорение и угловая скорость системы $s'$ разумеется являются инвариантами преобразования Лоренца. Учитывая значения характеристик систем отсчёта $s$ \eqref{15}, \eqref{16} и уравнения \eqref{80}, \eqref{81} можно записать, что в лабораторной системе отсчёта $S$ и системе $S^*$  собственное ускорение и угловая скорость системы $s'$ равны
\[W'^{\alpha } =a^{ \alpha\beta } \left(\frac{1}{\sqrt{1-v^{2} } } \dot{v}^{\beta } +\frac{1-\sqrt{1-v^{2} } }{v^{2} (1-v^{2} )}\mathbf{(v\dot{v})}v^{\beta } \right)=\] 
\begin{equation} \label{127}
=a^{* \alpha\beta } \left(\frac{1}{\sqrt{1-v^{*2} } } \dot{v}^{*\beta } +\frac{1-\sqrt{1-v^{*2} } }{v^{*2} (1-v^{*2} )} \mathbf{(v^*\dot{v}^*)}v^{*\beta } \right)=form-invariant 
\end{equation} 
\[\Omega '^{\alpha } =a^{\alpha\beta } \frac{1-\sqrt{1-v^{2} } }{v^{2} \sqrt{1-v^{2} } } \,\,e^{\beta \mu \nu } v^{\mu } \dot{v}^{\nu } +\omega '^{\beta } =a^{*\alpha \beta} \frac{1-\sqrt{1-v^{*2} } }{v^{*2} \sqrt{1-v^{*2} } }\,\, e^{\beta \mu \nu } v^{*\mu } \dot{v}^{*\nu } +\omega '^{*\beta }=\]
\begin{equation} \label{128} 
=form-invariant 
\end{equation}

\subsubsection* {Доказательство того, что связь между $\boldsymbol{\rho}$ и $\mathbf{r}$ является поворотом}
Возводя \eqref{108} в квадрат получим
\begin{equation} \label{109} 
\boldsymbol{\rho} ^{2} =\mathbf{A}^{2} -(\mathbf{vA})^{2} ,                                          
\end{equation} 
Вычислим $\mathbf{A}^{2} $ и $(\mathbf{vA})^{2} $ в явном виде. Возводя уравнение \eqref{103} в квадрат и приводя подобные члены имеем
\[\mathbf{A}^{2} =\mathbf{r}^{2} +\frac{(\mathbf{v^*r})^{2} }{(1-u^{2} )(1-v^{*2} )} +\frac{(\mathbf{ur})^{2} }{1-u^{2} } +\frac{(1-\sqrt{1-v^{*2} } )^{2} \mathbf{(uv^*)}^{2} (\mathbf{v^*r})^{2} }{v^{*4} (1-v^{*2} )(1-u^{2} )} -\frac{2\mathbf{(v^*r)(ur)}}{(1-u^{2} )\sqrt{1-v^{*2} } } -\] 
\begin{equation} \label{110} 
-\frac{2(1-\sqrt{1-v^{*2} } )}{v^{*2} (1-u^{2} )(1-v^{*2} )} \mathbf{(uv^*)(v^*r)}^{2} +\frac{2(1-\sqrt{1-v^{*2} } )}{v^{*2} \sqrt{1-v^{*2} } (1-u^{2} )} \mathbf{(uv^*)(v^*r)(ur)} 
\end{equation} 
Аналогично используя \eqref{111}, раскрывая скобки и приводя подобные члены получим
\[\mathbf{(vA)}^{2} =\frac{\mathbf{(v^*r)}^{2} }{(1-u^{2} )(1-v^{*2} )} +\frac{\mathbf{(ur)}^{2} }{1-u^{2} } +\frac{(1-\sqrt{1-v^{*2} } )^{2} \mathbf{(uv^*)}^{2} \mathbf{(v^*r)}^{2} }{v^{*4} (1-v^{*2} )(1-u^{2} )} -\frac{2\mathbf{(v^*r)(ur)}}{(1-u^{2} )\sqrt{1-v^{*2} } } -\] 
\begin{equation} \label{112} 
-\frac{2(1-\sqrt{1-v^{*2} } )}{v^{*2} (1-u^{2} )(1-v^{*2} )} \mathbf{(uv^*)(v^*r)}^{2} +\frac{2(1-\sqrt{1-v^{*2} } )}{v^{*2} \sqrt{1-v^{^*2} } (1-u^{2} )}\mathbf{(uv^*)(v^*r)(ur)} 
\end{equation} 
Следовательно, из \eqref{109} получим, учитывая \eqref{110} и \eqref{112}, что
\begin{equation} \label{113} 
\boldsymbol{\rho} ^{2} =\mathbf{r}^{2} ,                                       
\end{equation} 
а это и означает, что линейное преобразование $\boldsymbol{\rho} \to \mathbf{r} $: $\rho ^{\alpha } =b^{ \beta \alpha } r^{\beta }$  
является поворотом.

\subsubsection* {Матрица и угол поворота Вигнера}
Вычислим матрицу этого преобразования $b^{\beta\alpha } $ в явном виде. Подставляя в \eqref{108} значение $\mathbf{A}$ из  \eqref{103} и используя \eqref{111}, а также разложение
\begin{equation} \label{115} 
\mathbf{(u-v^*)}^2 - \mathbf{(u\times v^*)}^2= \left(1+\sqrt{1-u^{2}}\sqrt{1-v^{*2}}-\mathbf{uv^*}\right)\left(1-\sqrt{1-u^{2}}\sqrt{1-v^{*2}}-\mathbf{uv^*}\right)
\end{equation}
 получим
\[\boldsymbol{\rho} =\mathbf{r}+\frac{1-\sqrt{1-u^{2} } }{u^{2} \sqrt{1-u^{2} } }\mathbf{(ur)u}+\frac{1-\sqrt{1-v^{*2} } }{v^{*2} \sqrt{1-v^{*2} } }\mathbf{(v^*r)v^*}+\frac{(1-\sqrt{1-v^{*2} } )(1-\sqrt{1-u^{2} } )}{v^{*2} u^{2} \sqrt{1-v^{*2} } \sqrt{1-u^{2} } }\mathbf{(uv^*)(v^*r)u}-\] 
\[-\frac{\mathbf{(v^*r)u}}{\sqrt{1-u^{2} } \sqrt{1-v^{*2} } } -  \frac{1}{1+\sqrt{1-u^{2} } \sqrt{1-v^{*2} } - \mathbf{uv^*}} \cdot \left(\sqrt{1-u^{2} }\mathbf{v^*-u}+\frac{(1-\sqrt{1-u^{2} } )\mathbf{(uv^*)u}}{u^{2} } \right) \times \] 
\begin{equation} \label{116} 
\times\left(\frac{\mathbf{v^*r}}{\sqrt{1-u^{2} } \sqrt{1-v^{*2} } }- \frac{\mathbf{ur}}{\sqrt{1-u^{2} } } -\frac{(1-\sqrt{1-v^{*2} } )\mathbf{(uv^*)(v^*r)}}{v^{*2} \sqrt{1-v^{*2} } \sqrt{1-u^{2} } } \right)
\end{equation} 
Отсюда матрица преобразования $b^{ \beta \alpha} $ будет равна
\[b^{\beta \alpha} =\delta ^{\alpha \beta } +\frac{1-\sqrt{1-v^{*2} } }{v^{*2} \sqrt{1-v^{*2} } } v^{*\alpha } v^{*\beta } +\frac{1-\sqrt{1-u^{2} } }{u^{2} \sqrt{1-u^{2} } } u^{\alpha } u^{\beta } +\frac{(1-\sqrt{1-v^{*2} } )(1-\sqrt{1-u^{2} } )}{u^{2} v^{*2} \sqrt{1-v^{*2} } \sqrt{1-u^{2} } } \mathbf{(uv^*)}u^{\alpha } v^{*\beta } -\] 
\[-\frac{u^{\alpha } v^{*\beta } }{\sqrt{1-v^{*2} } \sqrt{1-u^{2} } } -\frac{1}{1+\sqrt{1-u^{2} } \sqrt{1-v^{*2} } - \mathbf{uv^*}} \cdot \left(\sqrt{1-u^{2} } v^{*\alpha} -u^{\alpha } +\frac{(1-\sqrt{1-u^{2} } )\mathbf{(uv^*)}u^{\alpha } }{u^{2} } \right)\times \] 
\begin{equation} \label{117} 
\times \left(\frac{v^{*\beta} }{\sqrt{1-u^{2} } \sqrt{1-v^{*2} } }  -\frac{u^{\beta } }{\sqrt{1-u^{2} } } -\frac{1-\sqrt{1-v^{*2} } }{v^{*2} \sqrt{1-v^{*2} } \sqrt{1-u^{2} } } \mathbf{(uv^*)}v^{*\beta } \right) 
\end{equation} 
 
Собственное вращение с матрицей поворота $b^{\alpha\beta}$ следует называть собственным вращением Вигнера. \footnote[6]{Обычно вращением Вигнера называется поворот спина при выполнении буста относительно лабораторной системы $S$. Поскольку это вращение подразумевается внутренним и так, как из раздела 4 известно, что собственное вращение  не является в лабораторной системе отсчёта чистым вращением, данный поворот правильнее рассматривать исключительно в системе $s$.}Любую матрицу поворота обычно записывают в стандартном виде в угловых координатах: ось поворота $\mathbf{n}$, угол поворота $\varphi$. Для транспонированной матрицы поворота она имеет вид \cite[с.23, нижняя формула (7)]{25}
 \begin{equation} \label{130}
b^{\beta\alpha}=\delta^{\beta\alpha}\cos\varphi+n^{\beta}n^{\alpha}(1-\cos\varphi)-e^{\alpha\gamma\beta}n^{\gamma}\sin\varphi
  \end{equation} 
  Заменим угол $\varphi$, соответствующий транспонированной матрице поворота, на угол  $\phi$, на который происходит поворот согласно $\varphi=-\phi$ \cite[с.29]{25}. В результате получим
  \begin{equation} \label{131}
b^{\beta\alpha}=\delta^{\beta\alpha}\cos\phi+n^{\beta}n^{\alpha}(1-\cos\phi)+e^{\alpha\gamma\beta}n^{\gamma}\sin\phi
  \end{equation} 
   В связи с этим встаёт задача определения угла поворота соответствующего матрице $b^{\alpha\beta} $.
 Наиболее просто угол $\phi$, на который происходит поворот можно получить исходя из формулы \eqref{121} векторным перемножением векторов $\mathbf{v'}$ и $\mathbf{v}$
 \begin{equation} \label{150}
 \mathbf{n}\sin\phi=\frac{\mathbf{v'\times v}}{v^2}
 \end{equation} 
 Можно однако поступить по другому и определить этот угол по матрице $b^{\beta\alpha} $ двумя способами. Очевидно, что свёртка $b^{\alpha \alpha}$ равна $1+2\cos \phi$. Следовательно, в результате некоторых вычислений можно получить следующее выражение
\begin{equation} \label{129}
 \cos \phi=\frac{b^{\alpha\alpha}-1}{2}=\frac{\sqrt{1-u^2}+\sqrt{1-v^{*2}}-\mathbf{uv^*}+\frac{(1-\sqrt{1-u^2})(1-\sqrt{1-v^{*2}})}{u^2v^{*2}}\mathbf{(uv^*)}^2} {1+\sqrt{1-u^2}\sqrt{1-v^{*2}}-\mathbf{uv^*}} 
 \end{equation} 
 С другой стороны, угол $\phi$  соответствующий матрице поворота  $b^{\alpha\beta}$ будет равен из \eqref{131}
  \begin{equation} \label{143}
n^{\gamma}\sin\phi=\frac{e^{\gamma\beta\alpha}b^{\beta\alpha}}{2}
  \end{equation}
Результат вычислений есть
\begin{equation} \label{132}
\mathbf{n}\sin \phi=\frac{\mathbf{u\times v^*}}{1+\sqrt{1-v^{*2}}\sqrt{1-u^2}-\mathbf{uv^*}}\cdot 
\left(1-\frac{(1-\sqrt{1-u^2})(1-\sqrt{1-v^{*2}})}{u^2v^{*2}}\mathbf{\,\,uv^*}\right)
  \end{equation} 
Формулы \eqref{129}, \eqref{132} можно записать проще, если воспользоваться формулой для тангенса половинного угла. В результате получим
  \begin{equation} \label{132,3}
\mathbf{n}\,\tg \frac{\phi }{2}=\mathbf{n}\frac{\sin \phi }{1+\cos \phi }=\frac{\mathbf{u}\times \mathbf{v^*}}{\left( 1+\sqrt{1-{{u}^{2}}} \right)\left( 1+\sqrt{1-{{v}^{*2}}} \right)-\mathbf{uv^*}}
\end{equation}
Данный угол совпадает с давно вычисленным углом Вигнера для обычного преобразования Лоренца \cite{28}, \cite[формула (20)]{29}.  Но в формулах \eqref{129}, \eqref{132}, \eqref{132,3} угол вращения Вигнера для неинерциального движения вообще говоря зависит от времени из-за переменности $\mathbf{v^*}(t)$.

\subsubsection*{О связи между собственной прецессией Томаса и вращением Вигнера.} 

Собственное вращение Вигнера не следует смешивать с собственной прецессией Томаса, это хотя и тесно связаные между собой, но неэквивалентные понятия. Суть открытия Томаса заключается в том, что если жёсткая система отсчёта $s$ движется таким образом, что в каждый момент лабораторного времени её координатные оси (вблизи начала отсчёта) совпадают с мгновенно сопутствующей ей инерциальной системой отсчёта, то система $s$ обладает некоторой собственной прецессией, которая и является прецессией Томаса. Собственное же вращение Вигнера дополнительно к прецессии Томаса, как видно из уравнений \eqref{126}, \eqref{128}, определяется матрицей $b^{\alpha\beta}$ и появляется при любом бусте. 

Всё вышесказанное можно пояснить следующим примером. Пусть в некоторой лабораторной системе $S^{*}$ начало жёсткой системы отсчёта $s$ движется со скоростью $\mathbf{v}^{*}$. Данная система отсчёта обладает собственной прецессией Томаса

       \begin{equation}  \label{649.10} 
\mathbf{\Omega}^{*}_{T}=\frac{1-\sqrt{1-v^{*2} } }{v^{*2} \sqrt{1-v^{*2} } }\mathbf{v^{*}\times \dot{v}^{*}} 
\end{equation}

Произведём теперь буст со скоростью $\mathbf{u}$. В новой лабораторной системе отсчёта $S$ начало отсчёта $s$ имеет уже скорость $\mathbf{v}$, связанную с $\mathbf{v}^{*}$ и $\mathbf{u}$ законом вычитания скоростей \eqref{100}, а оси координат $s$ испытывают дополнительное вигнеровское вращение \eqref{117}. Следовательно, угловая скорость собственного вращения $s$ в новой системе складывается  согласно формуле \eqref{81} из двух составляющих:
          \begin{equation} \label{649.11}
 \Omega^{*\alpha }_{T} = b^{ \alpha \beta}\Omega _{T}^{\beta } +\omega'^{\alpha }_{W} 
\end{equation}
                                             
Здесь  вектор $\mathbf{\Omega}_{T}$ является частотой собственной прецессии Томаса системы $s$ \eqref{16} рассматриваемой в новой системе $S$, $ b^{ \alpha \beta}$ есть матрица собственного вращения Вигнера \eqref{117}, а $\omega'^{\alpha }$ 
есть вигнеровская угловая скорость вращения, выраженная в системе координат $s$, причём она связана с вигнеровской матрицей соотношением аналогичным \eqref{79,3}:
\begin{equation} \label{649.12}
\omega'^{\gamma }_{W}=\frac {1}{2}\,e^{\mu \gamma\beta }b^{ \mu\alpha } \frac{db^{ \beta \alpha } }{dt}
\end{equation}                                

Если подставить в \eqref{649.11} значение $\mathbf{\Omega}_{T}$  выраженное через $\mathbf{v}^{*}$ и $\mathbf{u}$, а также значение матрицы вигнеровского вращения \eqref{117}, то \eqref{649.11} обратится в тождество. 

Рассмотрим отдельно частный случай \eqref{649.11}, когда производятся последовательные (в близкие моменты времени) бусты в мгновенно сопутствующие $s$ инерциальные системы отсчёта. В этом случае новая прецессия Томаса системы $s$ (первый член в правой части \eqref{649.11}) будет мала и вся собственная прецессия Томаса $\mathbf{\Omega}^{*}_{T}$ будет относиться на счёт вращения Вигнера. Это обстоятельство нетрудно доказать. Действительно, пусть в момент $T^{*}+\Delta T^{*}$ скорость системы $s$ стала $\mathbf{v}^{*}+\Delta \mathbf{v}^{*}$. Перейдём в систему отсчёта двигающуюся со скоростью $\mathbf{u=v}^{*}$.  Подставив данные значения в \eqref{132,3} и учитывая, что угол поворота Вигнера $\phi$ мал, в результате получим, что он равен
  \begin{equation} \label{133}
  \mathbf{n}\,\phi =\frac{1-\sqrt{1-v^{*2}}}{v^{*2}\sqrt{1-v^{*2}}}\,\mathbf{v}\times \Delta \mathbf{v^*}.
  \end{equation}
Очевидно, этот угол совпадает с собственной частотой прецессии Томаса умноженной на промежуток собственного времени. Это обстоятельство иногда рассматривается как подтверждение того, что прецессия Томаса и вращение Вигнера являются разными названиями одного и того же физического явления. Данное мнение является разумеется ошибочным, поскольку для произвольного буста проводить различие между вращением Вигнера и прецессией Томаса совершенно необходимо.

В качестве ещё одного примера подтверждающего справедливость \eqref{649.11} можно рассмотреть ещё его частный случай, когда система $s$ в первоначальной лабораторной системе двигалась прямолинейно и равноускоренно. В этом случае её собственная прецессия Томаса равна нулю. Тогда из \eqref{649.11} следует, что в новой лабораторной системе вращение Вигнера и прецессия Томаса системы $s$ должны быть противоположны и компенсировать друг друга. Данный вывод будет проверен далее, в разделе 10.

\subsection*  {8. Преобразование ЛМН в 4-мерном виде}

 Форминвариантность величин \eqref{127}, \eqref{128} наводит на мысль, что данные характеристики системы отсчёта являются 4-инвариантами. Данное предположение можно проверить записав общее преобразование ЛМН в 4-мерном виде. Для этого введём следующие векторы
 \begin{equation} \label{333.1}
  {{\Lambda }^{0i}}=\left( {{\Lambda }^{00}},{{\Lambda }^{0\alpha }} \right)=\left( \frac{1}{\sqrt{1-{{v}^{2}}}},\frac{{{v}^{\alpha }}}{\sqrt{1-{{v}^{2}}}} \right)
 \end{equation}
 \begin{equation} \label{333.12}
 {{\Lambda }^{\alpha i}}=\left( {{\Lambda }^{\alpha 0}},{{\Lambda }^{\alpha \beta }} \right)=\left( \frac{{{v}^{\gamma }}{{a}^{\alpha \gamma }}}{\sqrt{1-{{v}^{2}}}},{{a}^{\alpha \beta }}+\frac{1-\sqrt{1-{{v}^{2}}}}{{{v}^{2}}\sqrt{1-{{v}^{2}}}}{{v}^{\beta }}{{v}^{\mu }}{{a}^{\alpha \mu }} \right)
  \end{equation}
где греческие индексы как обычно пробегают значения 1,2,3. Первый индекс у каждого 4-вектора отвечает за его номер, а второй - за его компоненту. Смысл временоподобного 4-вектора заключается в том, что он является единичным и касательным к мировой линии начала отсчёта системы $s'$. Смысл же пространственноподобного 4-вектора под номером $\alpha$ заключается в том, что он является соответствующим ортом декартовой системы координат $s'$. В этом нетрудно убедиться заметив, что $\Lambda ^{0i},\Lambda^{\beta i}$ удовлетворяют равенствам ортонормированности
\begin{equation} \label{334} 
	{{\Lambda }^{0i}}\Lambda ^{0}_{\ \ i}=1,\,\,\,{{\Lambda }^{0i}}\Lambda ^{\alpha }_{\ \ i}=0,\,\,\,{{\Lambda }^{\alpha i}}\Lambda ^{\beta }_{\ \ i}={-\,{\delta }^{\alpha \beta }},  
\end{equation}  
 где латинские индексы есть 0,1,2,3. Такой локальный ортонормированный комплекс четырёх 4-векторов $\Lambda ^{0 i},\Lambda ^{\beta i}$, один из которых (временоподобный,  $\Lambda ^{0 i}$) расположен вдоль касательной к мировой линии начала отсчёта системы координат $s'$, а другие - пространственноподобные ($\Lambda ^{\beta i}$,  $\beta=1,2,3$) являются направляющими ортами системы $s'$, как известно называется тетрадой.  Тогда наиболее общее преобразование ЛМН \eqref{12}, \eqref{13} в систему $s'$ будет выглядеть в простом 4-мерном виде
\begin{equation} \label{332} 
{{X}^{i}}=\Lambda ^{\alpha i}{{{x}'}^{\alpha }}+\int\limits_{0}^{{{x}^{0}}}{\Lambda^{0 i}d{{x}^{0}}}.  
\end{equation}
где ${x}'^{\alpha },\,d{{x}^{0}}$ представляют собой коэффициенты при 4-векторе. Дифференцируя \eqref{332} получим
  \begin{equation} \label{335} 
   d{{X}^{i}}=\Lambda ^{\alpha i}d{{{x}'}^{\alpha }}+{{{x}'}^{\alpha }}d\Lambda ^{\alpha i}+\Lambda ^{0i}d{{x}^{0}}
\end{equation} 
Если теперь продифференцировать компоненты $\Lambda ^{\alpha 0} $ из \eqref{333.12}, заменить в получившемся выражении производные $da^{\alpha \beta }/dt$ согласно уравнению \eqref{79,2} и в итоге учесть выражение $\omega '^{\alpha}$ из \eqref{81}, то получим, что 
\begin{equation} \label{335.12} 
\frac{d{{\Lambda }^{\alpha 0}}}{d{{x}^{0}}}=\frac{{{a}^{\alpha \nu }}{{{\dot{v}}}^{\nu }}-{{e}^{\alpha \xi \eta }}{{a}^{\xi \nu }}{{a}^{\eta \beta }}\Omega _{T}^{\beta }{{v}^{\nu }}}{\sqrt{1-{{v}^{2}}}}+\frac{{{a}^{\alpha \nu }}{{v}^{\nu }}(\mathbf{v\dot{v}})}{{{\sqrt{1-{{v}^{2}}}}^{3}}}+{{e}^{\alpha \beta \gamma }}{{{\Omega }'}^{\gamma }}\frac{{{v}^{\nu }}{{a}^{\beta \nu }}}{\sqrt{1-{{v}^{2}}}}
 \end{equation} 
Далее в первом и втором члене в правой части этого равенства учтём равенство "уничтожения" \eqref{76}, значение прецессии Томаса \eqref{18} и везде подставим вместо производных $\dot{v}^{\nu}$ их значения согласно \eqref{17}. После этого необходимо учесть уравнение \eqref{80}. После всех вычислений получим, что 
\begin{equation} \label{336} 
  \frac{d\Lambda ^{\alpha 0}}{d{{x}^{0}}}={{W'}^{\alpha }}\frac{1}{\sqrt{1-{{v}^{2}}}}+{{e}^{\alpha \beta \gamma }}{{\Omega' }^{\gamma }}\frac{{{v}^{\nu }}{{a}^{\beta \nu }}}{\sqrt{1-{{v}^{2}}}}
 \end{equation} 
 Действуя аналогично, для производной компоненты $\Lambda ^{\alpha \mu}$ можно найти её следующее значение 
 \begin{equation} \label{337} 
   \frac{d\Lambda^{\alpha \mu }}{d{{x}^{0}}}={{W'}^{\alpha }}\frac{{{v}^{\nu }}{{a}^{\mu \nu }}}{\sqrt{1-{{v}^{2}}}}+{{e}^{\alpha \beta \gamma }}{{\Omega'}^{\gamma }}{{a}^{\beta \mu }}+{{e}^{\alpha \beta \gamma }}{{\Omega' }^{\gamma }}\frac{1-\sqrt{1-{{v}^{2}}}}{{{v}^{2}}\sqrt{1-{{v}^{2}}}}{{v}^{\mu }}{{v}^{\nu }}{{a}^{\beta \nu }}. 
 \end{equation} 
Уравнения \eqref{336} и \eqref{337} можно записать в единой форме как 
\begin{equation} \label{338} 
	\frac{d\Lambda ^{\alpha i}}{d{{x}^{0}}}={{W'}^{\alpha }}\Lambda ^{0 i}+{{e}^{\alpha \beta \gamma }}{{\Omega' }^{\gamma }}{{\Lambda }^{\beta i}} 
\end{equation} 
Из \eqref{334} и \eqref{338} очевиден тетрадный смысл характеристик систем отсчёта

\begin{equation} \label{339} 
    {{{W}'}^{\alpha }}=\Lambda ^{0}_{\;\;i}\frac{d\Lambda ^{\alpha i}}{d{{x}^{0}}}
\end{equation} 
\begin{equation} \label{340} 
	\Lambda _{\ \ i}^{\mu }\frac{d{{\Lambda }^{\alpha i}}}{d{{x}^{0}}}=-{{e}^{\alpha \mu \gamma }}{{{\Omega }'}^{\gamma }},\,\,\,\, 
{{{\Omega }'}^{\nu }}=-\frac{1}{2}\,{{e}^{\alpha \mu \nu }}\Lambda _{\;\;i}^{\mu }\frac{d{{\Lambda }^{\alpha i}}}{d{{x}^{0}}}
\end{equation}

Сумма в правой части \eqref{338} является скоростью изменения триады 4-векторов $\Lambda ^{i\alpha } $ относительно локальной невращающейся системы. Первый член в ней отвечает за перенос Ферми -- Уолкера \cite[c. 218, формула (6.14)]{16}, а второй член отвечает за наличие собственного вращения \cite[c. 221, формула (6.22)]{16}. Используя это соотношение, можно получить в \eqref{235} разложение вектора $dX^{i} $ по векторам тетрады 
\begin{equation} \label{341} 
d{{X}^{i}}=\Lambda ^{\beta i}(d{{{x}'}^{\beta }}+{{{x}'}^{\alpha }}{{e}^{\alpha \beta \gamma }}{{{\Omega }'}^{\gamma }}d{{x}^{0}})+\Lambda ^{0i}d{{x}^{0}}(1+{{{x}'}^{\alpha }}{{{W}'}^{\alpha }}) 
\end{equation} 
Возводя равенство \eqref{341} в квадрат и пользуясь равенствами \eqref{334} можно получить интервал в виде \eqref{3}. Таким образом равенство \eqref{341} представляет собой как-бы результат операции взятия квадратного корня из 4-интервала. Это полностью согласуется с аналогичным соотношением в работе \cite [формула (8)]{11}.

\subsection*  {9. Обратная задача релятивистской кинематики}

       В классическом случае порядок решения обратной задачи кинематики был такой. Чтобы найти параметры движения произвольной жёсткой системы $s'$ достаточно было по известной собственной угловой скорости найти углы поворота относительно поступательно двигающейся системы $s$  как функции времени. Эти углы задают матрицу собственного вращения. Затем, подставляя эту матрицу в уравнение для собственного ускорения, вычисляли ускорение относительно $s$, и далее решая соответствующее дифуравнение, находили скорость начала отсчёта. Тем самым все параметры движения были полностью определены. Факт же существования собственной прецессии Томаса кардинально изменяет данное решение обратной задачи кинематики. Дело в том, что собственная угловая скорость (при фиксированном вращении системы $s'$ относительно $s$) в релятивистском случае зависит от параметра $\mathbf{v}(t)$ определяющего переносное движение $s$. В связи с этим, найти матрицу собственного вращения без уравнения для собственного ускорения невозможно.

Прямая задача кинематики решается следующим образом. Система отсчёта $s'$ двигающаяся поступательно имеет собственное ускорение и угловую скорость равные согласно \eqref{15}, \eqref{16}, \eqref{80}, \eqref{81} 
       \begin{equation} \label{546}
       {{{W}'}^{\alpha }}={{a}^{\alpha \beta }}\left( \frac{1}{\sqrt{1-{{v}^{2}}}}{{{\dot{v}}}^{\beta }}+\frac{1-\sqrt{1-{{v}^{2}}}}{{{v}^{2}}(1-{{v}^{2}})}(\mathbf{v\dot{v}}){{v}^{\beta }} \right)                         
      \end{equation}                                                     
         \begin{equation} \label{547}
      {{{\Omega }'}^{\alpha }}={{a}^{\alpha \beta }}\frac{1-\sqrt{1-{{v}^{2}}}}{{{v}^{2}}\sqrt{1-{{v}^{2}}}}{{e}^{\beta \mu \nu }}{{v}^{\mu }}{{\dot{v}}^{\nu }}+{{{\omega }'}^{\alpha }}.                            
      \end{equation}                                                          
    Удобно совершить в этих уравнениях подстановку 
            \begin{equation} \label{548}
                            {{v}^{\alpha }}={{a}^{\beta \alpha }}{{{v}'}^{\beta }}  
                          \end{equation}   
 Эта подстановка приводит  \eqref{546}, \eqref{547}  к простому векторному виду. Действительно, производя подстановку в \eqref{546}, дифференцируя и учитывая, что скалярное произведение двух векторов при повороте не изменяется, а также, что $v^2=v'^2$ получим
         \begin{equation} \label{549}
         {{{W}'}^{\alpha}}={{a}^{\alpha \beta }}\left( \frac{{{{\dot{a}}}^{\gamma \beta }}{{{{v}'}}^{\gamma }}}{\sqrt{1-{{{{v}'}}^{2}}}}+\frac{a^{\gamma \beta }\dot{v}'^{\gamma }}{\sqrt{1-{{{{v}'}}^{2}}}}+\frac{1-\sqrt{1-{{{{v}'}}^{2}}}}{{{{{v}'}}^{2}}(1-{{{{v}'}}^{2}})}(\mathbf{{v}'{\dot{v}}'}){{a}^{\gamma \beta }}{{{v}'}^{\gamma }} \right) 
         \end{equation}  
         
Раскрыв скобки учтём в первом члене первое из равенств \eqref{79,3}, а в остальных членах равенство \eqref{74}. В результате получим
 \begin{equation} \label{550}
{{{W}'}^{\alpha }}=\frac{{{e}^{\alpha \nu \gamma }}{{{{\omega }'}}^{\nu }}{{{{v}'}}^{\gamma }}}{\sqrt{1-{{{{v}'}}^{2}}}}+\frac{\dot{v}'^{\alpha }}{\sqrt{1-{{{{v}'}}^{2}}}}+\frac{1-\sqrt{1-{{{{v}'}}^{2}}}}{{{{{v}'}}^{2}}(1-{{{{v}'}}^{2}})}(\mathbf{{v}'{\dot{v}}'}){{{v}'}^{\alpha }}
\end{equation} 
            В векторной форме это уравнение выглядит в виде
           \begin{equation} \label{551}          \mathbf{{W}'}=\frac{{\mathbf{{\dot{v}}'}}}{\sqrt{1-{{{{v}'}}^{2}}}}+\frac{{\boldsymbol{\omega} }'\times \mathbf{{v}'}}{\sqrt{1-{{{{v}'}}^{2}}}}+\frac{1-\sqrt{1-{{{{v}'}}^{2}}}}{{{{{v}'}}^{2}}(1-{{{{v}'}}^{2}})}(\mathbf{{v}'{\dot{v}}'})\mathbf{{v}'}   
              \end{equation}
      В уравнении \eqref{547} после подстановки \eqref{548} рассмотрим первый член в правой части без дробного множителя. В результате цепочки вычислений получим
     
     \[{{a}^{\alpha \beta }}{{e}^{\beta \mu \nu }}{{a}^{\lambda \mu }}{{{v}'}^{\lambda }}\left( {{a}^{\eta \nu }}{\dot{v}'^{\eta }}+{{{\dot{a}}}^{\sigma \nu }}{{{{v}'}}^{\sigma }} \right)={{a}^{\alpha \beta }}{{{v}'}^{\lambda }}
\left( {{e}^{\beta \mu \nu }}{{a}^{\lambda \mu }}{{a}^{\eta \nu }}{\dot{v}'^{\eta }}+{{e}^{\beta \mu \nu }}{{a}^{\lambda \mu }}{{{\dot{a}}}^{\sigma \nu }}{{{{v}'}}^{\sigma }} \right)=\]
\[={{a}^{\alpha \beta }}{{{v}'}^{\lambda }}\left( {{e}^{\beta \mu \nu }}{{a}^{\lambda \mu }}{{a}^{\eta \nu }}{\dot{v}'^{\eta }}+{{e}^{\beta \mu \nu }}{{a}^{\lambda \mu }}({{e}^{\eta \varepsilon \sigma }}{{{{\omega }'}}^{\varepsilon }}{{a}^{\eta \nu }}){v'^{\sigma }} \right)={{a}^{\alpha \beta }}{{{v}'}^{\lambda }}( {{e}^{\beta \mu \nu }}{{a}^{\lambda \mu }}{{a}^{\eta \nu }}{\dot{v}'^{\eta }}+\]
\[+{{e}^{\beta \mu \nu }}{{a}^{\lambda \mu }}({{e}^{\eta \varepsilon \sigma }}{{{{\omega }'}}^{\varepsilon }}{{a}^{\eta \nu }}){{{{v}'}}^{\sigma }} )={{a}^{\alpha \beta }}{{{v}'}^{\lambda }}({{e}^{\sigma \lambda \eta }}{{a}^{\sigma \beta }}{\dot{v}'^{\eta }}+{{e}^{\psi \lambda \eta }}{{a}^{\psi \beta }}{{e}^{\eta \varepsilon \sigma }}{{{\omega }'}^{\varepsilon }}{{{v}'}^{\sigma }})=\]
   \begin{equation} \label{551,5} 
   ={{{v}'}^{\lambda }}({{e}^{\alpha \lambda \eta }}{\dot{v}'^{\eta }}+{{e}^{\alpha \lambda \eta }}{{e}^{\eta \varepsilon \sigma }}{{{\omega }'}^{\varepsilon }}{{{v}'}^{\sigma }})={{e}^{\alpha \lambda \eta }}{{{v}'}^{\lambda }}({\dot{v}'^{\eta }}+{{e}^{\eta \varepsilon \sigma }}{{{\omega }'}^{\varepsilon }}{{{v}'}^{\sigma }})  
  \end{equation}
  Следовательно
    \begin{equation} \label{552} 
   {{{\Omega }'}^{\alpha }}=\frac{1-\sqrt{1-{{v}^{2}}}}{{{v}^{2}}\sqrt{1-{{v}^{2}}}}{{e}^{\alpha \lambda \eta }}{{{v}'}^{\lambda }}({\dot{v}'^{\eta }}+{{e}^{\eta \varepsilon \sigma }}{{{\omega }'}^{\varepsilon }}{{{v}'}^{\sigma }})+{\omega '^{\alpha }}    
    \end{equation}    
или в векторной форме
\begin{equation} \label{553}
            \mathbf{\Omega }'={\boldsymbol \omega }'+\frac{1-\sqrt{1-{{{{v}'}}^{2}}}}{{{{{v}'}}^{2}}\sqrt{1-{{{{v}'}}^{2}}}}\mathbf{{v}'}\times (\mathbf{{\dot{v}}'}+{\boldsymbol\omega }'\times \mathbf{{v}'})
                                    \end{equation}    
Мы видим таким образом, что в формулах для собственных характеристик неинерциальной системы отсчёта нет матрицы поворота, так, что их возможно записать в векторном виде. Уравнение \eqref{553} несложно решить относительно ${\boldsymbol\omega }'$. Получим
\begin{equation} \label{554}
         {\boldsymbol\omega}'=\sqrt{1-{{{{v}'}}^{2}}}{\mathbf{\Omega} }'+\frac{1-\sqrt{1-{{{{v}'}}^{2}}}}{{{{{v}'}}^{2}}}({\mathbf{\Omega} }'\mathbf{{v}'})\mathbf{{v}'}-\frac{1-\sqrt{1-{{{{v}'}}^{2}}}}{{{{{v}'}}^{2}}}\mathbf{{v}'}\times \mathbf{{\dot{v}}'}           \end{equation}  
Подставим теперь \eqref{554} в \eqref{551} и получим, что
        \begin{equation} \label{555}         
        \mathbf{{W}'}=\mathbf{{\dot{v}}'}+\frac{1}{1-{{{{v}'}}^{2}}}(\mathbf{{v}'{\dot{v}}'})\mathbf{{v}'}+{\mathbf{\Omega} }'\times \mathbf{{v}'}
         \end{equation} 
Умножая это уравнение на $\mathbf{{v}'}$ получим отсюда, что 
\begin{equation} \label{556}   
       \mathbf{v\dot{v}}=(1-{{{v}'}^{2}})\mathbf{{W}'{v}'}                                             \end{equation} 
Подставив это уравнение обратно \eqref{555} получим окончательно простое дифференциальное уравнение первого порядка
\begin{equation} \label{557}
 \mathbf{{\dot{v}}'}=\frac{d\,\mathbf{{v}'}}{dt}=\mathbf{{W}'}-(\mathbf{{W}'{v}'})\mathbf{{v}'}-\mathbf{\Omega}'\times \mathbf{{v}'}  
      \end{equation} 
Это есть векторное уравнение Риккати. Подставим теперь в \eqref{554} уравнение \eqref{557}, тогда получим
\begin{equation} \label{558}       
       {\boldsymbol\omega }'=\mathbf{\Omega}'-\frac{1-\sqrt{1-{{{{v}'}}^{2}}}}{{{{{v}'}}^{2}}}\mathbf{{v}'}\times \mathbf{{W}'},                               \end{equation}
где ${\boldsymbol\omega }'$ удовлетворяет \eqref{79,3}. 

       Таким образом схема решения обратной задачи кинематики выглядит в следующем виде. Сначала необходимо найти решение  нелинейного векторного дифференциального уравнения \eqref{557} первого порядка. Подстановка этого решения в \eqref{558} приводит  к системе дифуравнений \eqref{79,3}, т. е. к задаче нахождения 3 углов поворота по известной угловой скорости ${\boldsymbol\omega }'$. Знание этих углов тем самым даёт матрицу вращения. Наконец подстановка уже известных величин в \eqref{548} определяет параметр $\mathbf{v}(t)$. Тем самым обратная задача релятивистской кинематики решена. К этому надо ещё добавить, что одновременно был найден и порядок решения задачи релятивистской динамики о движении частицы под действием собственной силы, изменяющейся по некоторому закону ${\bf{f'}}(t) = m{\bf{W'}}(t)$.

\subsection*  {10. Криволинейное равноускоренное движение}
\subsubsection*{Определение 4-силы через величины, выраженные в собственной системе отсчёта} 

Равноускоренным движением естественно называть движение системы $s'$ с постоянным в её системе координат собственным ускорением. Релятивистский закон такого движения удовлетворяет уравнению релятивистской механики \cite[c. 128, формула (19.2)-(19.4)]{20}
\begin{equation} \label{634}
m\frac{d{{u}^{i}}}{ds}=\tilde{F}^{i},
\end{equation}
где 
\begin{equation} \label{634.1}
\tilde{F}^{i}  =\left( {\tilde{F}^{0} ,\tilde{F}^{\alpha} } \right)= \left({\frac{{{\bf{Fv}}}}{{\sqrt {1 - v^2 } }},\frac{{\bf{F}}}{{\sqrt {1 - v^2 } }}} \right),
\end{equation}
а $\bf{F}$ - есть трёхмерная сила в системе $S$. 
Возникает вопрос о связи данной силы с силой $\bf{f'}$ действующей на частицу в собственной системе. На этот вопрос нетрудно ответить с помощью уравнений \eqref{557}, \eqref{558}. Так, импульс частицы равен очевидному соотношению
 \begin{equation} \label{12.10}
{{P}^{\alpha }}={{a}^{\beta \alpha }}\frac{m{{{{V}'}}^{\beta }}}{\sqrt{1-{{{{V}'}}^{2}}}}
 \end{equation} 
Поскольку мы рассматриваем точечную частицу, произведём замену ${\bf{V}'} = {\bf{v}'}$ в \eqref{12.10} и продифференцируем его по времени лабораторной системы. Получим
\begin{equation} \label{12.12}
{{F}^{\alpha }}=m\frac{d}{dT}\left( \frac{{{a}^{\beta  \alpha }}{{{{v}'}}^{\beta  }}}{\sqrt{1-{{{{v}'}}^{2}}}} \right)=m\sqrt{1-{{{{v}'}}^{2}}}\frac{d}{dt}\left( \frac{{{a}^{\beta  \alpha }}{{{{v}'}}^{\beta  }}}{\sqrt{1-{{{{v}'}}^{2}}}} \right)=m\left( {{{\dot{a}}}^{\beta  \alpha }}{{{{v}'}}^{\beta }}+{{a}^{\beta  \alpha }}{{{\dot{v}}}^{\beta  }}+\frac{{{a}^{\beta  \alpha }}(\mathbf{{v}'{\dot{v}}'}){{{{v}'}}^{\beta  }}}{1-{{{{v}'}}^{2}}} \right)
\end{equation} 
Далее заменим производную ${\dot{a}}^{\beta \alpha}$ согласно уравнению \eqref{79,2} и вынесем матрицу поворота за скобки. Получим
 \begin{equation} \label{12.13}
{{F}^{\alpha }}=m{{a}^{\xi \alpha }}\left( {{e}^{\xi \eta \beta }}{{{{\omega }'}}^{\eta }}{{{{v}'}}^{\beta }}+{{{\dot{v}}}^{\xi }}+\frac{(\mathbf{{v}'{\dot{v}}'}){{{{v}'}}^{\xi }}}{1-{{{{v}'}}^{2}}} \right)
\end{equation} 
Данное уравнение можно переписать следующим образом 
\begin{equation} \label{12.14}
{{F}^{\alpha }}={{a}^{\xi \alpha }}{{{F}'}^{\xi}},
\end{equation} 
где
\begin{equation} \label{12.15}
{{{F}'}^{\xi }}=m\left( {{e}^{\xi \eta \beta }}{{{{\omega }'}}^{\eta }}{{{{v}'}}^{\beta }}+{{{\dot{v}}}^{\xi }}+\frac{(\mathbf{{v}'{\dot{v}}'}){{{{v}'}}^{\xi }}}{1-{{{{v}'}}^{2}}} \right),
\end{equation} 
или в векторном виде
\begin{equation} \label{12.16}
\mathbf{{F}'}=m\left( \boldsymbol{{\omega }}'\times \mathbf{{v}'}+\mathbf{{\dot{v}}'}+\frac{(\mathbf{{v}'{\dot{v}}'})\mathbf{{v}'}}{1-{{{{v}'}}^{2}}} \right).
\end{equation}
Теперь осталось только подставить в это уравнение вместо $\mathbf{\dot{v}}'$ и $\boldsymbol{{\omega }}'$ их значения согласно уравнениям соответственно \eqref{557} и \eqref{558}. Это приводит к уравнению
\begin{equation} \label{12.17}
\mathbf{{F}'}=m\left( \sqrt{1-{{{{v}'}}^{2}}}\mathbf{{W}'}+\frac{1-\sqrt{1-{{{{v}'}}^{2}}}}{{{{{v}'}}^{2}}}(\mathbf{{v}'{W}'})\mathbf{{v}'} \right)=  \sqrt{1-{{{{v}'}}^{2}}}\mathbf{{f}'}+\frac{1-\sqrt{1-{{{{v}'}}^{2}}}}{{{{{v}'}}^{2}}}(\mathbf{{v}'{f}'})\mathbf{{v}'},
\end{equation}
где $\mathbf{f'}$ есть
\begin{equation} \label{12.11}
\mathbf{f'}=m\mathbf{W'}.
 \end{equation} 
и является собственной силой в системе $s'$. Подставив теперь это выражение в \eqref{634.1} и учитывая \eqref{12.14}, а также, что $\mathbf{Fv}=\mathbf{{F}'{v}'}$ получим, что 4-сила выглядит в виде 
\begin{equation} \label{12.18}
\tilde{F}^{i}  =\left( {\tilde{F}^{0} ,\tilde{F}^{\alpha} } \right) = \left( {\frac{{{\bf{v'f'}}}}{{\sqrt {1 - v'^2 } }},a^{\xi \alpha } \frac{{v'^2 \sqrt {1 - v'^2 } f'^\xi   + (1 - \sqrt {1 - v'^2 } )({\bf{v'f'}})v'^\xi  }}{{v'^2 \sqrt {1 - v'^2 } }}} \right)
\end{equation}
где функция ${\bf{v'}}$ является решением \eqref{557} с $\mathbf{W'=f'}/m$, а матрица вращения удовлетворяет системе уравнений \eqref{79,3} с $\boldsymbol{\omega}'$ равной \eqref{558}. 
Если уравнение  \eqref{12.17} умножить на матрицу вращения согласно \eqref{12.14}, то считая, что ${{f}^{\alpha }}={{a}^{\xi \alpha }}{{{f}'}^{\xi}}$, получим 
\begin{equation} \label{634.2}
{\bf{F}} = \sqrt {1 - v^2 } {\bf{f}} + \frac{{1 - \sqrt {1 - v^2 } }}{{v^2 }}({\bf{vf}}){\bf{v}},
\end{equation}
т.е. известное правило преобразования силы из сопутствующей системы отсчёта в лабораторную систему \cite[c. 325, задача 4.42]{19}, что подтверждает все ранее выполненные расчёты. Если же теперь поступить обратным образом и определить равноускоренное движение в $S$ как движение под действием силы \eqref{12.18}, то разумеется придём к обратной задаче кинематики  \eqref{557} и \eqref{558}. 

Равноускоренная система $s'$, (в которой постоянна сила $\bf{f'}$)  связанная с частицей испытывает вращение Вигнера относительно системы $s$, которая в свою очередь имеет собственную прецессию Томаса, и оси которой всегда параллельны осям лабораторной системы $S$. Это означает, что матрица $a^{\xi \alpha }$, являющаяся матрицей Вигнера \eqref{117} $a^{\xi \alpha } = b^{\xi \alpha }$, зависит от времени. В силу этой причины и постольку, поскольку в процессе движения  направление скорости, вообще говоря, изменяется относительно осей и системы $s'$, и системы $S$, то постоянная в собственной системе $s'$ сила $\bf{f}'$ не будет постоянной в $s$ и в $S$ и наоборот, постоянная в в лабораторной системе $S$ сила $\bf{F}$ будет изменяться в собственной системе $s'$. Таким образом, следует различать задачу о равноускоренном движении и задачу о движении заряженной частицы в однородном электрическом поле в $S$. Только в том случае, когда движение частицы является прямолинейным, решения обеих задач совпадают. 

Движение частицы в однородном поле иначе называют гиперболическим и оно хорошо известно. Различие между равноускоренным и гиперболическим движением видно ещё и по тому, что движение под действием постоянной силы в системе $S$ определяется всего двумя начальными постоянными: скоростью в направлении силы и в направлении перпендикулярно силе. Равноускоренное же движение $s'$ характеризуется, кроме этих постоянных, ещё и углом определяющим начальную ориентацию $s'$ относительно $S$.

\subsubsection*{Определение равноускоренного движения с помощью буста.}
Очевидно, что общее преобразование в равноускоренную систему отсчёта с $\mathbf{W}'=const$  является преобразованием ЛМН, неизвестны лишь его параметры. Эти параметры наиболее общего преобразования в криволинейно движущуюся равноускоренную систему отсчёта будем искать сначала чисто кинематически, с помощью буста из некоторой инерциальной системы, относительно которой  равноускоренная система отсчёта движется прямолинейно, а потом - с помощью решения общих дифуравнений \eqref{557} и \eqref{558}. 

Итак, условимся, что некая инерциальная система отсчёта $S^*$ ориентирована так, чтобы прямолинейное равноускоренное движение системы отсчёта $s'$ осуществлялось вдоль её оси 1, а совершаемый буст производится в плоскости осей 1 и 3. Тогда переход в систему $s'$ из $S^*$, описывается преобразованием ЛМН с параметром $\mathbf{v^*}$ согласно \cite[формула (13)]{4} равным
\begin{equation} \label{637}
\mathbf{v}^*=(th(Wt+k),0,0),
\end{equation}
где $thk$ есть начальная скорость начала отсчёта $s'$ относительно $S^*$. Перейдём теперь из системы $S^*$ в новую инерциальную систему отсчёта $S$ ориентированную также как и $S^*$ и двигающуюся со скоростью $\mathbf{u}=({{u}_{1}},0,{{u}_{3}})$ относительно $S^*$. При таком бусте у системы $s'$ в новой системе отсчёта появляется начальная компонента скорости вдоль оси 3. Таким образом движение системы $s'$ в системе отсчёта $S$, которую можно принять за лабораторную систему, будет наиболее общим равноускоренным движением. Преобразование в равноускоренную систему отсчёта имеет стандартный вид общего преобразования ЛМН, где новая скорость $s$ относительно  $S$ будет равна согласно формуле \eqref{100}
\begin{equation} \label{638}
v_{1}=\frac{\sqrt{1-{{u}^{2}}}th(Wt+k)-{{u}_{1}}}{1-{{u}_{1}}th(Wt+k)}+\frac{\left( 1-\sqrt{1-{{u}^{2}}} \right)u_{1}^{2}th(Wt+k)}{{{u}^{2}}(1-{{u}_{1}}th(Wt+k))}
\end{equation}
\begin{equation} \label{639}
v_{2}=0
\end{equation}
\begin{equation} \label{640}
v_{3}= -\frac{{{u}_{3}}}{1-{{u}_{1}}th(Wt+k)}+\frac{\left( 1-\sqrt{1-{{u}^{2}}} \right){{u}_{1}}{{u}_{3}}th(Wt+k)}{{{u}^{2}}(1-{{u}_{1}}th(Wt+k))},
\end{equation}
Выраженная в системе координат $s'$ эта скорость будет равна согласно \eqref{102}
\begin{equation} \label{640,2}
{{{v}'}_{1}}=\frac{v_{1}^{*}-{{u}_{1}}}{1-{{u}_{1}}v_{1}^{*}}=\frac{th(Wt+k)-{{u}_{1}}}{1-th(Wt+k){{u}_{1}}}
\end{equation}
\begin{equation} \label{640,15}
v'_{2}=0
\end{equation}
\begin{equation} \label{640,3}
{{{v}'}_{3}}=-\frac{\sqrt{1-{{v}^{*2}}}{{u}_{3}}}{1-{{u}_{1}}v_{1}^{*}}=-\frac{{{u}_{3}}}{ch(Wt+k)-{{u}_{1}}sh(Wt+k)}
\end{equation}
При этом в общем преобразовании ЛМН вращение системы $s'$ относительно $s$ описывается уравнениями
\begin{equation} \label{640,1}
r_1=r'_1\cos \phi +r'_3\sin \phi,\,\,\,\,
r_2={r'}_{2},\,\,\,\,
r_3=-r'_1 \sin \phi +r'_3\cos \phi,
\end{equation}
где собственный угол поворота Вигнера $\phi$, который  осуществляется в плоскости осей 1 и 3 в направлении $3\to 1$ согласно формулам \eqref{129}, \eqref{132}, \eqref{132,3} удовлетворяет соотношениям
\begin{equation} \label{641}
\sin {{\phi }}=\frac{{{u}_{1}}sh(Wt+k)-\frac{1-\sqrt{1-{{u}^{2}}}}{{{u}^{2}}}{{u}_{1}}\left( ch(Wt+k)-1 \right)}{ch(Wt+k)-{{u}_{1}}sh(Wt+k)+\sqrt{1-{{u}^{2}}}}
\end{equation}

\begin{equation} \label{642}
\cos {{\phi }}=\frac{\left( \sqrt{1-{{u}^{2}}}+\frac{1-\sqrt{1-{{u}^{2}}}}{{{u}^{2}}}u_{1}^{2} \right)ch(Wt+k)-{{u}_{1}}sh(Wt+k)+1-\frac{1-\sqrt{1-{{u}^{2}}}}{{{u}^{2}}}u_{1}^{2}}{ch(Wt+k)-{{u}_{1}}sh(Wt+k)+\sqrt{1-{{u}^{2}}}}
\end{equation}
\begin{equation} \label{642,1}
tg\frac{\phi}{2}=\frac{{{u}_{3}}sh\,(Wt+k)}{\left( 1+\sqrt{1-{{u}^{2}}} \right)\left( ch\,(Wt+k)+1 \right)-{{u}_{1}}sh\,(Wt+k)}
\end{equation}
Зная параметры общего преобразования ЛМН $\mathbf{v}(t)$ и $\phi_B(t)$, тем самым равноускоренное движение полностью задано. Данные функции времени $t$ определяются в отличие от динамического решения не 2, а 3 параметрами $k$, $u_1$, $u_3$. Подставляя $t=0$ в уравнения \eqref{638}, \eqref{640}  получим

\begin{equation} \label{647}
v_{1}(0)=\frac{\sqrt{1-{{u}^{2}}}th\,k-{{u}_{1}}}{1-{{u}_{1}}th\,k}+\frac{\left( 1-\sqrt{1-{{u}^{2}}} \right)u_{1}^{2}th\,k}{{{u}^{2}}(1-{{u}_{1}}th\,k)}
\end{equation}

\begin{equation} \label{648}
v_{3}(0)=-\frac{{{u}_{3}}}{1-{{u}_{1}}th\,k}+\frac{\left( 1-\sqrt{1-{{u}^{2}}} \right){{u}_{1}}{{u}_{3}}th\,k}{{{u}^{2}}(1-{{u}_{1}}th\,k)}
\end{equation}

Начальный угол вращения Вигнера вокруг оси 2 из \eqref{642,1} равен

\begin{equation} \label{649}
\phi(0)=2\arctg{\frac{{{u}_{3}}sh\,k}{\left( 1+\sqrt{1-{{u}^{2}}} \right)\left( ch\,k+1 \right)-{{u}_{1}}sh\,k}}
\end{equation}
Смысл постоянных $k$, $u_1$, $u_3$ в лабораторной системе отсчёта определяется из системы трёх независимых уравнений \eqref{647}-\eqref{649}. Таким образом вместо $k$, $u_1$, $u_3$ можно пользоваться двумя произвольными постоянными начальными скоростями $v_{1}(0)$ и $v_{3}(0)$ в направлении осей 1 и 3 и начальным собственным поворотом Вигнера $\phi (0)$. Этот угол определяет начальную ориентацию системы $s'$ относительно $S$ и его без потери общности можно положить равным нулю. Этот выбор соответствует значению $k=0$. Тогда \eqref{638}, \eqref{640} будут определяться 2 параметрами $u_1=-v_1(0)$, $u_3=-v_3(0)$ и будут соответственно равны
\begin{equation} \label{649,1}
v_1  = \frac{{\sqrt {1 - v^2(0) } thWt + v_1(0) }}{{1 + v_1(0) thWt}} + \frac{{\left( {1 - \sqrt {1 - v^2(0) } } \right)v_1^2(0) thWt}}{{v^2(0) \left( {1 + v_1(0) thWt} \right)}}
\end{equation}
\begin{equation} \label{649,2}
v_3  =   \frac{{v_3(0) }}{{1 + v_1(0) thWt}} + \frac{{\left( {1 - \sqrt {1 - v^2(0) } } \right)v_1(0) v_3(0) thWt}}{{v^2(0) \left( {1 + v_1(0) thWt} \right)}}
\end{equation}

Данное преобразование ЛМН в равноускоренную систему отсчёта задаваемое его параметрами 
$\mathbf{v}(t)$ \eqref{638}-\eqref{640} и углом Вигнера \eqref{641}-\eqref{642,1} будет служить образцом для сравнения решения полученного с помошью обратной задачи кинематики. 

\subsubsection*{Взаимная компенсация вращения Вигнера и прецессии Томаса для равноускоренного движения}

Поскольку равноускоренное движение не обладает собственным вращением, то вращение Вигнера и прецессия Томаса должны компенсировать друг друга. Покажем это прямым расчётом.

Угловую скорость вращения Вигнера можно получить дифференцируя уравнение \eqref{641}. Получим
\begin{equation} \label{650}
{{\omega }}\cos {{\phi }}=\frac{W{{u}_{3}}\left[ \left( \sqrt{1-{{u}^{2}}}+\frac{1-\sqrt{1-{{u}^{2}}}}{{{u}^{2}}}u_{1}^{2} \right)ch(Wt+k)-{{u}_{1}}sh(Wt+k)+1-\frac{1-\sqrt{1-{{u}^{2}}}}{{{u}^{2}}}u_{1}^{2} \right]}{{{\left[ ch(Wt+k)-{{u}_{1}}sh(Wt+k)+\sqrt{1-{{u}^{2}}} \right]}^{2}}}
\end{equation}
Поделив это выражение на \eqref{642} получим
\begin{equation} \label{651}
{{\omega }}=\frac{W{{u}_{3}}}{ch(Wt+k)-{{u}_{1}}sh(Wt+k)+\sqrt{1-{{u}^{2}}}}
\end{equation}
Вычислим сейчас угловую скорость прецессии Томаса. Первый множитель в её определении будет равен
\begin{equation} \label{652}
\frac{1-\sqrt{1-{{v}^{2}}}}{{{v}^{2}}\sqrt{1-{{v}^{2}}}}=\frac{1}{\sqrt{1-{{u}^{2}}}}\frac{{{\left( ch(Wt+k)-{{u}_{1}}sh(Wt+k) \right)}^{2}}}{ch(Wt+k)-{{u}_{1}}sh(Wt+k)+\sqrt{1-{{u}^{2}}}}
\end{equation}
Дифференцируя параметр скорости получим
\begin{equation} \label{653}
\dot{v}_{1}=\frac{W\sqrt{1-{{u}^{2}}}}{{{\left( ch(Wt+k)-{{u}_{1}}sh(Wt+k) \right)}^{2}}}\left( 1-\frac{\left( 1-\sqrt{1-{{u}^{2}}} \right)u_{1}^{2}}{{{u}^{2}}} \right)
\end{equation}
\begin{equation} \label{654}
\dot{v}_{2}=0
\end{equation}
\begin{equation} \label{655}
\dot{v}_{3}=-\frac{\sqrt{1-{{u}^{2}}}\left( 1-\sqrt{1-{{u}^{2}}} \right)}{{{u}^{2}}}\frac{W{{u}_{1}}{{u}_{3}}}{{{\left( ch(Wt+k)-{{u}_{1}}sh(Wt+k) \right)}^{2}}}
\end{equation}
Второй множитель является векторным произведением и равен учитывая \eqref{653}, \eqref{655}, \eqref{638} и  \eqref{640}
\begin{equation} \label{656}
{{({{\mathbf{v}}}\times {{\mathbf{\dot{v}}}})}_{2}}=v_{3}\dot{v}_{1}-v_{1}\dot{v}_{3}=-\frac{W\sqrt{1-{{u}^{2}}}{{u}_{3}}}{{{(ch(Wt+k)-{{u}_{1}}sh(Wt+k))}^{2}}}
\end{equation}
Следовательно перемножая \eqref{652} и \eqref{656} получим, что прецессия Томаса равна
\begin{equation} \label{657}
{{ {{\Omega }_{T}} }}=-\frac{W{{u}_{3}}}{ch(Wt+k)-{{u}_{1}}sh(Wt+k)+\sqrt{1-{{u}^{2}}}}
\end{equation}
Сравнивая формулы \eqref{651} и \eqref{657} видно, что в сумме собственная прецессия Томаса и собственное вращение Вигнера системы $s$ дают нуль как это и должно быть. В том же случае, если система $s$ обладает собственной угловой скоростью, прецессия Томаса и вращение Вигнера друг друга, разумеется, не компенсируют.

\subsubsection*{Определение равноускоренного движения с помощью решения обратной задачи кинематики} 
Рассмотрим теперь применение вышеизложенной схемы решения обратной задачи к криволинейному равноускоренному движению.      
  Уравнение \eqref{557} для оси 1 есть
 \begin{equation} \label{2.4}
  \dot{v\,'}_1=W(1-{v'}_{1}^{2})
  \end{equation}                                                 
 Интегрируя это уравнение получим
  \begin{equation} \label{2.5}     
 {{v'}_{1}}(t)=\ th(\,Wt+\kappa ), 
 \end{equation}
 где 
       \begin{equation} \label{2.6}
          th\kappa ={{v'}_{1}}(0) 
              \end{equation}
  Проецируя уравнение \eqref{557} на ось 3 перпендикулярную ускорению  получим с учётом \eqref{2.5}, что
              \begin{equation} \label{2.7}  
                {{\dot{v'}}_{3}}+W\ th\,(Wt+\kappa )\,{{v'}_{3}}=0,
                \end{equation}  
  Отсюда
  \begin{equation} \label{2.8}      
    {{v'}_{3}}(t)={{v'}_{3}}(0)\frac{ch\,\kappa }{ch\,(Wt+\kappa )},
     \end{equation}                                        
   \begin{equation} \label{2.12} 
 {{v'}^{2}}=\frac{\ {v'}_{3}^{2}(0)c{{h}^{2}}\kappa +s{{h}^{2}}(Wt+\kappa )}{c{{h}^{2}}(Wt+\kappa )}
  \end{equation} 
\begin{equation} \label{2.12,3}
\sqrt{1-{{{{v}'}}^{2}}}=\frac{\sqrt{1-{{{{v}'}}^{2}}(0)}\ ch\kappa}{ch\,(Wt+\kappa )}
\end{equation} 
Подставляя эти значения в выражение \eqref{558} для вращения Вигнера получим ($\mathbf{\Omega}'=0$)
 \begin{equation} \label{2.18} 
 \dot{\phi}={\omega' }=-\frac{1-\sqrt{1-{{{{v}'}}^{2}}}}{{{{{v}'}}^{2}}}{v'}_3 {W}=-\frac{Wch\kappa \ {{v'}_{3}}(0)}{\ ch\ (Wt+\kappa )+\sqrt{1-{v'}_{3}^{2}(0)\ {{ch}^{2}}\kappa }\ }
  \end{equation} 
 Интегрируя это выражение получим, считая, что начальный угол Вигнера равен нулю
    \begin{equation} \label{2.20,1}
   \phi =2arctg\,\frac{{{e}^{\,Wt\,+\kappa }}+\sqrt{1-{{{{v_3}'}}^{2}}(0) ch\kappa }}{{{{{v}'}}_{3}}(0)\, ch^2 \kappa }-2arctg\,\frac{{{e}^{\,\kappa }}+\sqrt{1-{{{{v_3}'}}^{2}}(0)ch^2 \kappa }}{{{{{v}'}}_{3}}(0)\,ch\kappa }.
      \end{equation}   
                                     
       \subsubsection*{Сравнение двух определений равноускоренного движения} 
Сравним теперь решения  \eqref{2.5}, \eqref{2.8} с уже известными соответственно \eqref{640,2}, \eqref{640,3} не предполагая заранее, что система $s'$ ориентирована  также как и $S$, т.е. не считая, что начальный угол поворота Вигнера равен нулю. Нетрудно заметить, что даже в таком общем случае эти решения совпадают (как это и должно быть) при следующем выборе постоянных
        \begin{equation} \label{2.19} 
       \kappa =k-arth\ {{u}_{1}}
       \end{equation} 
       \begin{equation} \label{2.20} 
       {{{v}'}_{3}}(0)=-\frac{{{u}_{3}}}{chk-{{u}_{1}}shk}
       \end{equation} 
   Можно также убедиться в тождественности \eqref{2.18} и \eqref{651} при условиях \eqref{2.19}, \eqref{2.20}. Следовательно и интегрирование \eqref{2.18} приводит с точностью до постоянной \eqref{649} определяющей начальную ориентацию системы $s'$ относительно $s$ к вигнеровскому углу являющемуся решением \eqref{641}-\eqref{642,1}. Таким образом два кинематических решения определяющих равноускоренное движение полностью совпадают.

   \subsection*  {11. Выражение полученных формул через параметр $\mathbf{v}'$}
  
  Найденные с помощью уравнений \eqref{557}, \eqref{558} параметры общего преобразования ЛМН $\mathbf{v}'$ и матрица вращения $a^{\alpha\beta}$ однозначно определяют это преобразование.  Возникает естественная задача о применении именно этих переменных вместо $\mathbf{v}$ ко всем ранее выведенным формулам.
Удобство данной замены состоит в том, что для параметра $\mathbf{v}'$, как было показано в разделе 9, существует простое по форме дифуравнение \eqref{557}. Простота  уравнений \eqref{557}, \eqref{558} для параметров неинерциальной системы отсчёта $s'$  даёт повод думать, что запись преобразования ЛМН  через параметр $\mathbf{v}'$ именно в форме \eqref{559}, \eqref{560} будет  более употребительна в теории. С учётом \eqref{548} общее преобразование ЛМН можно записать в виде
           \begin{equation} \label{559}       
                  T=\frac{{{{{v}'}}^{\beta }}{{{{r}'}}^{\beta }}}{\sqrt{1-{{{{v}'}}^{2}}}}+\int\limits_{0}^{t}{\frac{dt}{\sqrt{1-{{{{v}'}}^{2}}}}}                             
                  \end{equation}    
                  \begin{equation} \label{560}
                  {{R}^{\alpha }}={{a}^{\beta \alpha }}\left\{ {{{{r}'}}^{\beta }}+\frac{1-\sqrt{1-{{{{v}'}}^{2}}}}{{{{{v}'}}^{2}}\sqrt{1-{{{{v}'}}^{2}}}}{{{{v}'}}^{\beta }}{{{{v}'}}^{\gamma }}{{{{r}'}}^{\gamma }} \right\}+\int\limits_{0}^{t}{\frac{{{a}^{\beta \alpha }}{{{{v}'}}^{\beta }}dt}{\sqrt{1-{{{{v}'}}^{2}}}}}         
\end{equation}    
   Вычисляя дифференциалы от \eqref{559}, \eqref{560} получим
     \begin{equation} \label{11.1} 
     dT=\left\{ \frac{1+\mathbf{{W}'{r}'}+(\mathbf{{v}'}\times \mathbf{{\Omega }')}\ \mathbf{{r}'}}{\sqrt{1-{{{{v}'}}^{2}}}} \right\}\ dt+\frac{\mathbf{{v}'}d\mathbf{{r}'}}{\sqrt{1-{{{{v}'}}^{2}}}},
          \end{equation}
     \begin{equation} \label{11.2} 
     d\,{{R}^{\alpha }}={{a}^{\mu \alpha }}\delta \,{{{R}'}^{\mu }},
          \end{equation}
          где
\[\delta \,\mathbf{{R}'}=\left\{ \frac{(1+\mathbf{{W}'{r}'})\,\mathbf{{v}'}}{\sqrt{1-{{{{v}'}}^{2}}}}+\frac{1-\sqrt{1-{{{{v}'}}^{2}}}}{{{{{v}'}}^{2}}\sqrt{1-{{{{v}'}}^{2}}}}\left[ (\mathbf{{v}'}\times \mathbf{{\Omega }')}\ \mathbf{{r}'} \right]\,\mathbf{{v}'}+\mathbf{{\Omega }'}\times \mathbf{{r}'} \right\}dt+d\mathbf{{r}'}+\]
\begin{equation} \label{11.3}
+\frac{1-\sqrt{1-{{{{v}'}}^{2}}}}{{{{{v}'}}^{2}}\sqrt{1-{{{{v}'}}^{2}}}}(\mathbf{{v}'}d\mathbf{{r}'})\mathbf{{v}'} ,
  \end{equation}
 Преобразование обратное к \eqref{11.1}-\eqref{11.3} будет
 \begin{equation} \label{12.1} 
 dt=\frac{dT-\mathbf{{v}'}d\mathbf{{R}'}}{(1+\mathbf{{W}'{r}'})\sqrt{1-{{{{v}'}}^{2}}}} ,
   \end{equation}
 \[d\mathbf{{r}'}=\delta \mathbf{{R}'}+\frac{1-\sqrt{1-{{{{v}'}}^{2}}}}{{{{{v}'}}^{2}}\sqrt{1-{{{{v}'}}^{2}}}}(\mathbf{{v}'}\delta \mathbf{{R}'})\mathbf{{v}'}+\frac{(\mathbf{{\Omega }}'\times \mathbf{{r}'})(\mathbf{{v}'}\delta \mathbf{{R}'})}{(1+\mathbf{{W}'{r}'})\sqrt{1-{{{{v}'}}^{2}}}}-\]
  \begin{equation} \label{12.3} 
 -\left( \frac{{\mathbf{{v}'}}}{\sqrt{1-{{{{v}'}}^{2}}}}+\frac{\mathbf{{\Omega }}'\times \mathbf{{r}'}}{\sqrt{1-{{{{v}'}}^{2}}}(1+\mathbf{{W}'{r}'})} \right)\,dT \ ,
  \end{equation}
 где 
   \begin{equation} \label{12.2} 
 \delta {{{R}'}^{\alpha }}={{a}^{\alpha \mu }}d{{R}^{\mu }}.
   \end{equation}

Напишем теперь формулы преобразования энергии и импульса частицы, т.е. формулы, по которым можно определить данную величину в лабораторной системе, зная её же в собственной системе отсчёта. Эти формулы находятся непосредственно из общего преобразования ЛМН, в котором удобно предварительно заменить параметр $\mathbf{v}$ на $\mathbf{v}'$. 

  Используя уравнения \eqref{11.1}-\eqref{12.2} и правило вычисления  частных производных  при замене переменных, можно найти как будут преобразовываться при общем преобразовании ЛМН производные действия по координатам и времени, т.е. соответственно импульс и энергия. В результате получим, что, если в системе $S$ частица имела импульс $\mathbf{P}$ и энергию $E$, а в системе $s'$ - соответственно импульс $\mathbf{p'}$ и энергию $e'$, то эти величины связаны следующими соотношениями
  \begin{equation} \label{12.5} 
 E=\frac{{{e}'}}{(1+\mathbf{{W}'{r}'})\sqrt{1-{{{{v}'}}^{2}}}}+\left( \frac{{\mathbf{{v}'}}}{\sqrt{1-{{{{v}'}}^{2}}}}+\frac{{\mathbf{\Omega} }'\times \mathbf{{r}'}}{\sqrt{1-{{{{v}'}}^{2}}}(1+\mathbf{{W}'{r}'})} \right)\mathbf{{p}'},
     \end{equation}
  
   \begin{equation} \label{12.55} 
 {{{P}}^{\alpha }}={{a}^{\mu\alpha}}{{P}'^{\mu }},
   \end{equation}   
где  
  \begin{equation} \label{12.6}  \mathbf{{P}'}=\frac{\mathbf{{v}'}{e}'}{(1+\mathbf{{W}'{r}'})\sqrt{1-{{{{v}'}}^{2}}}}+\mathbf{{p}'}+\frac{1-\sqrt{1-{{{{v}'}}^{2}}}}{{{{{v}'}}^{2}}\sqrt{1-{{{{v}'}}^{2}}}}(\mathbf{{v}'{p}'})\mathbf{{v}'}+\frac{\left( (\mathbf{\Omega}'\times \mathbf{{r}'})\mathbf{{p}'} \right)\mathbf{{v}'}}{(1+\mathbf{{W}'{r}'})\sqrt{1-{{{{v}'}}^{2}}}},      \end{equation} 
  и   
 \[{e}'=\left\{ \frac{1+\mathbf{{W}'{r}'}+(\mathbf{{v}'}\times \mathbf{\Omega }')\mathbf{{r}'}}{\sqrt{1-{{{{v}'}}^{2}}}} \right\}E-\frac{(1+\mathbf{{W}'{r}'})\mathbf{{v}'{P}'}}{\sqrt{1-{{{{v}'}}^{2}}}}-\frac{1-\sqrt{1-{{{{v}'}}^{2}}}}{{{{{v}'}}^{2}}\sqrt{1-{{{{v}'}}^{2}}}}\left[ \mathbf{{r}'}(\mathbf{{v}'}\times \mathbf{\Omega }') \right]\mathbf{{v}'{P}'}-\]
 \begin{equation} \label{12.7}  
 -(\mathbf{\Omega }'\times \mathbf{{r}'})\mathbf{{P}'}, 
       \end{equation} 
  \begin{equation} \label{12.8}      
\mathbf{{p}'}=-\frac{\mathbf{{v}'}E}{\sqrt{1-{{{{v}'}}^{2}}}}+\mathbf{{P}'}+\frac{1-\sqrt{1-{{{{v}'}}^{2}}}}{{{{{v}'}}^{2}}\sqrt{1-{{{{v}'}}^{2}}}}(\mathbf{{v}'{P}'})\mathbf{{v}'},    
     \end{equation}   
 где
 \begin{equation} \label{12.9} 
 {{{P}}'^{\alpha }}={{a}^{\alpha\mu}}{{P}^{\mu }},
   \end{equation}   
         
Очевидно, что, если система $s'$ является инерциальной ($\mathbf{W}'=0$, $\mathbf{\Omega }'=0$ ), то формулы \eqref{12.5}-\eqref{12.8} переходят в формулы стандартного преобразования импульса-энергии. Убедиться в непротиворечивости данных соотношений можно и другим способом: подставляя в \eqref{12.5}, \eqref{12.6} значения энергии равной массе частицы  $e'=m$ и импульса $\mathbf{p'}=0$ покоящейся в начале системы $s'$ ($\mathbf{r'}=0$, $\mathbf{p'}=0$, $d\mathbf{r'}=0$) частицы. В этом случае в системе $S$ энергия и импульс частицы будут совпадать с общепринятыми значениями. 

Отсюда можно найти формулу для преобразования скоростей. Она имеет следующий вид
  \begin{equation} \label{11.4} 
{{U}^{\alpha }}={{a}^{\beta \alpha }}{{{U}'}^{\beta }}
       \end{equation}    
где
 \begin{equation} \label{11.5} 
\mathbf{{U}'}=\frac{\sqrt{1-{{{{v}'}}^{2}}}\mathbf{{u}'}+\frac{1-\sqrt{1-{{{{v}'}}^{2}}}}{{{{{v}'}}^{2}}}(\mathbf{{u}'{v}'})\mathbf{{v}'}+\mathbf{{v}'}+(\mathbf{{W}'{r}'})\mathbf{{v}'}+\sqrt{1-{{{{v}'}}^{2}}}\mathbf{{\Omega }'}\times \mathbf{{r}'}-\frac{1-\sqrt{1-{{{{v}'}}^{2}}}}{{{{{v}'}}^{2}}}\left[ \mathbf{{r}'}(\mathbf{{\Omega }'}\times \mathbf{{v}'}) \right]\mathbf{{v}'}}{1+\mathbf{{v}'{u}'}+\mathbf{{r}'{W}'}-\mathbf{{r}'}(\mathbf{{\Omega }'}\times \mathbf{{v}'})}
       \end{equation}
           Рассмотрим теперь движение точек пространства системы отсчёта $s'$ непосредственно прилегающих к началу системы координат относительно лабораторной системы $S$. Если $\mathbf{u'}=0$, то $\mathbf{U'}$ будет равно
   \begin{equation} \label{11.6} 
\mathbf{{U}'}=\frac{\mathbf{{v}'}+(\mathbf{{W}'{r}'})\mathbf{{v}'}+\sqrt{1-{{{{v}'}}^{2}}}\mathbf{{\Omega }'}\times \mathbf{{r}'}-\frac{1-\sqrt{1-{{{{v}'}}^{2}}}}{{{{{v}'}}^{2}}}\left[ \,\mathbf{{r}'}(\mathbf{{\Omega }'}\times \mathbf{{v}'}) \right]\,\mathbf{{v}'}}{1+\mathbf{{r}'}\,\mathbf{{W}'}-\mathbf{{r}'}(\mathbf{{\Omega }'}\times \mathbf{{v}'})}
       \end{equation}
       Обратное преобразование скорости к \eqref{11.6} есть
    \begin{equation} \label{11.16}                             \mathbf{{u}'}=(1+\mathbf{{W}'{r}'})\frac{\sqrt{1-{{{{v}'}}^{2}}}\mathbf{{U}'}+\frac{1-\sqrt{1-{{v}^{2}}}}{{{v}^{2}}}(\mathbf{{v}'}\,\mathbf{{U}'})\mathbf{{v}'}-\mathbf{{v}'}}{1-\mathbf{{v}'{U}'}}-\mathbf{{\Omega }'}\times \mathbf{{r}'} 
\end{equation}
Для малых  $\mathbf{r}'$ получим
        \begin{equation} \label{11.7} 
\mathbf{{U}'}=\mathbf{{v}'}+\sqrt{1-{{{{v}'}}^{2}}}\mathbf{{\Omega }'}\times \mathbf{{r}'}-\frac{\sqrt{1-{{{{v}'}}^{2}}}(1-\sqrt{1-{{{{v}'}}^{2}}})}{{{{{v}'}}^{2}}}\left[ (\mathbf{{\Omega }'}\times \mathbf{{r}'})\mathbf{{v}'} \right]\,\mathbf{{v}'}
  \end{equation}
Обобщение уравнения \eqref{37} на случай произвольного вращения системы отсчёта $s'$ даёт уравнение
       \begin{equation} \label{11.8} 
     	{{v}^{\alpha }}={{V}^{\alpha }}-\frac{{{V}^{\beta }}{{a}^{\gamma \beta }}{{{{r}'}}^{\gamma }}{{{\dot{V}}}^{\alpha }}}{\sqrt{1-{{V}^{2}}}}
\end{equation}
Умножая это уравнение на ${{a}^{\mu \alpha }}$ получим
     \[{{v'}^{\mu }}={{a}^{\mu \alpha }}{{v}^{\alpha }}={{a}^{\mu \alpha }}{{V}^{\alpha }}-\frac{{{V}^{\beta }}{{a}^{\mu \alpha }}{{a}^{\gamma \beta }}{{{{r}'}}^{\gamma }}{{{\dot{V}}}^{\alpha }}}{\sqrt{1-{{V}^{2}}}}={{{V'}}^{\mu }}-\frac{{{{{V'}}}^{\gamma }}{{{{r}'}}^{\gamma }}{{a}^{\mu \alpha }}({{{\dot{a}}}^{\varepsilon \alpha }}{{{{V'}}}^{\varepsilon }}+{{a}^{\varepsilon \alpha }}{{\dot{V'}}^{\varepsilon }})}{\sqrt{1-{{V}^{2}}}}={{{V}'}^{\mu }}-\]
       \begin{equation} \label{11.9} 
     -\frac{(\mathbf{{V'}{r}'}){{a}^{\mu \alpha }}({{e}^{\xi \eta \varepsilon }}{{{{\omega' }}}^{\eta }}{{a}^{\xi \alpha }}{{{{V'}}}^{\varepsilon }}+{{a}^{\varepsilon \alpha }}{{{\dot{V'}}}^{\varepsilon }})}{\sqrt{1-{{V}^{2}}}}={{V'}^{\mu }}-\frac{(\mathbf{{V'}{r'}})({{e}^{\mu \eta \varepsilon }}{{{{\omega' }}}^{\eta }}{V'^{\varepsilon }}+{\dot{V'}^{\mu }})}{\sqrt{1-{{V}^{2}}}}
\end{equation}
или в векторном виде
      \begin{equation} \label{11.10} 
     \mathbf{{v}'}=\mathbf{{V'}}-\frac{(\mathbf{{V'}{r}'})(\mathbf{{\dot{V'}}}+\boldsymbol{{\omega' }}\times \mathbf{{V}'})}{\sqrt{1-{{{{V'}}}^{2}}}}   
      \end{equation}
где ${V'^{\mu }}=a^{\mu \alpha }V^{\alpha }$.   
Заметим, что если учесть, что пренебрегая первыми степенями по  $\mathbf{r}'$ выполняется уравнение     
            \begin{equation} \label{11.11}                                                                \mathbf{\dot{v}'}=\frac{1}{\sqrt{1-{{V'}^{2}}}}\frac{d\,\mathbf{V'}}{dT},                                                    \end{equation}              
то из \eqref{557}, \eqref{558} следуют дифференциальные уравнения для решения обратной задачи кинематики для случая, когда собственные ускорение $\mathbf{{W}'}$ и угловая скорость $\mathbf{{\Omega }'} $ заданы как функции лабораторного времени $T$
       \begin{equation} \label{11.12}                                        \frac{d\,\mathbf{{V}'}}{dT}=\sqrt{1-{{V}^{2}}}\left[ {\mathbf{{W}'}} \right.-(\mathbf{{W}'{V}'})\mathbf{{V}'}-\mathbf{{\Omega }'}\times \left. {\mathbf{{V}'}} \right]                         \end{equation}                
   \begin{equation} \label{11.13} 
   \boldsymbol{{\omega }'}=\mathbf{{\Omega }'}-\frac{1-\sqrt{1-{{{{V}'}}^{2}}}}{{{{{V}'}}^{2}}}\mathbf{{V}'}\times \mathbf{{W}'}
         \end{equation}                 
Подставив тогда в \eqref{11.10} эти выражения получим, что
      \begin{equation} \label{11.14}
      \mathbf{{v}'}=\mathbf{{V}'}-(\mathbf{{V}'{r}'})\,\left[ \mathbf{{W}'}-\mathbf{{V}'}(\mathbf{{V}'}\,\mathbf{{W}'})+\frac{1-\sqrt{1-{{V}^{2}}}}{{{V}^{2}}\sqrt{1-{{V}^{2}}}}\mathbf{{V}'}\times (\mathbf{{V}'}\times \mathbf{{W}'})+\frac{1-\sqrt{1-{{V}^{2}}}}{\sqrt{1-{{V}^{2}}}}\mathbf{{\Omega }'}\times \mathbf{{V}'} \right]   \end{equation}  
Если теперь подставить это выражение в \eqref{11.7}, то получим окончательно
       \[\mathbf{{U}'}=\mathbf{{V}'}-(\mathbf{{V}'{r}'})\left[ \left( \frac{1-\sqrt{1-{{V}^{2}}}}{{{V}^{2}}\sqrt{1-{{V}^{2}}}}-1 \right)\mathbf{{V}'}(\mathbf{{V}'}\mathbf{{W}'})-\frac{1-2\sqrt{1-{{V}^{2}}}}{\sqrt{1-{{V}^{2}}}}\mathbf{{W}'}+\frac{1-\sqrt{1-{{V}^{2}}}}{\sqrt{1-{{V}^{2}}}}\mathbf{{\Omega }'}\times \mathbf{{V}'} \right]+\]
       \begin{equation} \label{11.15} 
+\sqrt{1-{{{{V}'}}^{2}}}\mathbf{{\Omega }'}\times \mathbf{{r}'}-\frac{\sqrt{1-{{{{V}'}}^{2}}}(1-\sqrt{1-{{{{V}'}}^{2}}})}{{{{{V}'}}^{2}}}\left[ (\mathbf{{\Omega }'}\times \mathbf{{r}'})\mathbf{{V}'} \right]\,\mathbf{{V}'}
\end{equation}
     
Выразим теперь обратное преобразование ЛМН через переменную $\mathbf{{V}'}$. Форма уравнения для временной координаты $t$ \eqref{60} при этом в силу её скалярности не изменится. Поэтому в правой части \eqref{60} достаточно везде поставить штрихи и заменить $\mathbf{\dot{V}'}$ на \eqref{11.12}. В результате получим 
\begin{equation} \label{11.19} 
t=\int_{0}^{T}{\sqrt{1-{{{{V}'}}^{2}}}}dT-\frac{\mathbf{{L}'{V}'}}{\sqrt{1-{{{{V}'}}^{2}}}}+\frac{(\mathbf{{L}'{V}'})(\mathbf{{L}'{W}'})}{1-{{{{V}'}}^{2}}}-\frac{(\mathbf{{L}'{V}'})(\mathbf{{L}'}(\mathbf{{\Omega }'}\times \left. \mathbf{{V}'}) \right])}{1-{{{{V}'}}^{2}}},
 \end{equation}
где 
 \begin{equation} \label{11.20} 
    {{{L}'}^{\alpha }}={{a}^{\alpha \beta }}\left( {{R}^{\beta }}-\int\limits_{0}^{T}{{{V}^{\beta }}dT} \right), \,\,\,    {{{V}'}^{\alpha }}= {a}^{\alpha \beta } {V}^{\beta }          
\end{equation}
Здесь $\mathbf{{L}'}$ и $\mathbf{{V}'}$ есть соответственно длина и скорость в новой системе координат лабораторной инерциальной системы отсчёта $S$:($T$,$\mathbf{R}$). Данная система координат в каждый момент лабораторного времени мгновенно совпадает с системой координат $R'^{\alpha}$ вращающейся системы отсчёта согласно условию $R'^{\alpha}={a}^{\alpha \beta }R^{\beta}$.

Уравнения для координаты $\mathbf{r'}$ можно получить из \eqref{61} аналогично уравнениям \eqref{551}, \eqref{553}, \eqref{11.10}. Для этого необходимо над всеми символами  поставить штрихи и произвести везде замену
     \begin{equation} \label{11.17}  
     \mathbf{\dot{V}'}\to \mathbf{{\dot{V}}'}+\boldsymbol{{\omega }'}\times \mathbf{{V}'}.
  \end{equation}                                                                                                               
После этого нужно подставить вместо $\mathbf{{\dot{V}}'}$ и $\boldsymbol{{\omega }'}$ их значения согласно соответственно \eqref{11.12}, \eqref{11.13}. В результате получим
\[\mathbf{{r}'}=\mathbf{{L}'}+\frac{1-\sqrt{1-{{{{V}'}}^{2}}}}{{{{{V}'}}^{2}}\sqrt{1-{{{{V}'}}^{2}}}}(\mathbf{{L}'{V}'})\mathbf{{V}'}+\frac{\left( 1-2\sqrt{1-{{{{V}'}}^{2}}} \right)\left( 1-\sqrt{1-{{{{V}'}}^{2}}} \right)}{{{{{V}'}}^{2}}{{\sqrt{1-{{{{V}'}}^{2}}}}^{\,3}}}(\mathbf{{L}'{V}'})(\mathbf{{L}'{W}'})\mathbf{{V}'}+\]
 \[+\frac{{{\left( 1-\sqrt{1-{{{{V}'}}^{2}}} \right)}^{2}}\left( 2-{{{{V}'}}^{2}}-4\sqrt{1-{{{{V}'}}^{2}}} \right)}{2{{{{V}'}}^{4}}{{(1-{{{{V}'}}^{2}})}^{2}}}{{(\mathbf{{L}'{V}'})}^{2}}(\mathbf{{W}'{V}'})\mathbf{{V}'}+\frac{{{(1-\sqrt{1-{{{{V}'}}^{2}}})}^{\,3}}}{2{{{{V}'}}^{2}}{{(1-{{{{V}'}}^{2}})}^{2}}}{{(\mathbf{L{V}'})}^{2}}\left( \mathbf{{\Omega }'}\times \mathbf{{V}'} \right)-\]
  \begin{equation} \label{11.18} 
-\frac{{{\left( 1-\sqrt{1-{{{{V}'}}^{2}}} \right)}^{2}}}{{{{{V}'}}^{2}}{{\sqrt{1-{{{{V}'}}^{2}}}}^{\,3}}}(\mathbf{{L}'{V}'})\left[ \mathbf{{L}'}(\mathbf{{\Omega }'}\times \mathbf{{V}'}) \right]\mathbf{{V}'}-\frac{{{(1-\sqrt{1-{{{{V}'}}^{2}}})}^{2}}\left( 1-2\sqrt{1-{{{{V}'}}^{2}}} \right)}{2{{{{V}'}}^{2}}{{(1-{{{{V}'}}^{2}})}^{2}}}{{(\mathbf{L{V}'})}^{2}}\mathbf{{W}'}
 \end{equation}
Для $\mathbf{{\Omega }'}=\mathbf{\Omega }_T$ данное общее обратное преобразование ЛМН \eqref{11.19}, \eqref{11.18} перейдёт в обратное специальное преобразование \eqref{60}, \eqref{61}. Аналогично специальному, общее обратное преобразование ЛМН \eqref{11.19}, \eqref{11.18} возможно пригодно также и для реальных систем отсчёта с максимальной жёсткостью. 

\subsection*{Итоги и выводы}

 Основные результаты данной статьи следующие.
   
В разделе 1 данной статьи было найдено преобразование скорости материальной точки \eqref{16}, \eqref{17} из лабораторной инерциальной системы отсчёта $S$  в движущуюся поступательно систему  $s$, собственное вращение которой равно частоте прецессии Томаса. Был показан также физический смысл параметра $\mathbf{v}\, (t)$ преобразования ЛМН \eqref{10}, \eqref{11}. Замечательно, что он оказался прост. Со скоростью $\mathbf{v}\, (t)$ относительно лабораторной системы отсчёта $S$ движутся точки системы координат невращающейся ($\mathbf{\Omega} =0$) системы отсчёта $k$, у которой начало и оси координат сопутствуют в каждый момент собственного времени и совпадают с началом и осями координат системы $s$. Также с её помощью $\mathbf{v}(t)$ вычисляется скорость \eqref{25}, \eqref{26} точек самой системы $s$. Точки жёсткой неинерциальной системы относительно $S$ двигаются неоднородно.

В разделе 2 предлагается  обратное специальное преобразование ЛМН (с точностью до квадратов собственных расстояний включительно) из лабораторной инерциальной системы $S$  в движущуюся систему отсчёта $s$ в двух формах \eqref{43}, \eqref{45} или \eqref{52}, \eqref{61}. Возможно, что это преобразование будет использоваться для расчётов других эффектов СТО возможных при жёстком неинерциальном движении. Данное преобразование применимо в случае, когда размеры идеально твёрдой системы отсчёта $s$ ограничены \eqref{38.83} и её движение предполагается достаточно плавным, без рывков. Оно имеет существенно нелинейный вид, зависящий от ускорения начала отсчёта. Прямыми и явными основными следствиями указанного преобразования являются а) рассинхронизация (в  уравнениях \eqref{58}, \eqref{60} она показана в фигурных скобках) в системе $s$  координатных часов этой системы отсчёта предварительно синхронизированных в лабораторной системе $S$ и б) нелинейное сокращение Лоренца в $S$ линейки системы $s$ \eqref{45}, \eqref{59}. Во втором порядке по собственным координатам эти эффекты зависят только от скорости и ускорения, но не от закона движения системы отсчёта $s$. Нетрудно заметить, что в случае собственного ускорения равного нулю обратное преобразование ЛМН переходит в обычное преобразование Лоренца.  Координатное время в системе отсчёта $s$ \eqref{58}, \eqref{60} вследствие относительности времени разумеется отличается от собственного времени начала отсчёта. Уравнения \eqref{45}, \eqref{59} противоречат обычно принятому лоренцеву сокращению, но хорошо согласуются с \cite{4}. Нелинейность сокращения Лоренца означает, что произвольно движущуюся систему отсчёта $s$  с точки зрения лабораторного наблюдателя  вообще говоря нельзя моделировать последовательностью мгновенно сопутствующих инерциальных систем отсчёта. Такое моделирование системы отсчёта $s$ возможно лишь для наблюдателя системы $s$. Качественно нелинейность сокращения Лоренца вполне объяснима тем, что точки жёсткой системы отсчёта двигаются неодинаково. Область оси координат системы $s$, прилегающая к её переднему концу, движется при прямолинейном разгоне $s$ с меньшей скоростью, чем скорость $V$ начала отсчёта. Вследствие этой причины область оси координат вблизи её передней точки относительно лабораторной системы отсчёта $S$ сокращается меньше чем область около начала отсчёта. Таким образом, общая длина ускоренного стержня, находящегося в процессе разгона вдоль своего направления должна быть несколько больше чем длина такого же, но инерциального стержня, все точки которого двигаются в данный момент времени с одинаковой скоростью $V$ \cite{4}. 

Кроме этих эффектов вычислены также неоднородность движения точек системы координат $s$ \eqref{63}, \eqref{64} и эффект неоднородности хода физического времени \eqref{66}, \eqref{67} в системе $s$ с точки зрения лабораторной системы $S$. Данные эффекты при условии \eqref{38.83} в первом порядке по координатам зависят также только от скорости и ускорения. Параметр  $\mathbf{v}\, (t)$ входящий в преобразование ЛМН связан со скоростью $\mathbf{V}$ начала отсчёта  приближённой общей формулой \eqref{37} с точностью до первой степени по собственной координате $\mathbf{r}$ включительно. Легко проверить, что для прямолинейного ускоренного движения данная общая формула переходит в известную из \cite{3}, \cite{4}.

В разделе 2 также было выдвинуто в некоторой степени обоснованное предположение,  что обратное преобразование ЛМН и формулы \eqref{37}, \eqref{53}, \eqref{62}, \eqref{63} пригодны для реальных систем отсчёта с максимально возможной в СТО жёсткостью. Под такой системой понимается система оси координат, которой изготовлены из материала с максимально возможной в СТО скоростью звука и начало которой двигается заданным образом, а остальные точки являются свободными. 

В разделе 3 показано, что матрица вращения $a^{\alpha \beta}$ имеет смысл собственной матрицы (т.е. жёсткое вращение $s'$ происходит относительно $s$, а не относительно лабораторной системы $S$). Она удовлетворяет условиям ортогональности \eqref{74}, равенствам "уничтожения"  \eqref{76} и условию \eqref{79,2} или  \eqref{79,3}, которое связывает матрицу вращения и вектор угловой скорости. Рассматриваемая из лабораторной системы $S$ эта матрица имеет вид \eqref{84} и подчиняется уравнению \eqref{89}, т.е. не является вращением. Собственная прецессия Томаса есть частный случай собственного вращения. Следовательно, можно сделать вывод, что не существует прецессии Томаса, понимаемой как чистое вращение относительно лабораторной системы отсчёта $S$, у тела, собственные размеры которого сохраняются. Простой векторный вид прецессия Томаса имеет только в сопутствующей системе отсчёта. Поэтому для сравнения различных уже имеющихся выражений для частоты прецессии Томаса необходимо выбирать не лабораторную, а собственную систему отсчёта. 

С помощью прямого и обратного закона сложения скоростей  \eqref{23}, \eqref{24} в разделах 4 и 5 было найдено преобразование собственной центроаффинной скорости (т.е. тензорного коэффициента между скоростью и координатой) измеряемой в системе $k$ в аффинную скорость измеряемую в лабораторной системе. В разделе 4 было вычислено прямое \eqref{232}, а в разделе 5 - обратное преобразование в систему $k$  \eqref{252}. Полученные формулы оказались, довольно громоздкими, но на практике интересна не сама аффинная скорость малой области, а её угловая скорость вращения главных осей и скорость её растяжения по главным осям. Для этих величин также были найдены законы преобразований \eqref{242}, \eqref{243} и \eqref{255}, \eqref{256}. Если в системе $s$ элемент сплошной среды не двигается, то в лабораторной системе $S$ его главные оси вращаются с угловой скоростью \eqref{275} и тензором скорости деформации \eqref{276}. Из \eqref{232} или \eqref{275}, \eqref{276} следует, что выражение дли прецессии Томаса в лабораторной системе отсчёта имеет не векторный, а общий (неантисимметричный) тензорный вид. Также в параграфе 4 было показано, что при неинерциальном движении и жёстком собственном вращении тела с частотой $\boldsymbol{\omega}$ в лабораторной системе существуют две плоскости, в которых это тело кинематически не деформируется: это плоскости перпендикулярные вектору скорости тела $\mathbf{V}$ и вектору $\boldsymbol{\omega} \times\mathbf{V}-\dot{\mathbf{V}}/(1-V^{2})$.

Полученные в разделах 4 и 5 преобразования аффинной скорости в разделе 6 были применены к вращению твёрдого тела. При этом оказалось, что если первоначально вращение тела относительно наблюдателя в лабораторной системе являлось жёстким, то при положении наблюдателя на периферии тела, оно двигается, вообще говоря, нежёстко \eqref{262}. Справедливо и обратное. Жёсткость движения тела сохраняется только для случая равномерного вращения тела. Собственная угловая скорость вращения главных осей   малой области твёрдого тела точек  для случая неравномерного вращения будет равна \eqref{260}. В случае, когда  тело ракручивается ($\mathbf{\dot{\Omega }}\|\mathbf{\Omega}$ ), локальная угловая скорость главных осей не зависит от его углового ускорения и равна \eqref{261}.  Первый член формулы \eqref{260} был найден  в  \cite{31} и в \cite[с. 59]{36}. Кроме того, если расстояние между двумя точками относительно наблюдателя находящегося в центре неравномерно вращающегося тела сохраняется, то при сдвиге наблюдателя это расстояние будет изменяться даже в ближней окрестности наблюдателя. Для твёрдого тела результирующий коэффициент собственного растяжения периферии \eqref{269} оказался совпадающим с известным значением. Совпадение рассчитанных локальных угловой скорости и относительного растяжения жёстко вращающегося твёрдого тела с уже известными значениями косвенно свидетельствует об истинности преобразования ЛМН и преобразования аффинной скорости \eqref{232}, \eqref{252}. Таким образом в п. 6 фактически было показано отсутствие парадокса Эренфеста, согласно которому в лабораторной системе периметр вращающегося диска якобы сокращается по Лоренцу и диск при увеличении радиальной скорости должен сжиматься. На самом деле кроме лоренцева сокращения имеется ещё и эйнштейновское растяжение материала диска, так, что комбинация этих эффектов приводит к тому, что относительно $S$ размеры диска не изменяются. 

В разделе 7 этой статьи, были доказаны единственность преобразования ЛМН и его форминвариантность при бусте. Преобразование ЛМН для идеально жёсткой системы отсчёта является единственным переходящим в обычное преобразование Лоренца, поскольку сводится к дифференциальному преобразованию Лоренца в случае, когда собственное ускорение и вращение относительно сопутствующей системы отсчёта  обладающей прецессией Томаса отсутствует. Кроме того, это преобразование также форминвариантно относительно произвольного буста, причём параметры преобразования изменяются согласно формулам \eqref{100},  \eqref{126}.  При таком бусте система отсчёта испытывает дополнительное собственное вращение Вигнера характеризующееся матрицей $b^{\alpha \beta}$ \eqref{117}. Угол поворота определяемый этой матрицей равен \eqref{129}, \eqref{132} или в другой форме \eqref{132,3}. Вычисление угла поворота Вигнера, полученного при доказательстве форминвариантности сошлось с уже известным (например \cite[формула (20)]{29}). 
Также показано, что вращение Вигнера и прецессия Томаса являются различными явлениями, хотя и тесно связанными \eqref{649.11}. 

В разделе 8 этой статьи была записана  4-мерная форма \eqref{332} общего преобразования ЛМН, где 4-векторы тетрады равны \eqref{333.1}. Характеристики жёсткой системы отсчёта $s'$ форминвариантны и имеют 4-мерный тетрадный смысл \eqref{339}, \eqref{340}. 

В 9 разделе были представлены дифференциальные уравнения \eqref{557}, \eqref{558} решающие обратную задачу релятивистской кинематики, т.е. восстановления параметров общего преобразования ЛМН по известным собственному ускорению и угловой скорости. Знание собственного ускорения и угловой скорости как функции собственного времени полностью определяет весь класс физически эквивалентных систем отсчёта путём решения простых по виду дифференциальных уравнений первого порядка. Таким образом, движение жёсткой системы $s'$ с известными характеристиками можно полностью определить решая обратную задачу кинематики, т.е. нелинейное векторное дифуравнение первого порядка \eqref{557} и систему \eqref{79,3}, где $\boldsymbol{\omega}'$ удовлетворяет \eqref{558}. К сожалению, данные дифференциальные уравнения решаются аналитически видимо для ограниченного класса неинерциальных систем отсчёта. Однако с развитием численных методов и технологий даже численное решение обратной задачи кинематики представляет интерес.
    
 В разделе 10 показано, что известное гиперболическое движение частицы под действием постоянной силы в $S$ не имеет отношения к задаче о равноускоренном движении. Находится также выражение 4-силы в системе $S$ через величины относящиеся к собственной системе отсчёта \eqref{12.18}. Наиболее общее преобразование в равноускоренную систему отсчёта можно найти как решением обратной задачи кинематики \eqref{557}, \eqref{558}, так и простым лоренцевским бустом, как было сделано в разделе 7. Это свидетельствует как о правильности предлагаемого порядка решения обратной задачи, так и о корректности преобразования ЛМН. Кроме того, были вычислены  параметры равноускоренной системы отсчёта \eqref{638}-\eqref{640}, \eqref{642,1}. В качестве независимой проверки уравнения \eqref{649.11} были вычислены частоты прецессии Томаса \eqref{657} и вращения Вигнера \eqref{651} равноускоренной системы. Оказалось, что эти частоты взаимно компенсируются и результирующее вращение равноускоренной системы $s'$ равно нулю, как и должно быть.

В 11 разделе представлена более удобная запись общего преобразования ЛМН \eqref{559}, \eqref{560} и закон преобразования скорости \eqref{11.4}, \eqref{11.5} из системы $S$ в $s'$. Удобство заключается в том, что характеристиками движения тела будут матрица вращения и параметр $\mathbf{v'}(t)$ являющийся решением дифференциального уравнения \eqref{557} для обратной задачи кинематики. Этот параметр в свою очередь был выражен через $\mathbf{V'}(t)$ (уравнение \eqref{11.14}), являющийся скоростью начала отсчёта в новой системе координат $S$, которая повёрнута относительно старой с матрицей поворота $a^{\alpha \beta}$. Кроме того, обратное преобразование ЛМН также выражено через этот параметр $\mathbf{V'}(t)$  и характеристики произвольной жёсткой системы отсчёта $s'$. Поэтому обратное преобразование ЛМН \eqref{11.19}, \eqref{11.18} записанное в такой форме можно считать общим.

 Подытоживая можно сделать общий вывод о том, что общее преобразование ЛМН является корректным преобразованием в жёсткую неинерциальную систему отсчёта. Вытекающие из него физические следствия нетривиальны, согласуются друг с другом и с результатами полученными другими авторами. Данное преобразование следует положить в основу изложения СТО в привилегированных системах 4-координат. 

Автор выражает большую признательность профессору Н. Г. Мигранову за полезные обсуждения и поддержку.

\newpage
%
%
\end {document}